\documentclass[pra,a4paper,preprintnumbers,aps,superscriptaddress,twocolumn,floatfix,showpacs,showkeys]{emulateapj}
\usepackage{latexsym}
\usepackage{epsfig}
\usepackage{amssymb}
\usepackage{graphicx} 
\usepackage{natbib} 

\begin{document}
\title{ISW-LSS cross-correlation in coupled Dark Energy models with massive neutrinos}
% or
%\title{Cosmology with an Anisotropic Equation of State and the CMB Anomalies}
% or
%\title{Dark Energy with an Anisotropic Equation of State}
% or
%\title{Cosmology with an Anisotropic Equation of State}
%\date{\today}

\author{Roberto Mainini$^1$ and David F. Mota$^2$}
\affil{$^1$ INAF -- Osservatorio Astronomico di Roma, via Frascati 33, 00040, Monte 
Porzio Catone, (RM), Italy\\
$^2$ Institute of Theoretical Astrophysics, University of Oslo, Box 1029, 
0315 Oslo, Norway} 

\begin{abstract}
We provide an exhaustive analysis of the Integrated Sach--Wolfe effect (ISW)
in the context of coupled Dark Energy cosmologies where a component of massive 
neutrinos is also present. We focus on the effects of both the 
coupling between Dark Matter and Dark Energy and of the neutrino mass on the 
cross-correlation between
galaxy/quasar distributions and ISW effect. 
We provide a simple expression to appropriately rescale the galaxy bias when
comparing different cosmologies. 
Theoretical predictions of the cross--correlation function are then compared 
with observational data. 
We find that, while it is not possible to distinguish among the models at low 
redshifts, discrepancies between coupled models and $\Lambda$CDM 
increase with $z$.
In spite of this, current data alone seems not able to distinguish 
between coupled models and $\Lambda$CDM. 
However, we show that upcoming galaxy surveys will permit tomographic analysis
which allow to better discriminate among the models.
We discuss the effects on cross-correlation measurements of ignoring 
galaxy bias evolution, $b(z)$, and magnification bias correction and
provide fitting formulae for $b(z)$ for the cosmologies considered.
We compare three different tomographic schemes and investigate how the 
expected signal to noise ratio, $snr$, of the ISW--LSS cross--correlation 
changes when increasing the number of tomographic bins. The dependence of 
$snr$ on the area of the survey and the survey shot noise is also discussed.

\end{abstract}

\maketitle

\section{Introduction}
\label{sec:intro}
Several observations made over the recent years, related to a large extension 
to Large Scale Structures (LSS) and
anisotropies of the Cosmic Microwave Background (CMB)
as well as the magnitude--redshift relation for type Ia Supernovae  
have given us a convincing picture of the energy and matter density in
the Universe \citep{per99,rie98,spe03,teg04,lar10}.

Baryonic matter accounts for no more than 30$\%$ of the mass in galaxy 
clusters while the existence of a large clustered component of 
Dark Matter (DM) seems now firmly established, although its nature is 
still unknown.
However, they 
contribute to the total energy density of the Universe with
only a few percent and about 25$\%$ respectively.

No more than another few percent could be accounted for by massive neutrinos,
but only in the most favorable, but unlikely case. According to \citet{kri07}
(see also \citet{elg02}) 
the total mass of neutrinos cannot exceed the limit of 1.43 $eV$ 
(see, however,
\citet{lav09a,lav09b,kri10} for a recent analysis
on neutrino mass limits in coupled dark energy models). 
A very small part $(10^{-4})$ of the total energy density is due to massless
neutrinos and CMB radiation. 

The model suggested by observations is only viable if the remaining 75$\%$
is ascribed to the so--called Dark Energy (DE) responsible for the present
day cosmic acceleration. 

Although strongly indicated by the observations, the existence of DE is
 even more 
puzzling than DM. It can be identified with a cosmological constant $\Lambda$
or with a yet unknown dynamical component with negative pressure. 
On the other hand, its manifestation can be interpreted as a geometrical
property of the gravity on large scales resulting from a failure of 
General Relativity (GR) on those scales (see \citet{cop06} for 
a review). 
   
Within the context of GR, as an alternative to the cosmological constant,
DE is usually described as a scalar field $\phi$, 
self--interacting through a suitable potential $V(\phi)$, or a cosmic fluid
with negative pressure \citet{pee03} (see \citet{koi1,koi2,li} for alternatives and references therein).

Scalar fields naturally arise in particle physics. Furthermore,
if they are {\it
tracker} fields \citep{ste99}, {\it fine tunings} associated to the small value of the
present DE energy density can be significantly alleviated unlike 
the cosmological constant case. 

In addition to self--interaction, a scalar field can in principle be 
coupled to any other field present in nature. 
However, in order to drive the cosmic acceleration, 
its present time mass is expected to be, at least on large scales,  
$m_\phi \sim H_0 \sim 10^{-33} eV$ ($H_0$ being the present Hubble parameter).
Such a tiny mass gives rise to long--range interactions which could be tested
with fifth--force type experiments. 
Couplings to ordinary particles are strongly constrained by such a kind of 
experiments but limits on the DM coupling are looser (constraints on coupling
for specific models were obtained in 
\citet{mac04,ame03,oli05,lee06,guo07,mai07} from CMB,
N-body simulations and matter power spectrum analysis).

A possible common origin of DM and DE and/or a their direct coupling
\citep{wet95,ame99,gas02,bar00,chi03,rho03,far04}.
would ease one of the most critical problems in modern cosmology, 
the so--called {\it coincidence problem}:
why expansion started to accelerate just at the eve of our cosmic epoch, 
after decelerating during all epochs after inflation? Why DE and DM have
similar densities just now?
Because of the coupling DM and DE densities keep similar 
values during a long period and the only peculiar feature of the present epoch
is the recent overtaking of DM density by DE density.
  
If present, DM--DE coupling could have a relevant role in 
the cosmological evolution affecting not only the overall cosmic expansion
but also modifying the DM particles dynamics with  relevant
consequences on the growth of the matter density perturbations in both
linear and nonlinear regime (e.g., on halo density profiles, 
cluster mass function and its evolution, see
\citet{wan98,mai03a,mai03b,kly03,dol04,mac04,per04,mot04,oli06,nun05,motv04,
mao05,nun05,wan06,man06,nun06,dut07,mot07,mai08,sha08,mot08,mai09,bal09,win10,bal10}.
LSS is then a powerful probe of DE nature which permit to put 
significant constraints on DE parameters. Constraints often become even more
stringent when data from other probes are simultaneously taken into account.

CMB is another powerful probe of DE nature. 
In principle, by joining anisotropy 
and polarization data, DE parameters can be significantly constrained.
CMB and LSS probe the universe at different epochs and are therefore 
complementary to each other. Future data from high resolution CMB experiment
like PLANCK \footnote{http://www.rssd.esa.int/index.php?project=planck} and 
LSS surveys (EUCLID \footnote{http://www.euclid-imaging.net/}, LSST 
\footnote{http://www.lsst.org/lsst}, 
DES \footnote{https://www.darkenergysurvey.org/}, 
JDEM \footnote{http://jdem.gsfc.nasa.gov/}, etc)  will allow to 
constrain DE up to an unprecedented accuracy.

In this paper we will focus on the Integrated Sachs Wolfe (ISW) effect 
\citep{sac67}.
ISW effect
is a secondary anisotropy of the CMB and a direct signature of DE. 
The effect arises when a photon from the last scattering surface passes
through a 
time--dependent gravitational potential changing its energy so that additional
temperature anisotropies are generated. 
Decay of gravitational potentials may occur through cosmic curvature, 
in the presence of DE or in alternative gravity models.

Assuming General Relativity is the correct theory of gravity and 
that the Universe is spatially flat,
large--scale gravitational potentials do not evolve
significantly in the matter era. Cosmic acceleration, however, causes the 
gravitational potentials to decay making the ISW effect highly sensitive to
the presence of DE.

Though difficult to detect directly in the CMB,
ISW signal can be measured by cross--correlating the CMB with 
tracers of LSS and has recently been detected using WMAP data of CMB in 
combination with several LSS surveys at the 
the $\sim 3-4 \sigma$ confidence level providing
independent evidence for the existence of the DE (see \citet{gia08a,xia09}
and references therein).

Cross-correlation then 
provides a powerful method to discriminate among different DE models and,
in particular, to detect a possible interaction between DE and DM other than
investigate the clustering properties of DE on large scales.
If present, DM--DE coupling changes  both the 
scaling of the DM energy density and the growth rate of matter perturbations 
yielding a significant evolution of the metric potentials even 
in the matter era.

The aim of this paper is to provide an exhaustive analysis of the ISW 
effect in the contest of the {\it so called} coupled DE cosmologies 
\citep{ame00} mainly 
focusing on the effects of the DM--DE coupling on the cross--correlation 
between galaxy/quasar distributions and ISW effect.
  
Such models can be motivated in the contest of scalar--tensor theories of 
gravity or describe the low energy limit of a more fundamental theory beyond
the standard model of particle physics, e.g. string theory.

The models which we aim to investigate differ from the standard 
$\Lambda$CDM in three different
aspects: (i) {DE is a self--interacting scalar field $\phi$ rather
than a cosmological constant $\Lambda$. We shall consider a class of 
self--interaction potentials $V(\phi)$
admitting {\it tracker solutions}}. (ii) {A linear DM--DE coupling is
allowed.} (iii) {We allow neutrinos to be massive.}

The effects of massive neutrinos in cosmology have been studied
thoroughly for many years (for a review see \citet{les06}). Cosmological
 observations are mostly
sensitive to the sum of neutrino masses, $M_\nu$. Currently, the strongest 
upper limit on neutrino mass scale comes from cosmology.
One of the effects of massive neutrinos is to induce a small 
decay of the gravitational 
potentials during both matter and DE domination so that, in principle, 
ISW effect would also provide information on their mass.
Furthermore, as recently outlined in \citet{lav09a,lav09b,kri10} the effects
that massive neutrinos have on the angular power spectrum of the CMB 
anisotropies, $C_l$, and 
matter power spectrum, $P(k)$, are almost opposite to
those of the DM--DE coupling, resulting in 
a strong degeneracy between the coupling strength $\beta$ and $M_\nu$. 
A recent analysis by means of Monte Carlo Markov Chain method has shown that a 
cosmology with significant $M_\nu$ and $\beta$ is statistically 
preferred to one with no coupling and almost massless neutrinos. Further,
when priors on the neutrino mass from
earth-based neutrino mass experiments (Heidelberg-Moscow neutrinoless
double $\beta$-decay, KATRIN 
 tritium $\beta$-decay \footnote{http://www-ik.fzk.de/katrin/publications/documents/Design Report2004-12Jan2005.pdf}) are added to the analysis,
a $5-6 \sigma$ detection of a DM-DE coupling is found.

The plan of the paper is as follows: in Section 2 we describe our model
while the ISW effect theory is reviewed in Section 3 where we also discuss
how the ISW signal depends on the main parameters of the model.
In Section 4 we discuss galaxy bias and magnification bias. Comparison between
theoretical prediction and observation data is presented in Section 5 while
a tomographic analysis is performed in Section 6. Section 7 is devoted to the conclusions.

\section{The model}

We assume  a spatially flat Friedmann--Robertson--Walker (FRW) background 
with metric $ds^2 = a^2(\eta) \left( -d\eta^2 + dx^i dx_i \right)$
($\eta$ is the conformal time) filled with baryons, photons,  neutrinos, DM 
and a component of DE which will be ascribed to a scalar field $\phi$ 
self--interacting through a potential $V(\phi)$.
In the following, the indexes $b$, $c$, $\nu$ and $\phi$ will denote baryons, 
cold DM, massive neutrinos and DE. Photons and massless neutrinos will be 
referred as radiation and denoted by $r$.

In addition to self--interaction we also consider a possible  interaction 
between the scalar field and DM. Here we 
give only the equations for baryons, DM and DE 
being the equations for the other components the usual ones (see, e.g 
\citet{ma95}).

The Friedmann equation for the scale factor $a$, the continuity equations for 
baryons and DM 
and the evolution equation for the scalar field read: 

\begin{eqnarray}
 {\cal H}^2 = {8 \pi \over 3} G 
\left(\rho_b + \rho_c+ \rho_\phi \right) a^2 \\
\dot \rho_{b} + 3 {\cal H}\rho_{b} = 0 \\
\dot \rho_{c} + 3 {\cal H}\rho_{c}=  -C \dot \phi \rho_c \\
\ddot \phi +2{\cal H}\dot\phi+a^2V(\phi)=Ca^2\rho_c
\label{dens}
\end{eqnarray}
where $~\dot {}~$ denotes the derivative with respect to $\eta$, 
${\cal H}=\dot a/a$, $\rho_i$ is the energy density of the 
component $i=b, c, \nu, \phi, r$ and the constant $C$ parametrizes the 
DM--DE coupling strength. 

Working in the conformal Newtonian gauge the metric of a perturbed flat FRW 
Universe takes the form:
\begin{equation}
ds^2 = a^2(\tau) \left[ -\left(1 + 2\Phi \right) d\tau^2 + 
\left(1 - 2\Psi \right) dx^i dx_i \right]
\label{metric}
\end{equation}
where $\Phi$ plays the role of the Newtonian potential, $\Psi$ is the 
Newtonian spatial curvature, $|\Phi|,|\Psi| << 1$ and:
\begin{eqnarray}
\Phi= -{3 \over 2}{{\cal H}^2 \over k^2}\sum_i\left [ \Omega_i \delta_i+3{{\cal H}\over k^2}
(1+w_i)\Omega_i \theta_i \right]
\\
\Psi=\Phi-\sum_i{9 \over 2} {{\cal H}^2\over k^2}(1+w_i) \Omega_i \sigma_i
\end{eqnarray}
where $\Omega_i=\rho_i/\rho_{cr}$, $\delta_i=\delta\rho_i /\rho_i$, $\theta_i$,
 $\sigma_i$ and $w_i=p_i/\rho_i$ are the density parameter, density contrast, 
four--velocity divergence, shear and state parameter
of the component $i$ ($p_i$ and $\rho_{cr}=3{\cal H}^2/8\pi a^2$ being the 
pressure of the component $i$ and the critical density of the Universe).

Linear perturbation equations for DM and baryons read:
\begin{eqnarray}
\dot \delta_c+\theta_c-3\dot\Psi=-C\dot{\delta\phi} \\
\dot\theta_c+({\cal H}-C\dot\phi)\theta_c=k^2(\Phi-C\delta\phi)\\
\dot \delta_b+\theta_b-3\dot\Psi= 0\\
\dot\theta_b+{\cal H}\theta_b-c_s^2k^2\delta_b=k^2\Phi+ \Gamma_{phot-b}
\end{eqnarray}
where $c_s^2$ is the baryon  sound speed, $\Gamma_{phot-b}$ is the standard 
term describing momentum exchange with photons due to Thomson scattering 
(see, e.g Ma \& Bertschinger 1995) and $\delta\phi$
is the perturbation to the scalar field which evolves according to:
\begin{eqnarray}
\nonumber
\ddot {\delta\phi} +2{\cal H} \dot{\delta\phi}+(k^2+a^2 V'')\delta\phi - 4\dot\Phi\dot\phi + 2a^2\Phi V'=
\\
C\left(\rho_c \delta_c +2\rho_c\Phi\right)a^2
\end{eqnarray}

As we are interested in  the cross--correlation between ISW effect and galaxy 
distributions, above equations can be simplified. At late time radiation and 
massive neutrinos can be neglected so that no shear stresses are present and 
$\Phi=\Psi$. Furthermore, the cross--correlation signal comes from scales well
within the horizon, $\sim 100-200$ Mpc, so that the second term in (6) 
can be neglected.

The above equations then reduce to the usual Poisson equation for the gravitation potential:
\begin{equation}
\Phi= -{3 \over 2}{{\cal H}^2 \over k^2}\left [ \Omega_c\delta_c+
\Omega_b\delta_b+\Omega_\nu\delta_\nu \right], 
\end{equation}
a modified Jeans equation for DM and the usual one for baryons:
\begin{eqnarray}
\ddot\delta_c + ({\cal H} -C\dot\phi)\dot\delta_c &=&{3\over 2}{\cal H}^2\left[\left(1+{4 \over 3}\beta^2\right)\Omega_c\delta_c +\Omega_b \delta_b +\Omega_\nu\delta_\nu\right]
\nonumber \\
\ddot\delta_b+ {\cal H} \dot\delta_b &=&{3\over 2}{\cal H}^2\left[\Omega_c\delta_c +\Omega_b \delta_b +\Omega_\nu\delta_\nu\right]=0,
\end{eqnarray}
and a Poisson--like equation for the scalar field perturbation:
\begin{equation}
\delta\phi= {{\cal H}^2 \over k^2}\left [ \Omega_c\delta_c+
\Omega_b\delta_b+\Omega_\nu\delta_\nu \right] 
\end{equation} 

where we have defined the dimensionless coupling parameter:
$$\beta=\sqrt{3/16 \pi}~m_p C$$
($m_p=G^{-1/2}$ 
is the Planck mass). 

As clearly visible from the above equations and widely discussed in 
\citet{ame00},  coupling affects the dynamics of DM particles. 
As a consequence baryons and DM develop a bias $b^*$, i.e. 
$\delta_b=b ^*\delta_{dm}$. Notice that, this bias, which origin is to ascribe
to the coupling, is something completely different from the galaxy bias  
due to hydrodynamical effects, discussed in the 
subsequent sections.

It is also worth mentioning that, unlike the uncoupled case, in the presence of 
coupling, Universe goes through
an evolutionary phase named $\phi$--matter dominated era ($\phi$MDE) just after 
matter--radiation equivalence. In this period
the scalar field $\phi$ behaves as $\it stiff ~matter$ ($p_\phi / \rho_\phi=1$)
having a non--negligible kinetic energy which dominates over the potential one. 
After this stage, the usual matter era follows  before entering in the 
accelerated regime with a final De Sitter attractor. 
Notice also that, because of the $\phi$MDE and the non--usual scaling of the DM energy density, i.e. $\rho_c \propto a^{-3}e^{-C\phi}$, after equivalence the background expansion law will differ from the usual $a\propto \eta^2$.

\subsection{Potential}

\begin{figure}[t]
\begin{center}
\includegraphics[scale=0.3, angle=-90]{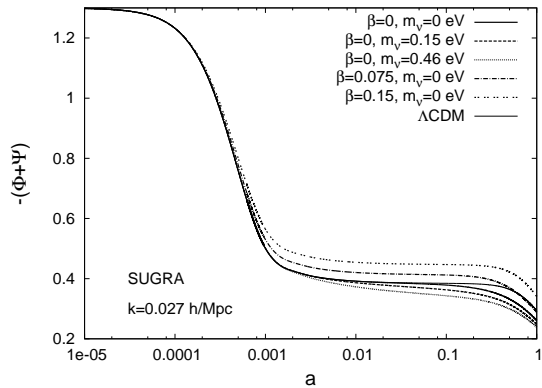}
\label{f1}
\caption{Evolution of the gravitational potentials as a function of the 
scale factor for different values of $\beta$ and $m_v$ for SUGRA model.
For comparison the $\Lambda$CDM case is also displayed}
\end{center}
\end{figure}

\begin{figure*}[t]
\begin{center}
\includegraphics[scale=0.27, angle=-90]{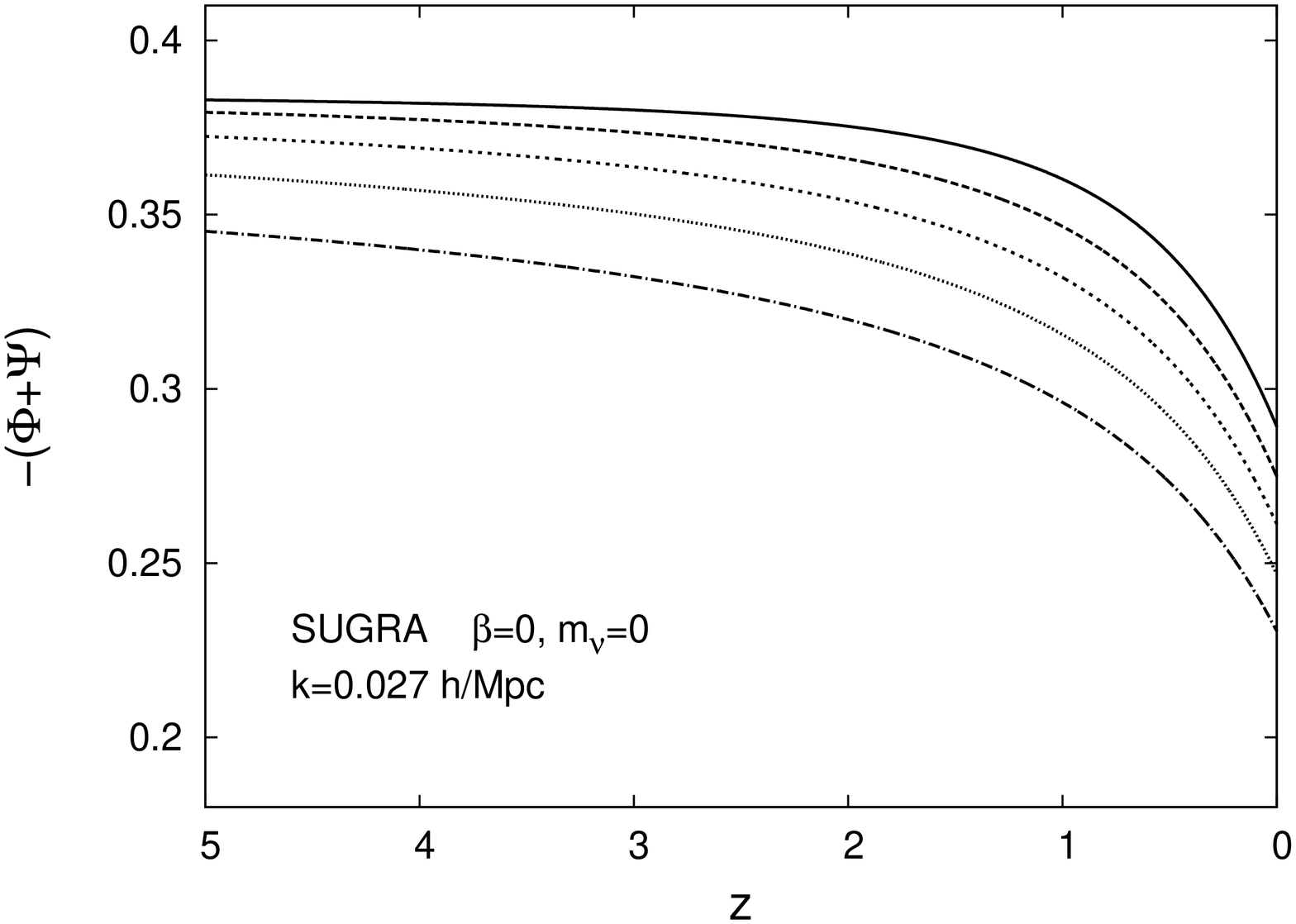}
\includegraphics[scale=0.27, angle=-90]{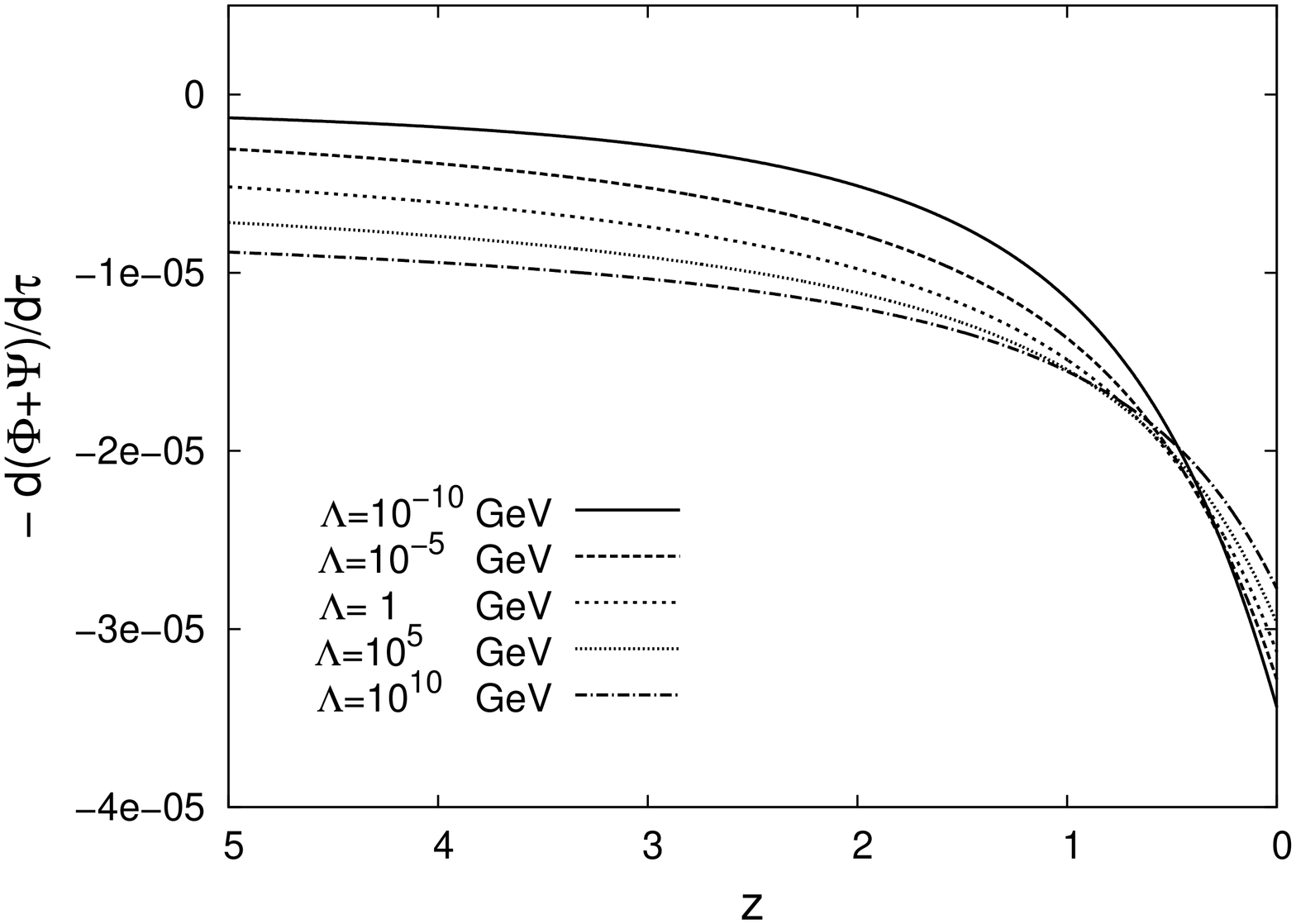}
\caption{Redshift evolution of $\Phi+\Psi$ (left) and its time derivative 
(right) for different values of $\Lambda$ in uncoupled SUGRA with massless 
neutrinos}
\label{f2}
\end{center}
\end{figure*}
\begin{figure*}[t]
\begin{center}
\includegraphics[scale=0.27, angle=-90]{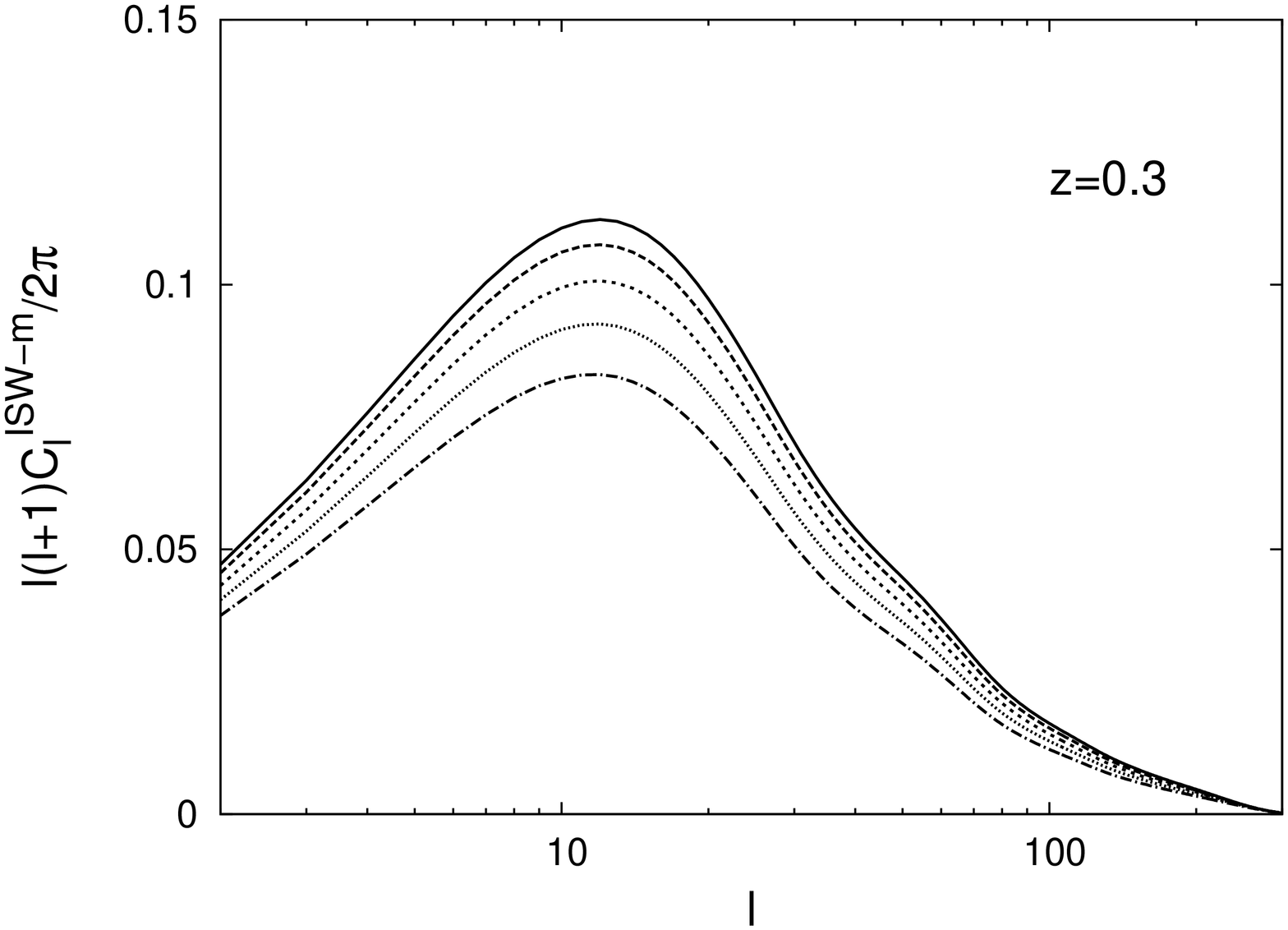}
\includegraphics[scale=0.27, angle=-90]{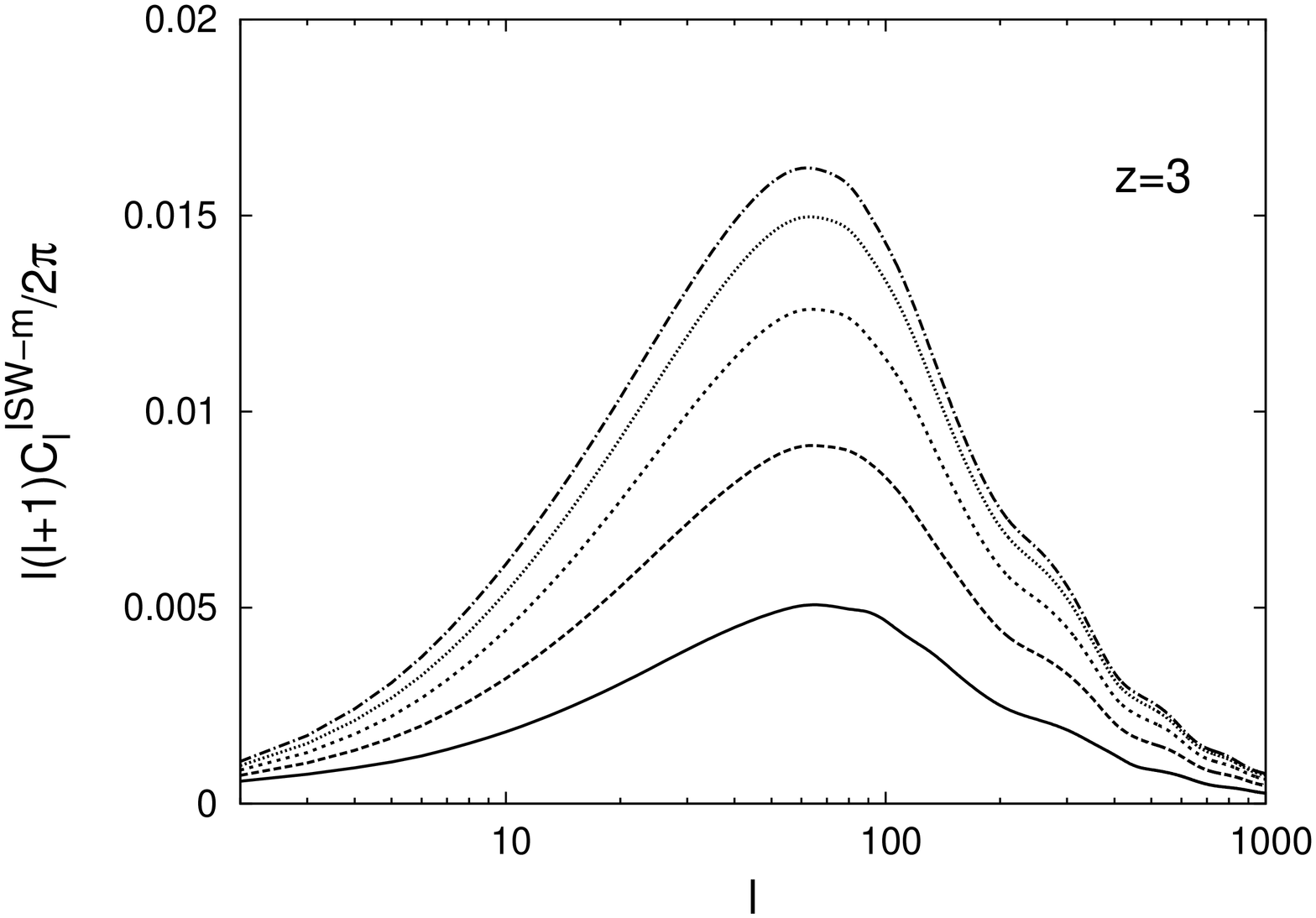}
\includegraphics[scale=0.27, angle=-90]{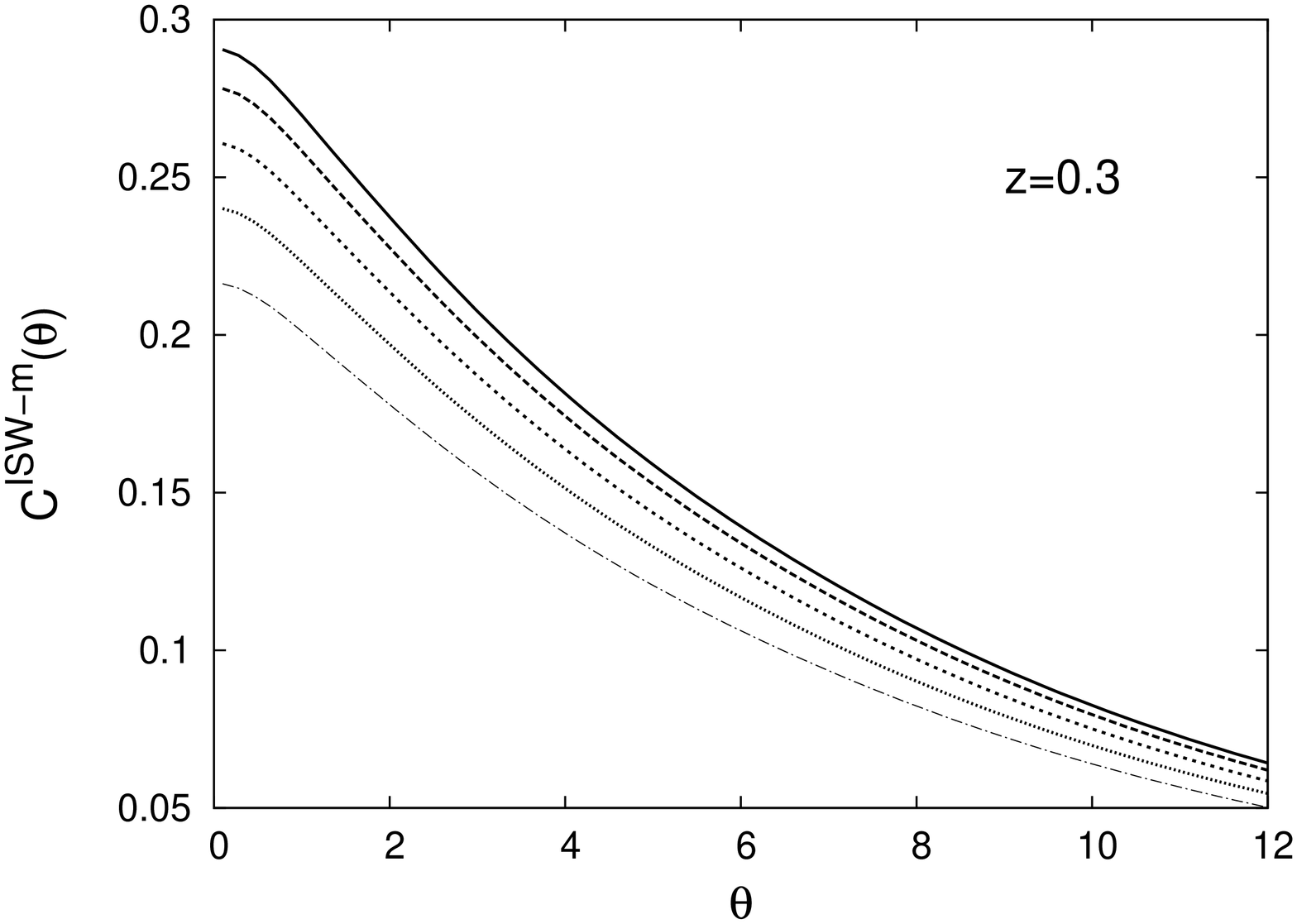}
\includegraphics[scale=0.27, angle=-90]{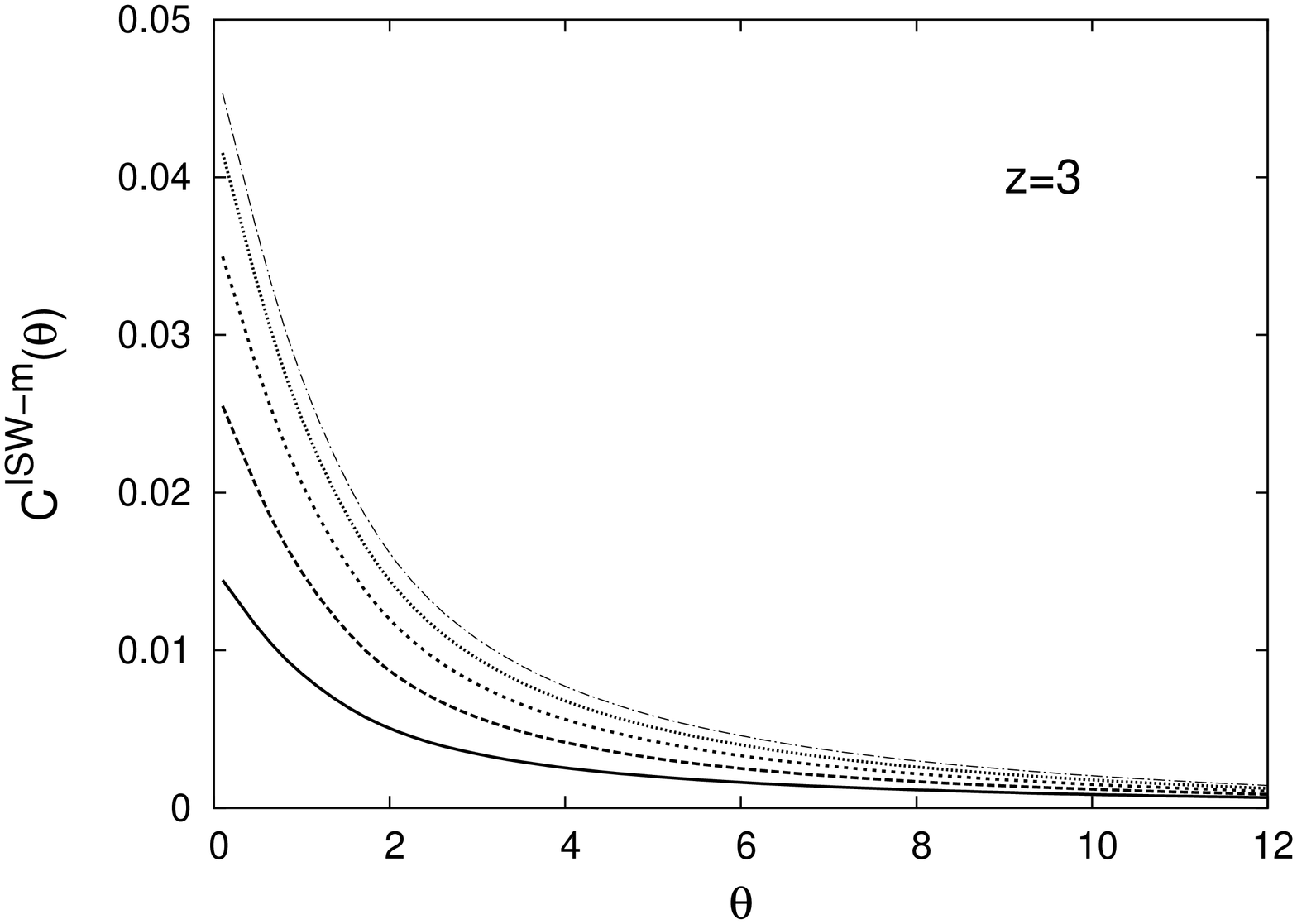}
\caption{ISW--matter cross--correlation power spectra (top) and functions  
(bottom). Their dependence on $\Lambda$ is shown at z=0.3 (left) 
and z=3 (right). Models are the  uncoupled SUGRA with massless 
neutrinos}
\label{f3}
\end{center}
\end{figure*}

\begin{figure*}[t]
\begin{center}
\includegraphics[scale=0.27, angle=-90]{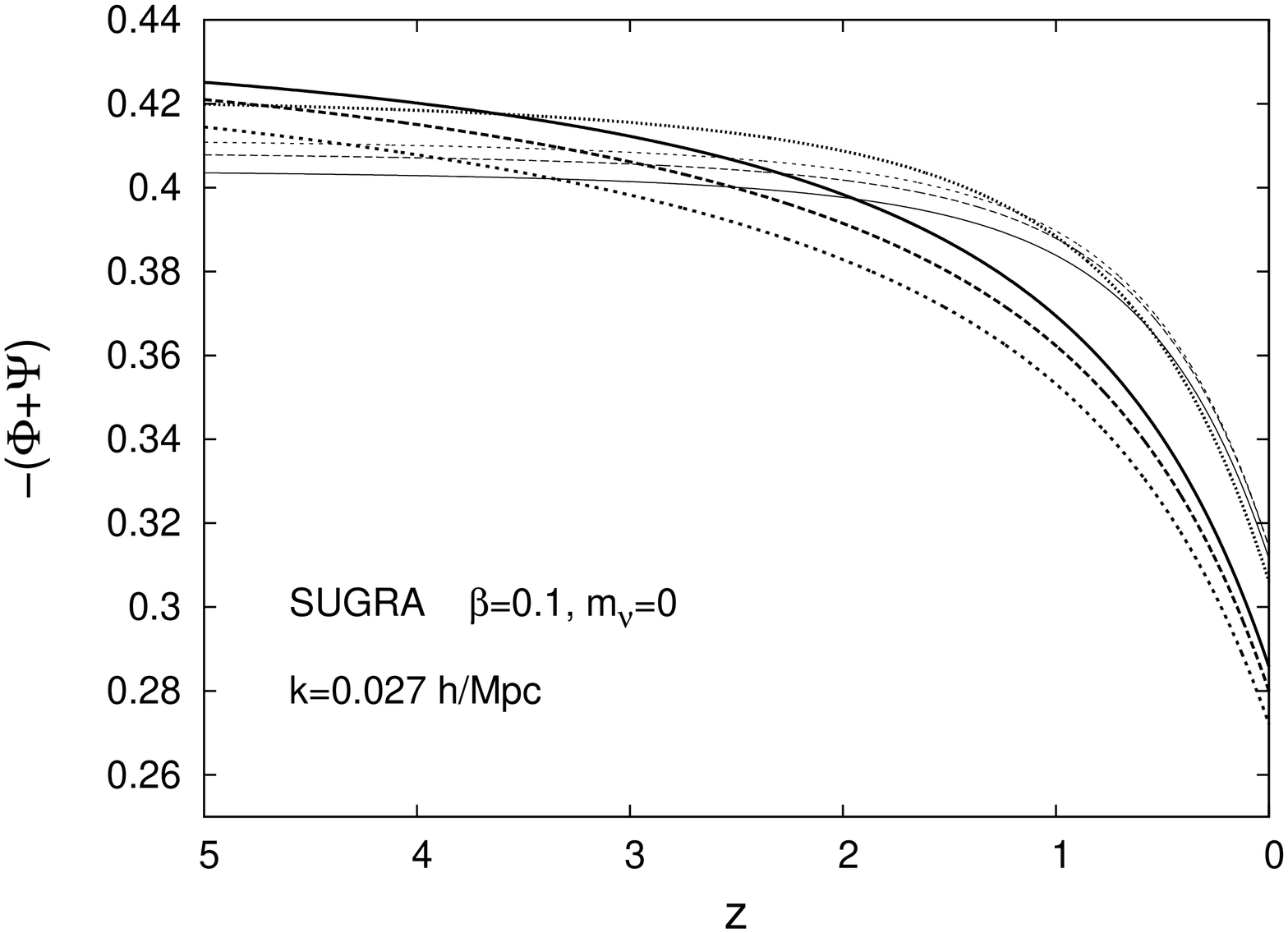}
\includegraphics[scale=0.27, angle=-90]{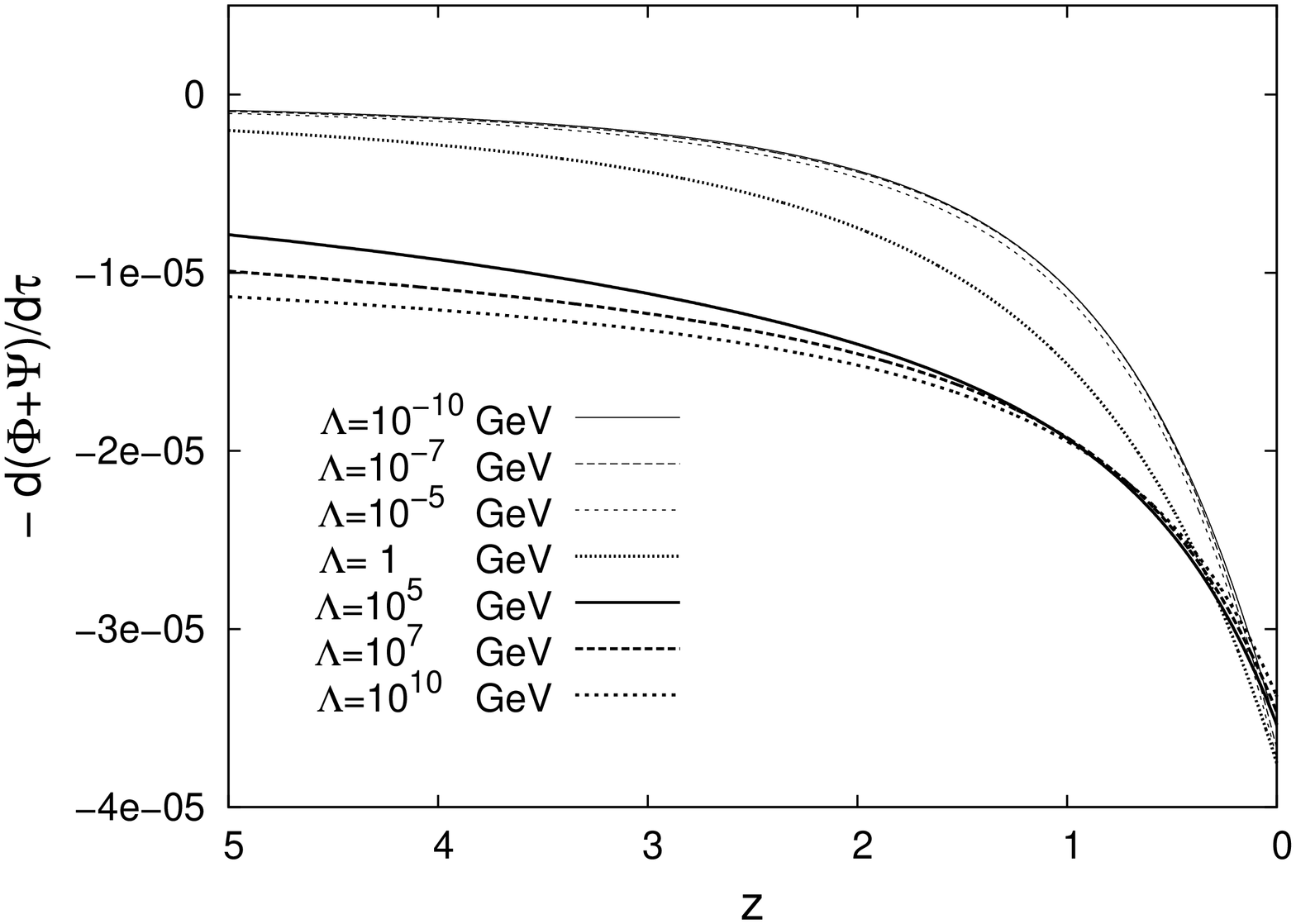}
\caption{Redshift evolution of $\Phi+\Psi$ (left) and its time derivative 
(right). Models are coupled SUGRA with massless neutrinos}
\label{f4}
\end{center}
\end{figure*}
\begin{figure*}[t]
\begin{center}
\includegraphics[scale=0.27, angle=-90]{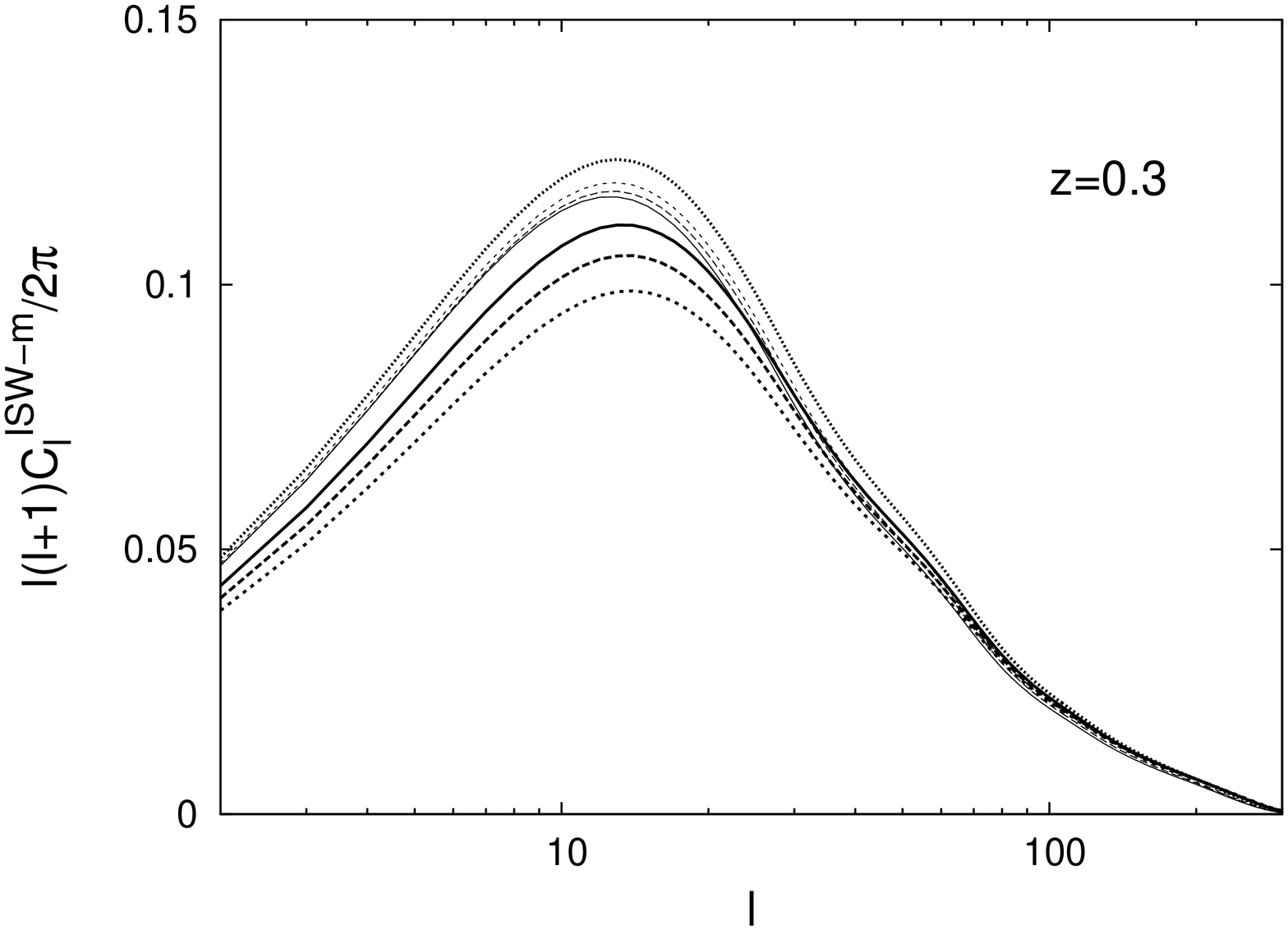}
\includegraphics[scale=0.27, angle=-90]{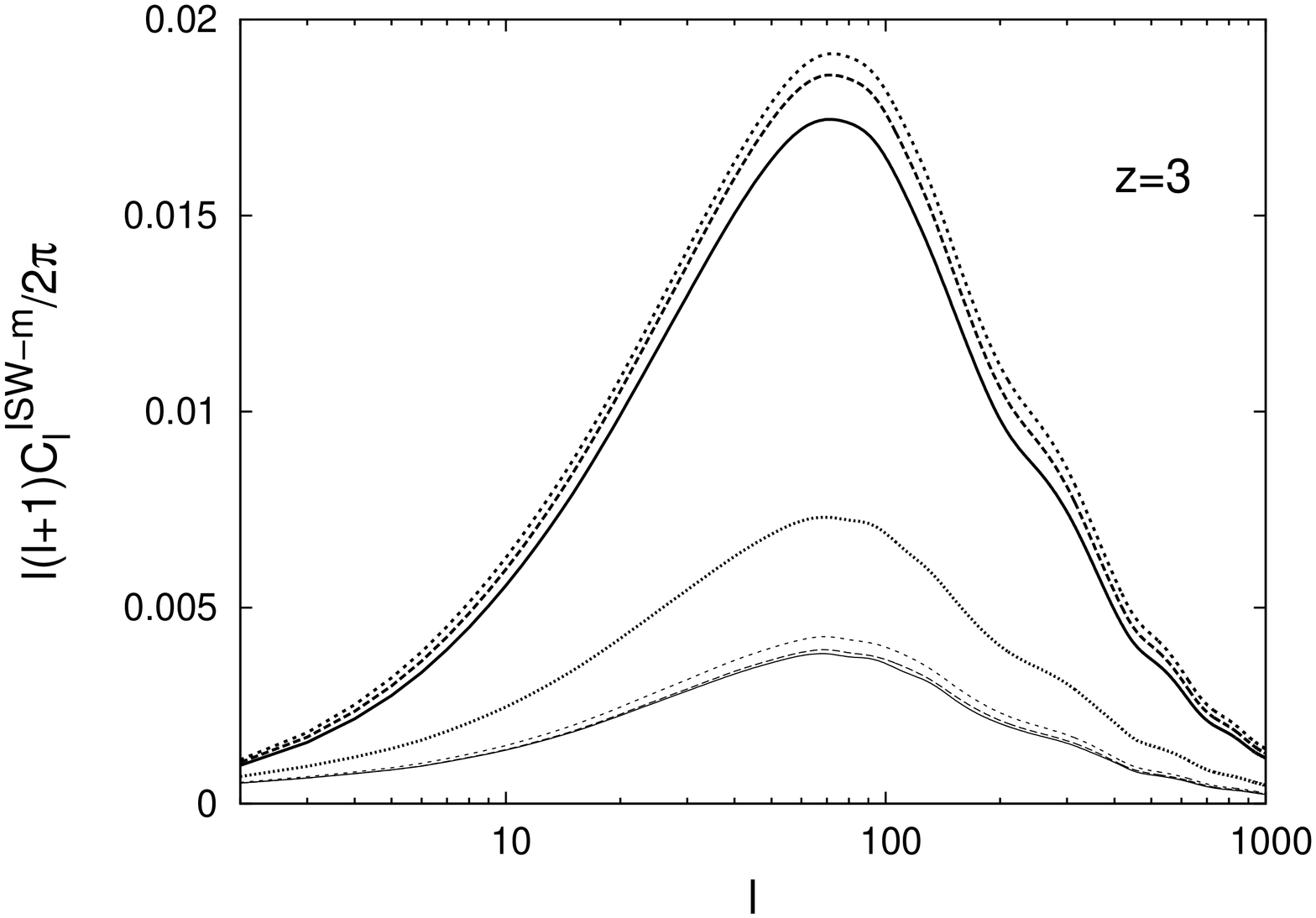}
\includegraphics[scale=0.27, angle=-90]{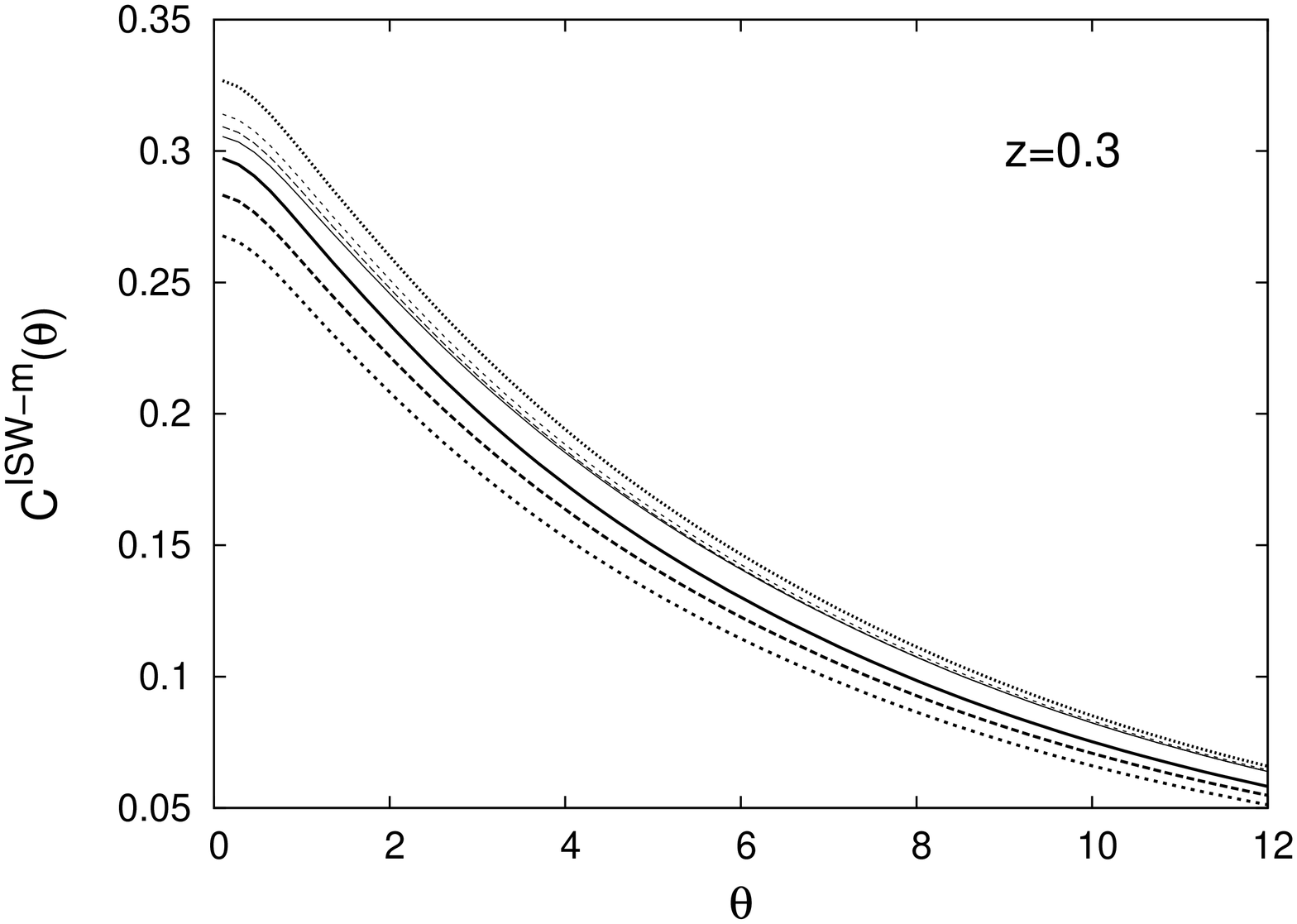}
\includegraphics[scale=0.27, angle=-90]{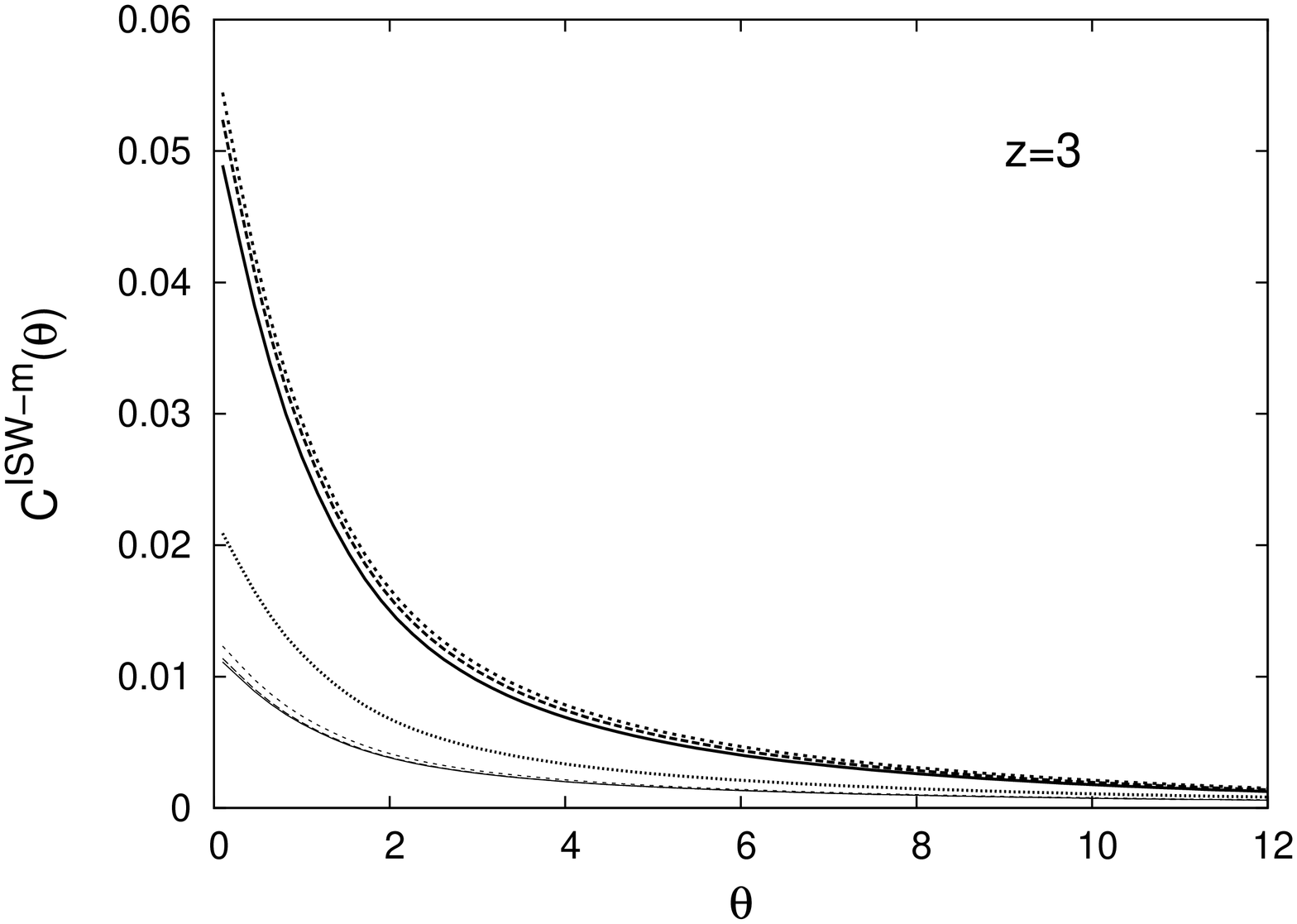}
\caption{ISW--matter cross--correlation power spectra (top) and functions  
(bottom). Their dependence on $\Lambda$ is shown at z=0.3 (left) 
and z=3 (right).  Models are coupled SUGRA with massless neutrinos}
\label{f5}
\end{center}
\end{figure*}

\begin{figure*}[t]
\begin{center}
\includegraphics[scale=0.27, angle=-90]{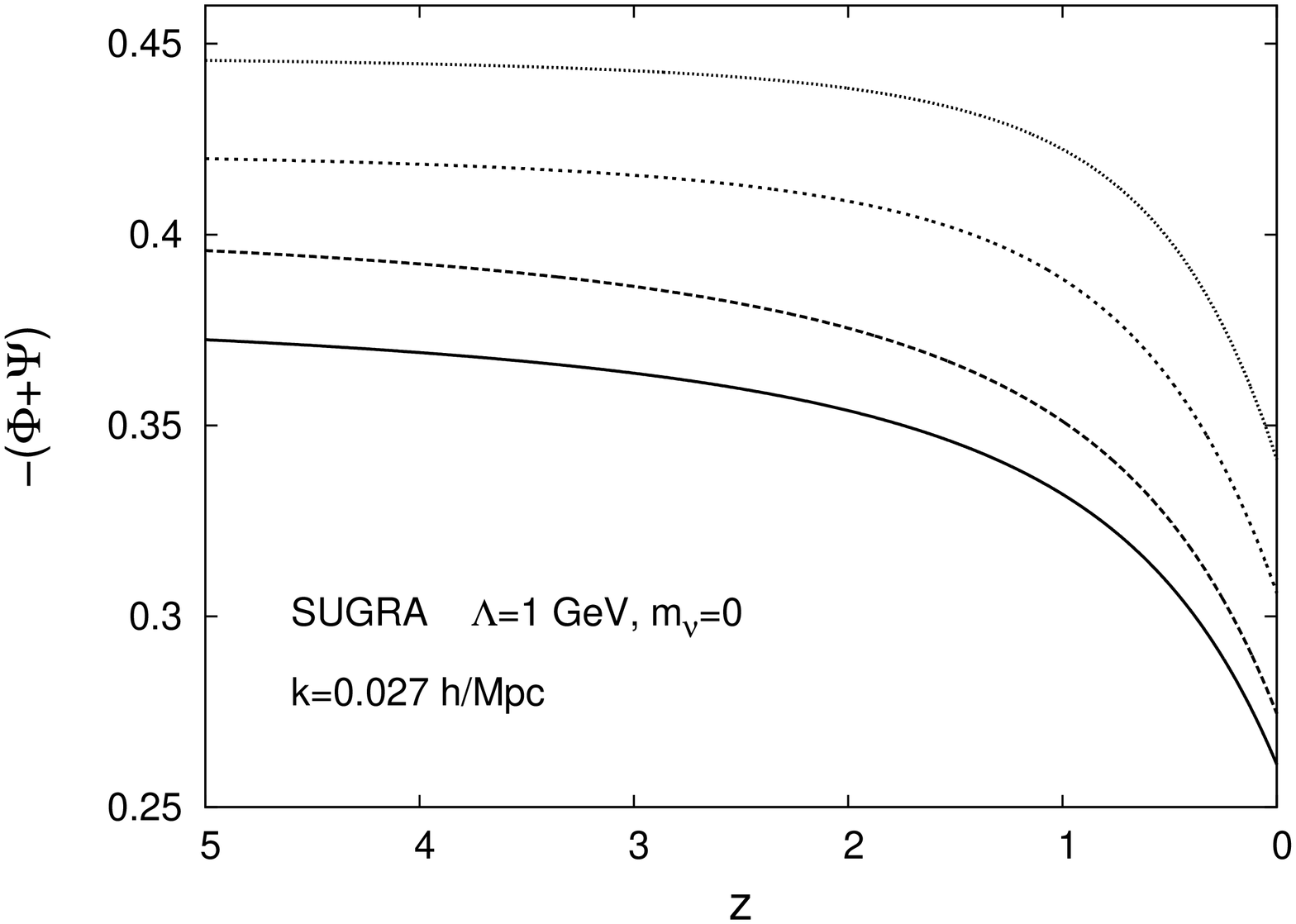}
\includegraphics[scale=0.27, angle=-90]{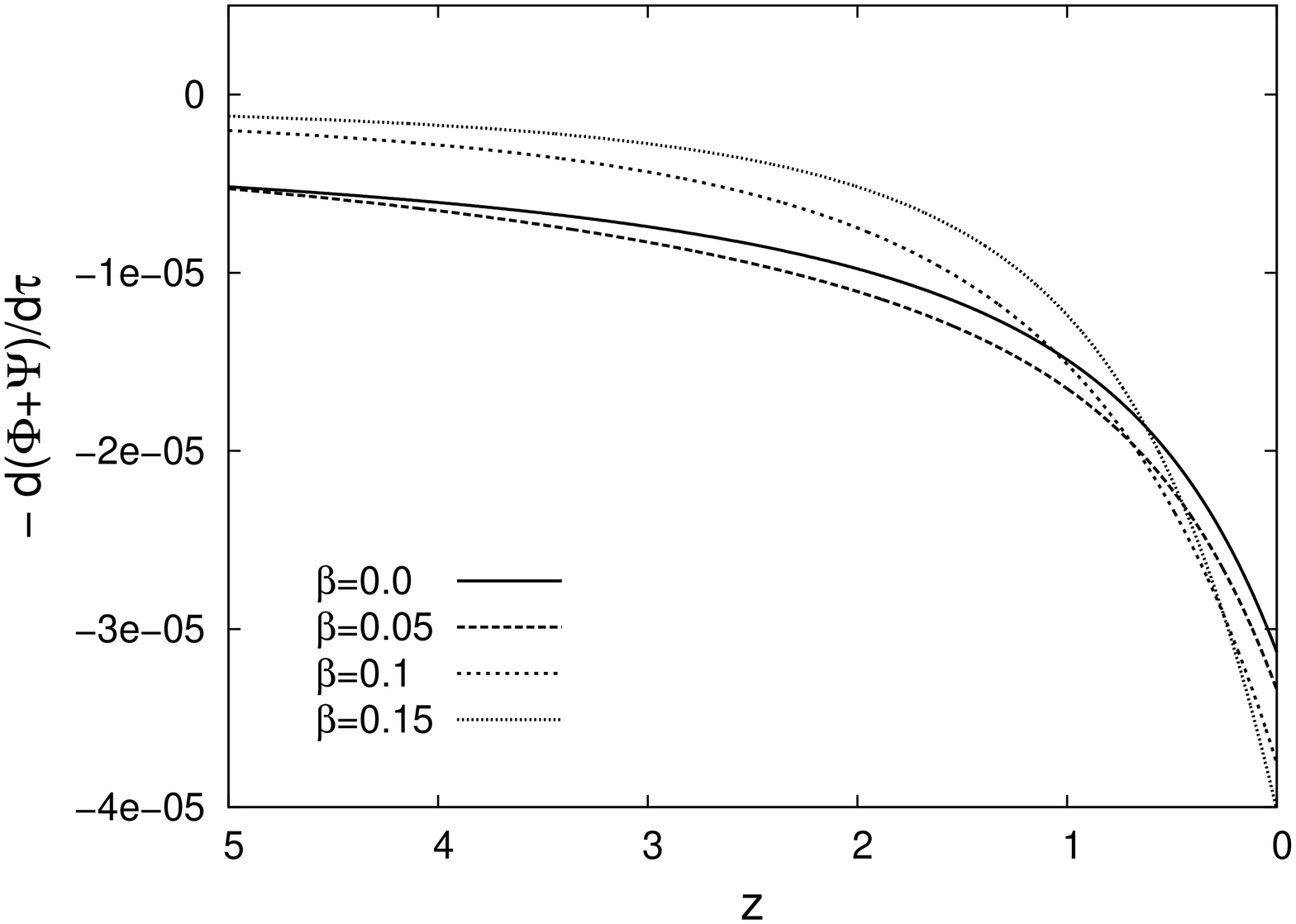}
\caption{Redshift evolution of $\Phi+\Psi$ (left) and its time derivative 
(right) for different values of $\beta$ in SUGRA models with massless 
neutrinos}
\label{f6}
\end{center}
\end{figure*}
\begin{figure*}[t]
\begin{center}
\includegraphics[scale=0.27, angle=-90]{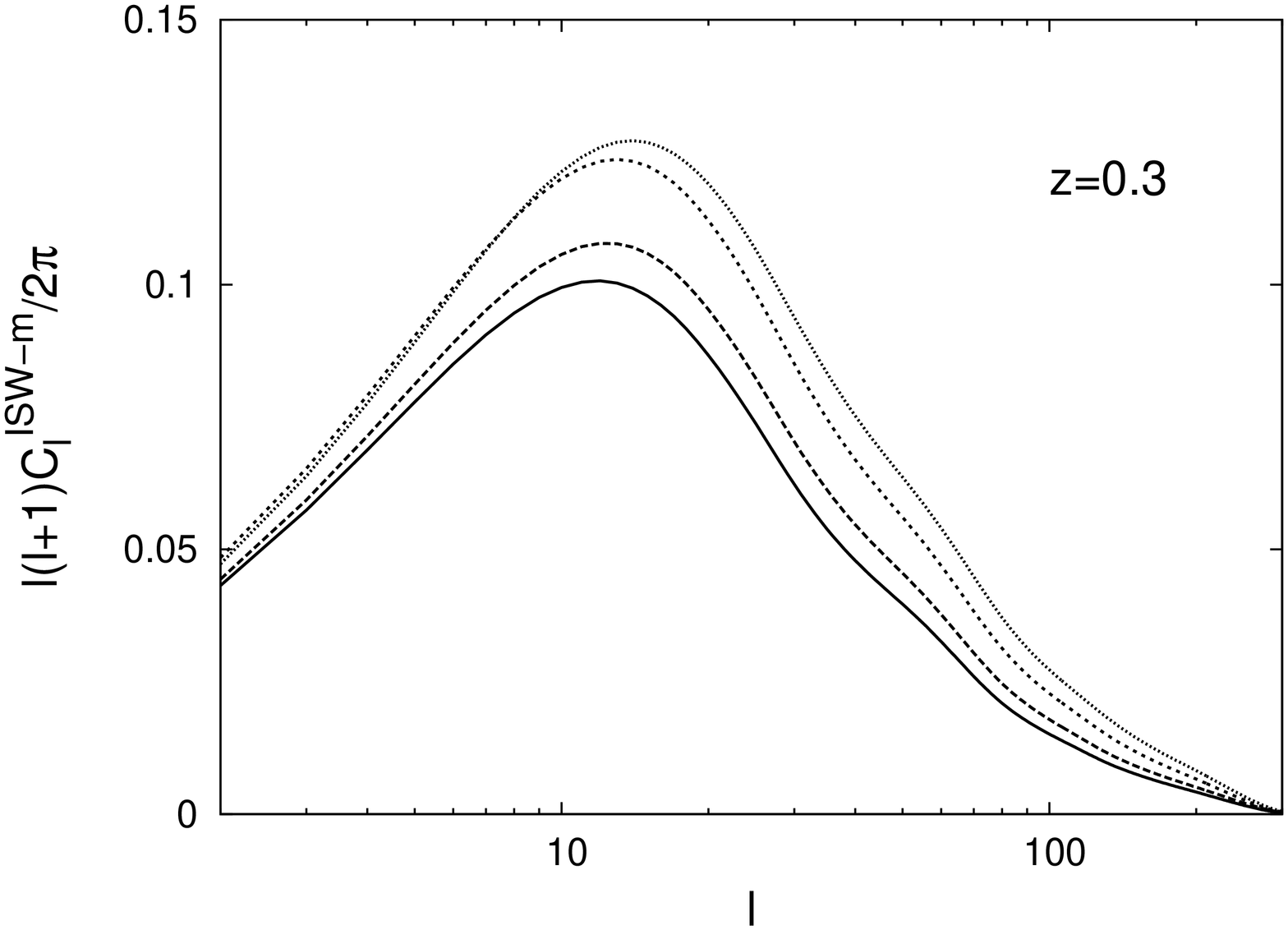}
\includegraphics[scale=0.27, angle=-90]{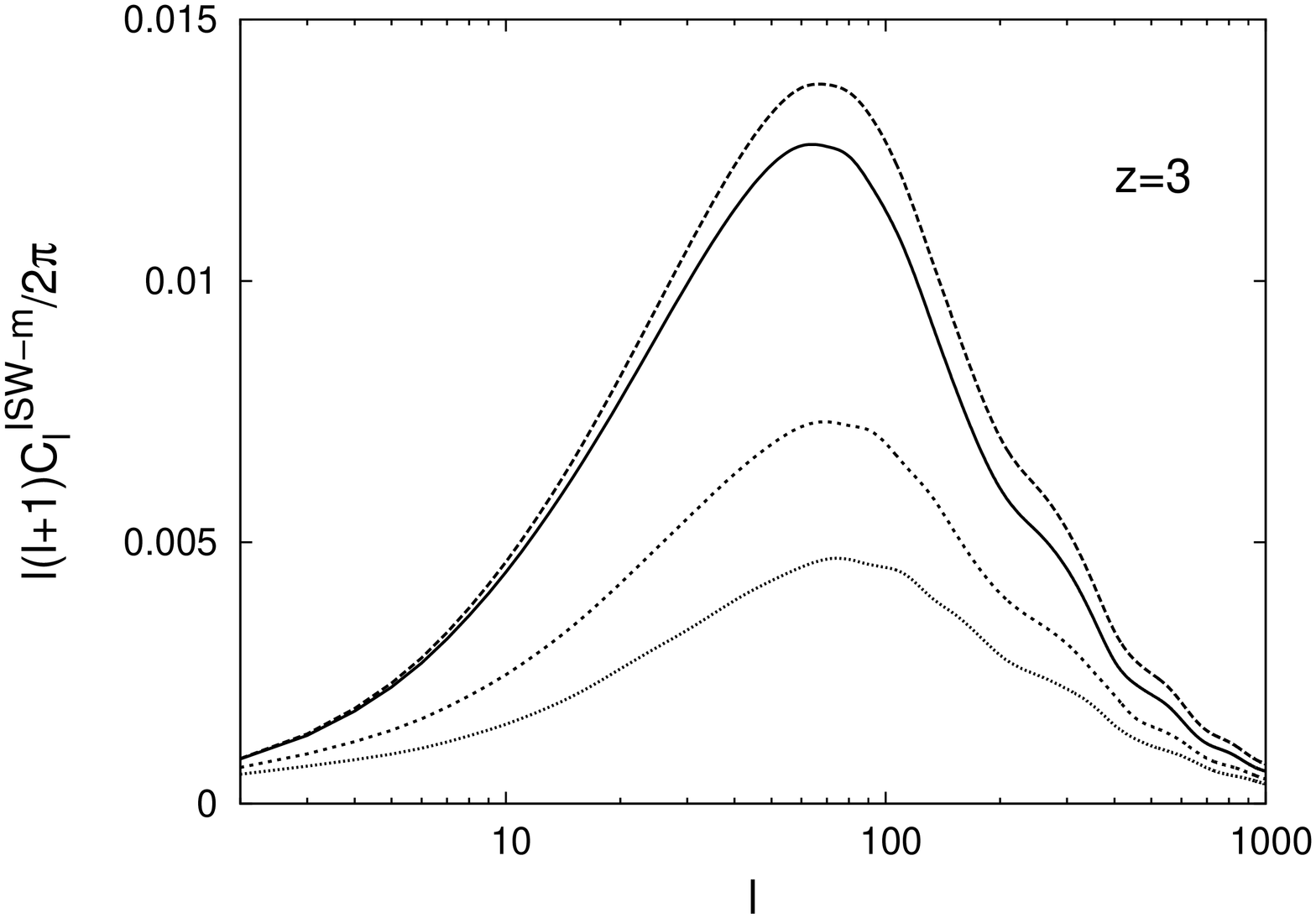}
\includegraphics[scale=0.27, angle=-90]{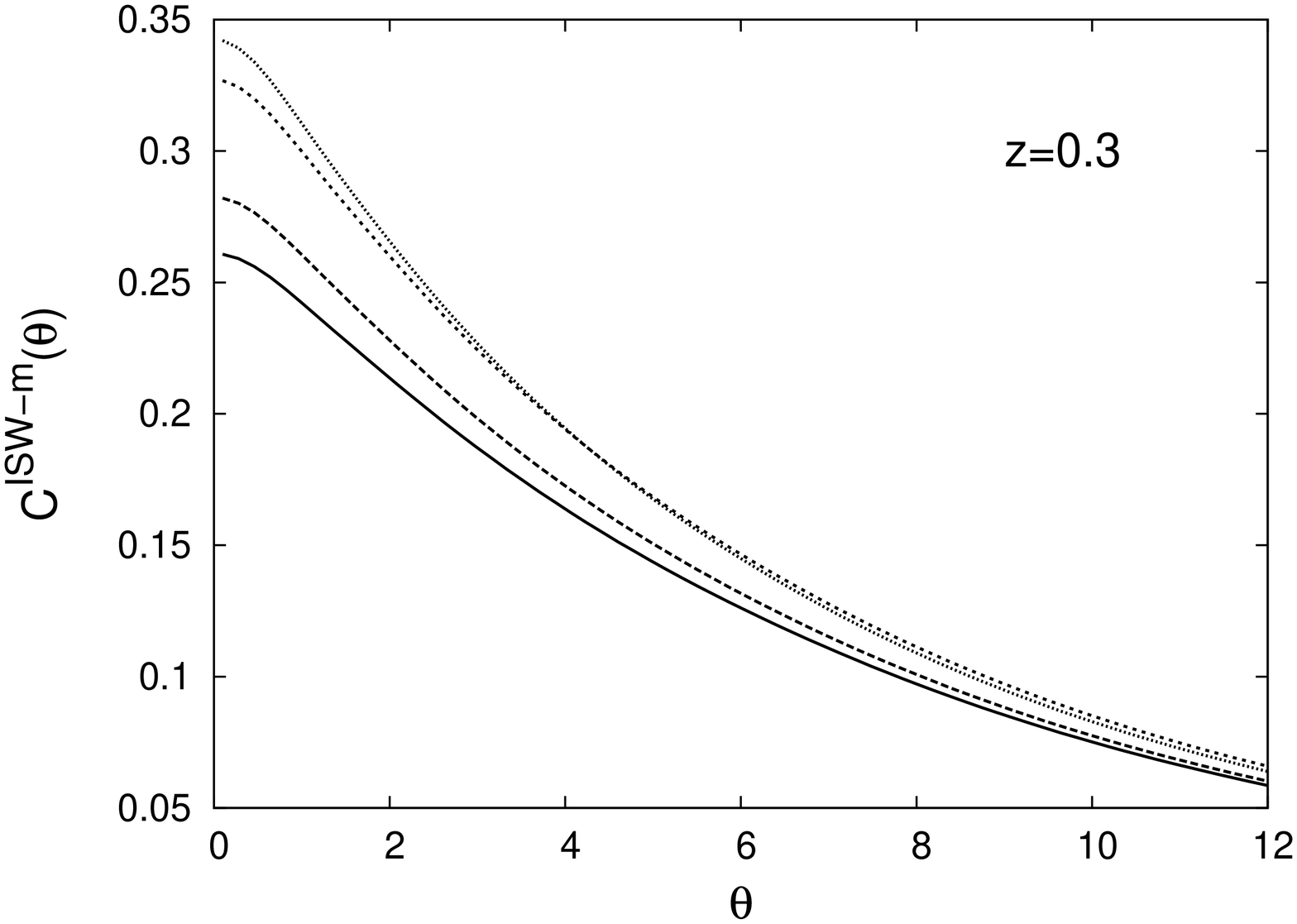}
\includegraphics[scale=0.27, angle=-90]{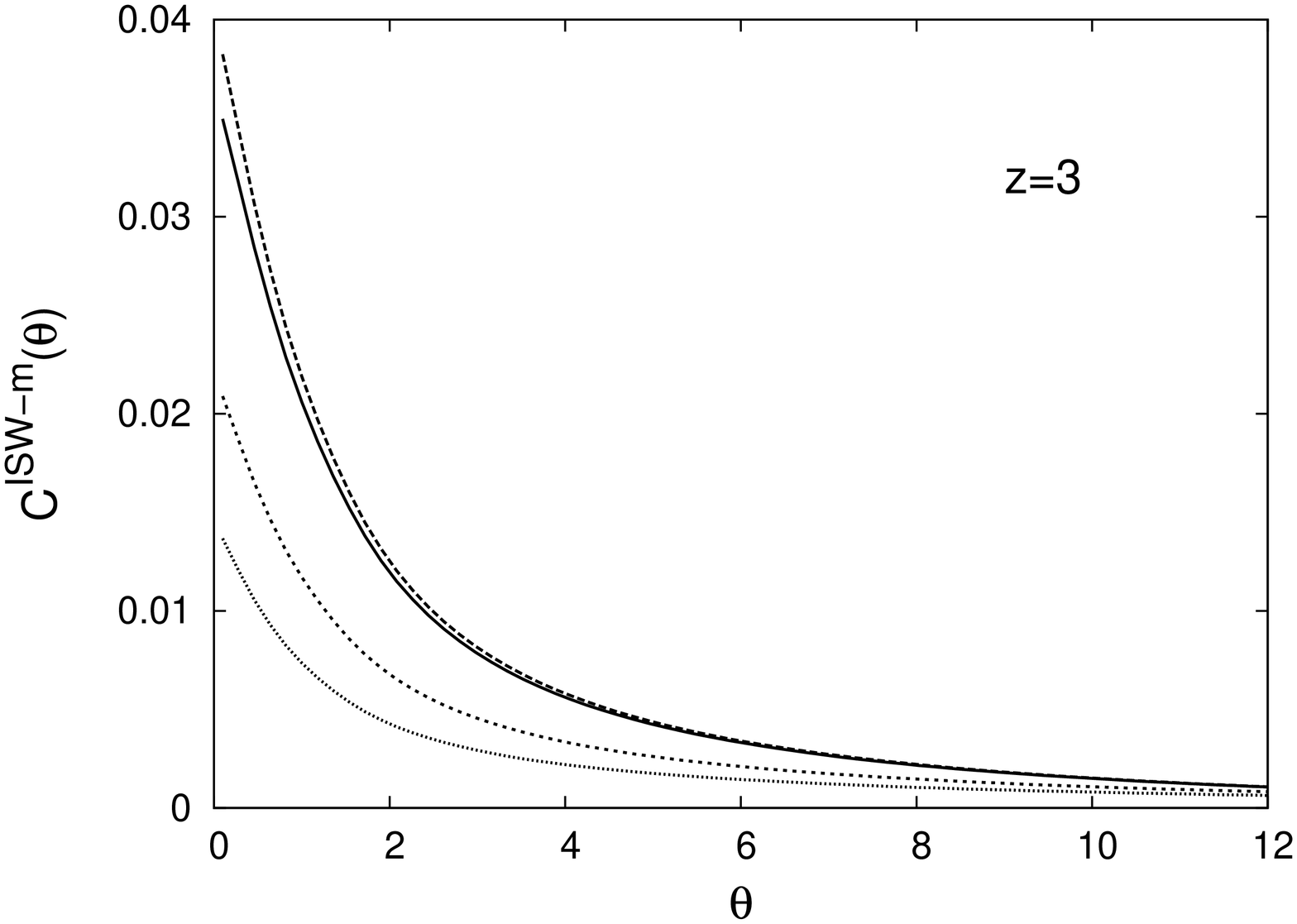}
\caption{ISW--matter cross--correlation power spectra (top) and functions  
(bottom). Their dependence on $\beta$ is shown at z=0.3 (left) 
and z=3 (right). Models are the  SUGRA models with massless 
neutrinos}
\label{f7}
\end{center}
\end{figure*}

We shall consider the Ratra--Peebles (RP) \citep{rat88}
and SUGRA self--interaction potentials \citep{bra00},
reading
\begin{equation}
\nonumber
V(\phi) = {\Lambda^{\alpha+4} \over \phi^\alpha}~~~~~~~~~~~~~~~~~~~~~~~~~~~~~~~~~~~~~~~~RP
\label{RP}
\end{equation}
and
\begin{equation}
\nonumber
V(\phi) = {\Lambda^{\alpha+4}\over\phi^\alpha} e^{4\pi {\phi^2 \over m_p^2}}
~~~~~~~~~~~~~~~~~~~~~~~~~~SUGRA
\label{sugra}
\end{equation}
respectively; they allow tracker 
solutions for any $\alpha >0$. 
For both potentials, once $\alpha$ and $\Lambda$ are assigned, the
present time DE density parameter $\Omega_{DE}$ is uniquely defined. 

Limits on these models without coupling between DE and CDM have
been studied in La Vacca \& Kristiansen 2009.
They find,  that only quite small $\lambda= \log \Lambda/GeV$ are allowed.
In the SUGRA case in particular only $\lambda \lesssim -3.5$ is allowed.
Such small values are well below the range motivated by particle
physics. Therefore the physical appeal of the SUGRA potential is
spoiled.

Let us however outline that, when the $\beta$ degree of freedom
is opened, $\Lambda$ values as large as 30$\, $GeV become allowed, at
the 1--$\sigma$ level, while, at the 2--$\sigma$ level, no significant
constraint on the energy scale $\Lambda$ remains. Even for the RP
potential, for which a limit $\lambda \lesssim -8.5$ held, in the absence
of coupling, values $\lambda \sim -2$ become allowed (La Vacca et al 2009).

In the absence of DM--DE coupling, RP yields quite a slowly varying 
$w_\phi(a)=p_\phi / \rho_\phi$ state parameter. On the contrary, 
SUGRA yields a fast varying $w_\phi$.
Although coupling causes a $w_\phi$ behavior significantly different from the 
uncoupled case, one could again consider these potentials as examples 
of rapidly or slowly varying $w_\phi$. 

The results shown in the next two sections are qualitatively the same for
RP and SUGRA models.
We will show them only for SUGRA cosmologies while the results for 
RP models will be shown only  when comparing the theoretical predictions 
with observational data and dealing with redshift tomography.

\section{ISW effect}
\label{sec:isw}

ISW effect arises when CMB photons from the last scattering surface pass through a time--dependent gravitational potential changing its energy so that additional temperature anisotropies are generated. The ISW temperature fluctuation, $\Delta T^{ISW}$, is given by:
\begin{equation}
\Delta T^{ISW}=T \int{e^{-\tau}(\dot{\Phi}+\dot{\Psi})d\eta}~.
\label{isweq}
\end{equation}
where $T$ is the CMB temperature and $e^{-\tau}$ is the visibility function of the photons.

As outlined in previous section, we will deal with scales within the horizon and redshifts such that radiation and  any anisotropic stress can be neglected ($\Phi=\Psi$) . In the matter era and in the absence of DM--DE coupling ($\beta=0$),  Poisson equation reads:
\begin{equation}
\Phi=-{3 \over 2} {H^2 \over k^2} \Omega_m \delta_m
\end{equation}
($\Omega_m$ and $\delta_m \propto a \propto \eta^2$ are the total matter density parameter and density contrast respectively)
from which one can appreciate that the gravitational potential stays constant, $\dot \Phi = 0$, and no ISW effect arises.  However, when DE starts to dominate the cosmic expansion, $\Phi$ is no longer constant and ISW effect generates
secondary anisotropies in the CMB.

On the other hand, as explained in the previous section, DM-DE coupling affects both the background and density perturbation evolution, resulting in a variation of $\Phi$ even during the matter domination.

Figure 1 shows how $\beta$ and $m_\nu$ affect the time evolution of the sum $\Phi+\Psi$ which time derivative forms the source of the ISW effect. Evolution of the gravitational potentials is obtained by a modified version of the public code CMBFAST integrating the fully relativistic equations and taking into account the contributions of all of the components, i.e, photons, DM, baryons, neutrinos and DE.

Notice how $\beta$ and $m_\nu$ affect $\Phi+\Psi$ in an opposite fashion. 

Performing a measurement of the ISW effect  is, however, a  difficult task because of its small signal compared with that of primary anisotropies ($\sim10$ times larger). Furthermore, while on small scales
the small differences in temperature tend to cancel out, the large scales, from which the most ISW effect contributes come from, are strongly affected by the cosmic variance.

%\begin{figure*}[t]
%\begin{center}
%\includegraphics[scale=0.3, angle=-90]{cliswm.SDSS.ps}
%\includegraphics[scale=0.3, angle=-90]{clmm.SDSS.ps}
%\includegraphics[scale=0.3, angle=-90]{cciswm.SDSS.ps}
%\includegraphics[scale=0.3, angle=-90]{ccmm.SDSS.ps}
%\caption{ISW--matter (left) power spectra correlation functions (right) 
%for the models in Fig. 1}
%\label{f8}
%\end{center}
%\end{figure*}
%\begin{figure*}[t]
%\begin{center}
%\includegraphics[scale=0.3, angle=-90]{cciswgal.SDSS.ps}
%\includegraphics[scale=0.3, angle=-90]{ccgalgal.SDSS.ps}
%\caption{Correlation functions of Fig. 8 after the bias rescaling. 
%See text for details}
%\label{f9}
%\end{center}
%\end{figure*}

The problem can be overcome by cross--correlating the ISW anisotropies with some tracers of the matter density, e.g.  astrophysical objects like galaxies. 

The observed galaxy density contrast in the direction $\hat {\bf n}_1$ is:
\begin{eqnarray}
\delta_{gal}(\hat{\bf n}_1)=\int{b(z)\frac{dN}{dz}(z)\delta_{\rm m}(\hat{\bf
n}_1,z)dz}
\end{eqnarray}
where ${dN/dz}$ is the normalized selection function of the galaxy survey and $b(z)$ is the galaxy bias, which will be discussed in the next section,  relating the galaxy density contrast to the inhomogeneities in the mass distribution, $\delta_{gal}=b \delta_m$. Since $\delta_m$ is related to the gravitational potential through the Poisson equation, the observed galaxy density will be correlated with the ISW anisotropies in the nearby direction $\hat{\bf n}_2$:
\begin{equation}
\Delta T^{ISW}(\hat{\bf n}_2)= 2T \int{e^{-\tau(z)}{d\Phi \over dz}(\hat{\bf n}_2,z)dz}
\end{equation}
The 2--point angular cross--correlation function (CCF) and auto--correlation function (ACF) in the harmonic space are then defined as:
\begin{eqnarray}
C^{ISW-gal}(\theta)&\equiv& \left \langle \Delta T (\hat{\bf n}_1)\delta_{gal}(\hat{\bf n}_2) \right \rangle
 \nonumber \\ &=&\sum^{\infty}_{l=2}\frac{2l+1}{4\pi}C^{ISW-gal}_lP_l[\cos(\theta)] \nonumber \\
C^{gal-gal}(\theta) &\equiv &\left\langle\delta_{gal}(\hat{\bf n}_1)\delta_{gal}(\hat{\bf n}_2)\right \rangle
 \nonumber \\ &=& \sum^{\infty}_{l=2}\frac{2l+1}{4\pi}C^{gal-gal}_lP_l[\cos(\theta)]
\label{ctheta}
\end{eqnarray}
where $\theta=|\hat{\bf n}_1-\hat{\bf n}_2|$, $P_l$'s are the Legendre polynomials and the cross-- and auto--correlation power spectra are given by:
\begin{eqnarray}
C^{ISW-gal}_l&=&4\pi\int{{dk\over k}\Delta^2(k)I^{ISW}_l(k)I^{gal}_l(k)} \nonumber \\
C^{gal-gal}_l&=&4\pi\int{{dk\over k}\Delta^2(k)[I^{gal}_l(k)]^2}
\label{cl}
\end{eqnarray}
where $\Delta^2$ is the primordial power spectrum of scalar perturbations  and the integrands $I^{ISW}$ and $I^{gal}$ are:
\begin{eqnarray}
I^{ISW}_l(k)=2T\int{e^{-\tau(z)}\frac{d\Phi}{dz}j_l[k\chi(z)]dz}~,\\
I^{gal}_l(k)=\int{b(z)\frac{dN}{dz}(z)\delta_{\rm
m}(k,z)j_l[k\chi(z)]dz}
\end{eqnarray}
(here $j_l(x)$ are the spherical Bessel functions, and $\chi$ is the
comoving distance).

\begin{figure}[t]
\begin{center}
\includegraphics[scale=0.3, angle=-90]{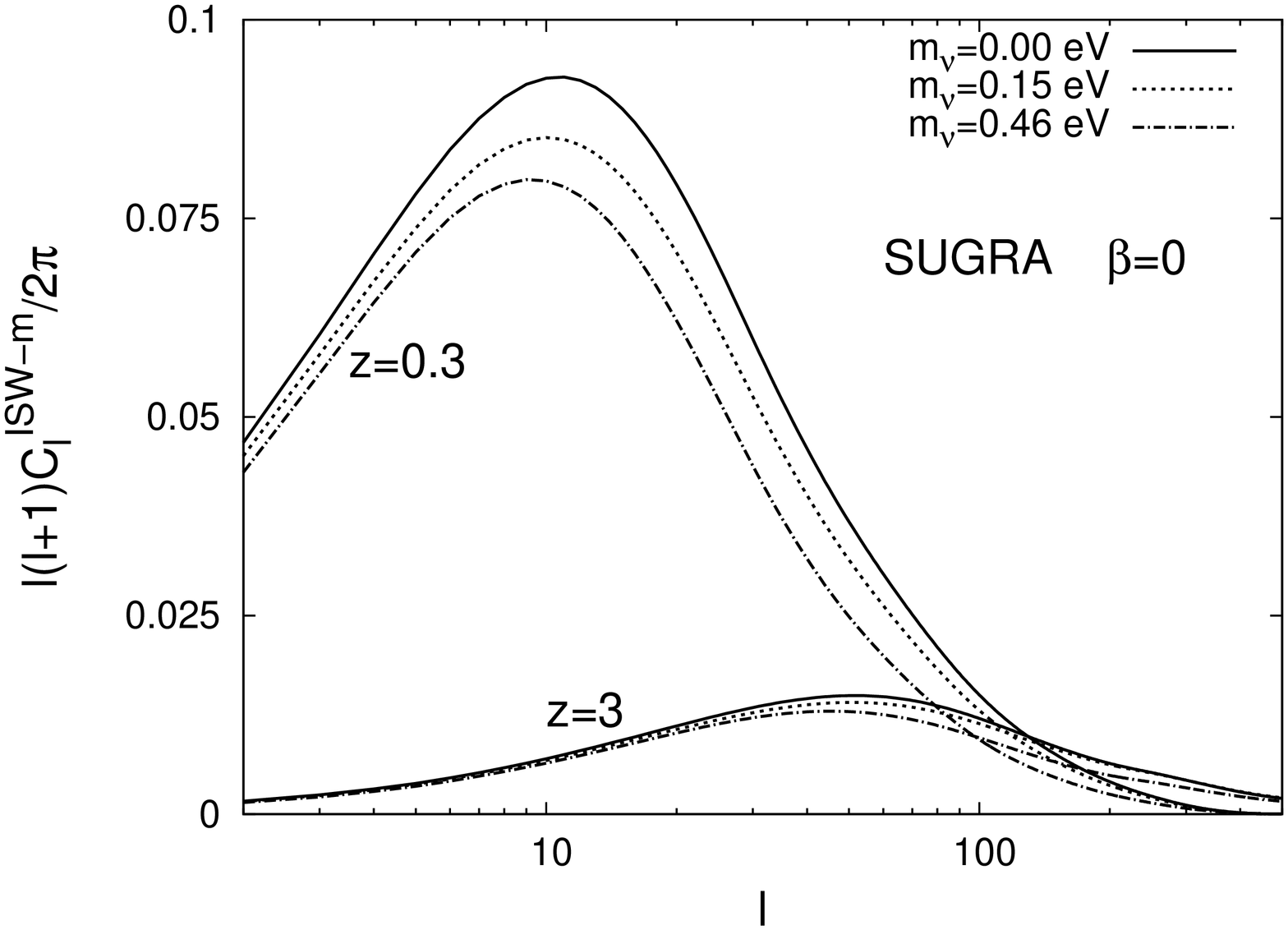}
\caption{Effect of massive neutrinos on ISW--matter cross correlation}
\label{fnu}
\end{center}
\end{figure}

In the following we use our modified CMBFAST code to calculate the theoretical
CCF and ACF.
In order to better understand the effects of $\Lambda$, $\beta$ and $m_\nu$ on them, we start to compute
the ISW-matter CCF and ACF, $C^{ISW-m}(\theta)$ and $C^{m-m}(\theta)$, and power spectra, $C^{ISW-m}_l$ and $C^{m-m}_l$, 
the values of which are obtained similar to (\ref{ctheta}) and (\ref{cl}) by replacing $\delta_{gal} (\hat{\bf n}_1)$ and $I^{gal}_l(k)$ with:
\begin{eqnarray}
\delta_{m}(\hat{\bf n}_1)=\int{\frac{dN}{dz}(z)\delta_{\rm m}(\hat{\bf
n}_1,z)dz}\\
I^{m}_l(k)=\int{\frac{dN}{dz}(z)\delta_{\rm
m}(k,z)j_l[k\chi(z)]dz}
\end{eqnarray}
where the only difference from $\delta_{gal}$ and $I^{gal}_l$ is the bias 
factor. Here, we model $dN/dz$ as a narrow Gaussian centered at two 
different redshifts, z=0.3 and z=3. This will permit to obtain some 
information about the time evolution of the correlations. Realistic 
selection functions will be considered in the next sections.
Cosmological parameters are assumed to be the same as in the WMAP 5 years best fit $\Lambda$CDM model \citep{kom09}.

\subsection{Dependence on $\Lambda$, $\beta$ and $m_\nu$}
Let us consider first the uncoupled case ($\beta=0$) and no massive neutrinos.
Figure \ref{f2} shows the redshift evolution of the gravitational potentials 
(left panel) and the source of ISW effect $\dot \Phi + \dot \Psi$ (right 
panel) for different values of $\Lambda$. The dependence on $\Lambda$ 
of the ISW-matter correlations $C_l^{ISW-m}$ and  
$C^{ISW-m}(\theta)$  is then displayed in Fig. \ref{f3} at two different redshifts,
$z=0.3$ and $z=3$.
When increasing $\Lambda$, both
$C_l^{ISW-m}$ and $C^{ISW-m}(\theta)$ show an opposite behavior at low 
and high redshifts. This reflects the behavior of $\dot\Phi + \dot\Psi$.

In the presence of coupling one can distinguish between two different behaviors
for small and large $\Lambda$'s. This is  shown in Figs. \ref{f4} and \ref{f5}
for $\beta=0.1$. In the first case, the evolution of the gravitational 
potentials and the cross--correlation signal are almost independent from 
$\Lambda$. It can be understood noticing that for small $\Lambda$, 
$\phi$MDE is very long and the 
usual tracker solution is (almost) never reached. In this phase, coupling 
terms in the field equations dominate so that the tracker solution is almost 
independent from $\Lambda$. 
When increasing $\Lambda$, $\phi$MDE becomes shorter and the behavior resemble
that of the uncoupled case.
The transition between the two above regimes occurs around $\Lambda=1$ GeV.

Dependence on $\beta$ is shown in Figs \ref{f6} and \ref{f7}. Again, the 
behavior of the cross--correlation  reflects that of the 
ISW source. However, while coupling can have opposite effects on the cross-correlation 
at different redshifts, i.e. it can increase or decrease the signal, 
massive neutrinos always decrease the signal.
It is shown in Fig. \ref{fnu} which displays the behavior of $C_l^{ISW-m}$
as a function of $m_\nu$ at two different redshifts. Cross-correlation signal
always decreases at the increasing of the neutrino mass.

\section{Galaxy bias and magnification bias}
The galaxy bias $b$ can, in general, evolve both in redshift or as a function of the scale. However,
on the large scales of interest for the ISW effect, the bias is usually assumed to be linear,  spatially constant and only redshift--dependent, i.e. $\delta_{gal}=b(z)\delta_m$. This assumption is fully consistent with results from numerical simulations, redshift surveys and semi--analytic calculation in the contest of the so--called halo--model (see \cite{bla99,per07}).

However, given a galaxy selection function $dN/dz$ picked at certain redshift $\bar z$, the bias can be approximated with an appropriated constant. In this case it will be 
\begin{equation}
C^{gal-gal}=b^2 C^{m-m}
\label{bias}
\end{equation}
Within the above approximation, given a particular survey and assumed a cosmological model, 
the bias is usually estimated by fitting the theoretical matter--matter correlation function, $C^{m-m}$, for the assumed cosmology, to the observed galaxy--galaxy correlation function, ${\hat C}^{gal-gal}$. 

Biases have been estimated for different surveys by several authors assuming the WMAP
best fit $\Lambda$CDM cosmology (see \citet{bou02,bou04,gia06,
mye06,ras07,bla07}). Since we are considering cosmological models different 
from a
$\Lambda$CDM, we need to appropriately rescale those biases to each of the our models.
Note, however, how the estimation of $b$ in (\ref{bias}) depends on the 
normalization of the power spectrum in $C^{m-m}$ (see (\ref{cl})).  
For a fixed normalization, taking into account (\ref{bias}), biases will 
be rescaled according to:
\begin{equation}
b^2_{model}=b^2_{{\Lambda}CDM} {<C^{m-m}_{{\Lambda}CDM}> \over <C^{m-m}_{model}>}
\label{rescal}
\end{equation} 
where $< ~>$ indicates the average on the angular scales $\theta$ of interest.

However, (\ref{rescal})
should be generalized when magnification bias effect due to gravitational 
lensing is non--negligible. Gravitational lensing by intervening matter 
changes the observed galaxy 
number density $\hat \delta_{gal}$, leading a correction term $\delta_\mu$ 
being added to the 
intrinsic galaxy fluctuation $\delta_{gal}$
$$
\hat \delta_{gal}=\delta_{gal}+\delta_{\mu}
$$
With this correction, the observed ACF becomes:
\begin{eqnarray}
\nonumber
\hat C^{gal-gal}&=&C^{gal-gal}+2C^{gal-\mu}+C^{\mu-\mu}\\
&=&b^2C^{m-m}+2bC^{m-\mu}+C^{\mu-\mu}
\end{eqnarray}
where $C^{x-y}=<\delta_x \delta_y>$ and the rescaled bias will then be 
the solution of:
\begin{eqnarray}
\nonumber
b^2_{model}<C^{m-m}_{model}>+2b_{model}<C^{m-\mu}_{model}>+\\
<C^{\mu-\mu}_{model}>-
<\hat C^{gal-gal}_{\Lambda CDM}>=0
\label{rescalmag}
\end{eqnarray}

Auto-- and cross--correlations corrected for magnification bias 
are obtained considering in (25) the function \citep{ho08}:

%\begin{eqnarray}
%\nonumber
\begin{equation}
f(z)=b(z){dN\over dz} \delta_m(k,z)+\int^\infty_z dz' W(z,z')
(\alpha(z')-1) {dN\over dz'}
\label{magb}
\end{equation}
%\end{eqnarray}
where $\alpha(z')$ is the slope of the number counts of galaxy number 
density as a function of the flux, $N(>F)\propto F^{-\alpha}$. It depends
on the choice of galaxy sample and is redshift dependent. 
The lensing window function (in a flat Universe) is:
$$
W(z,z')=k^2 \Phi(k,z) 
{\chi(z')-\chi(z) \over \chi(z')} \chi(z)
$$  

\begin{figure}[t]
\begin{center}
\includegraphics[scale=0.3, angle=-90]{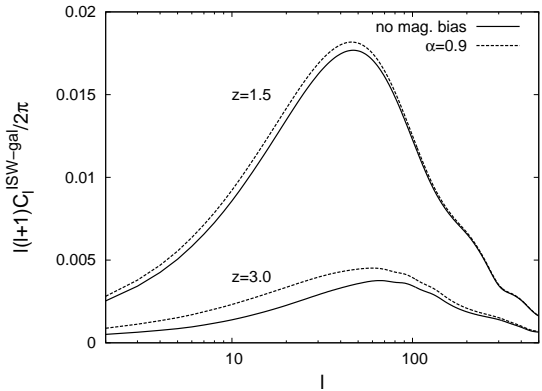}
\caption{Comparison between cross-correlation spectra with and without magnification bias at
z=1.5 and z=3.0}
\label{mag}
\end{center}
\end{figure}

Magnification bias increases with redshift and could be important 
when dealing with deep survey, e.g. quasars. This is shown in
Fig. \ref{mag} which compares the effect of the  magnification bias 
on the ISW-gal correlation at $z=1.5$ and $z=3$.
A detailed analysis on how magnification bias affects ACF and CCF 
can be found in \citet{lov07,hui07,lov08,hui08}.

\begin{figure}[b]
\begin{center}
\includegraphics[scale=0.3, angle=-90]{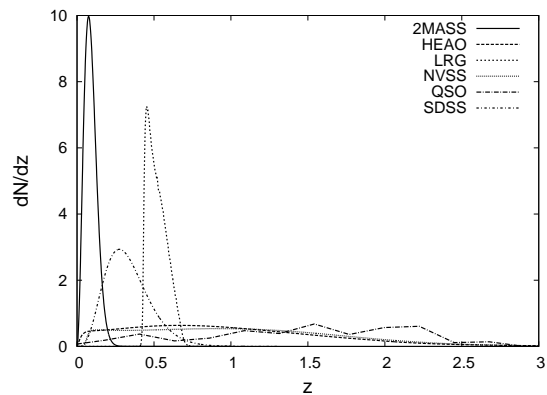}
\caption{Redshift distribution $dN/dz$ of all catalogues}
\label{f10}
\end{center}
\end{figure}

%%%%%%%%%%%%%%%%%%%%%%%%%%%%%%%%%%%%%%%%%%%%%%%%%%%%%%%%%%%%%%%%%%%%%
\begin{table}[b]
%\vskip -.2truecm
\begin{center}
\begin{tabular}{lcc}
\\
\hline
{\rm Parameter}& RP  & SUGRA  \\

\hline
%\hline
%\\
{$10^2\, \omega_{b}$} & 2.260 $\pm$ 0.061 & 2.260  $\pm$ 0.065\\ 
%\\

{$\omega_{c}$} & 0.1039 $\pm$ 0.0062      & 0.1042  $\pm$ 0.0084 \\ 
%\\

{$\tau$}& 0.087 $\pm$ 0.016 &        0.088 $\pm$ 0.017       \\   
%\\
$M_\nu$ (eV) (95\% C.L.) &  {$<$~1.13} & {$<$~1.17} \\   
%\\
$\beta$ (95\% C.L.) & {$<$0.17} & {$<$0.18} \\   
%\\
$\log_{10}(\Lambda/\textrm{GeV})(95\% C.L.)$ & {$<$~-4.2} & {$<$~6.3} \\   
%\\
{$n_s$} & 0.969  $\pm$ 0.015       & 0.970 $\pm$ 0.018      \\
%\\
{{\rm ln}$(10^{10}A_s)$}& 3.055 $\pm$ 0.040      & 3.057 $\pm$ 0.041      \\
%\\
$H_0$ {\rm (km/s/Mpc)} & 71.8  $\pm$ 2.5     & 71.9 $\pm$ 2.7      \\
%\\
\hline

\end{tabular}
\caption{Best fit values and $1- \sigma$ error bars for RP and SUGRA models. 
Only upper limits on $M_{\nu}$, $\beta$ and $\Lambda$ are
shown. }
\label{comparisontabble}
\end{center}
\end{table}
%%%%%%%%%%%%%%%%%%%%%%%%%%%%%%%%%%%%%%%%%%%%%%%
\begin{table}[b]
\begin{center}
\begin{tabular}{lccc}
\\
\hline 
  & $\Lambda$CDM & RP & SUGRA  \\
\hline
%\\
2MASS & 1.40 & 1.46  & 1.47 \\
%\\
SDSS gal & 1.00 & 1.03 & 1.04 \\
%\\
LRG & 1.80 & 1.83 & 1.84 \\
%\\
NVSS & 1.50 & 1.53 & 1.54 \\
%\\
HEAO & 1.06 & 1.09 & 1.09 \\
%\\
QSO & 2.30 & 2.33 & 2.33 \\
\hline
\\ \\
\end{tabular}
\caption{Galaxies biases for different catalogues and models. 
Biases are calculated according to (\ref{rescal}) using for 
$b_{{\Lambda}CDM}$ the values given in \citet{gia08a}.}
\end{center}
\end{table}

\begin{figure*}[]
\begin{center}
\includegraphics[scale=0.3, angle=-90]{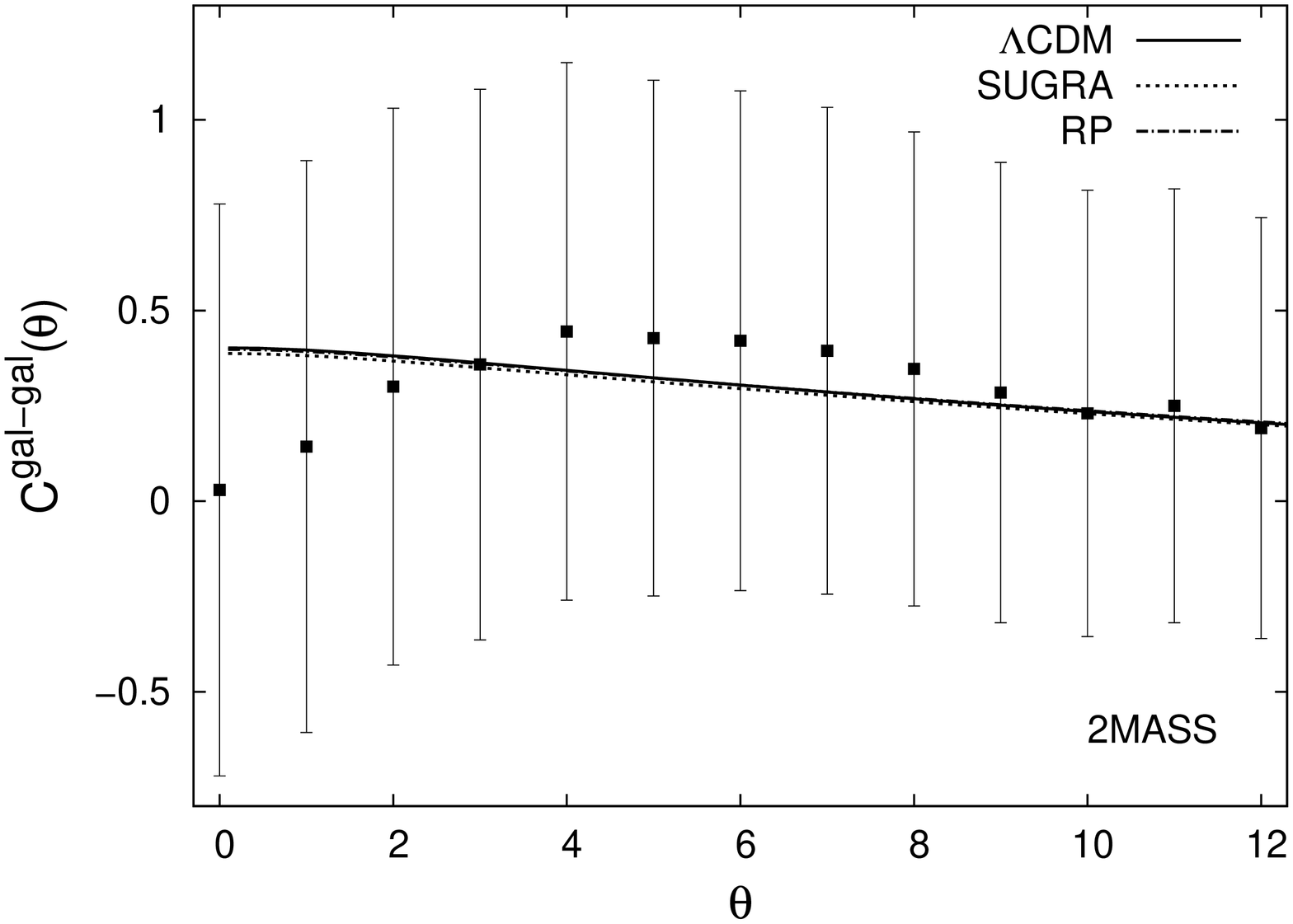}
\includegraphics[scale=0.3, angle=-90]{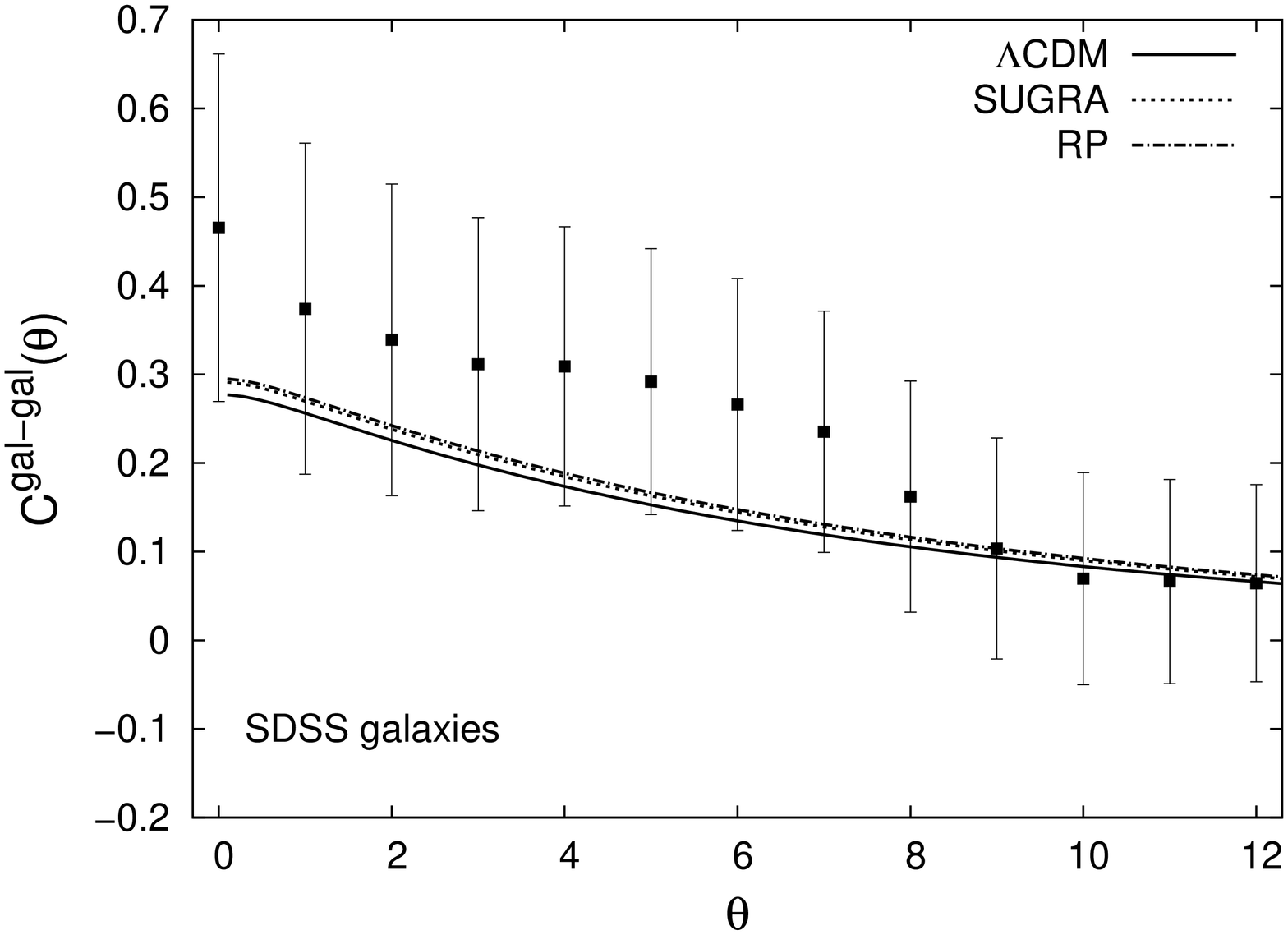}
\includegraphics[scale=0.3, angle=-90]{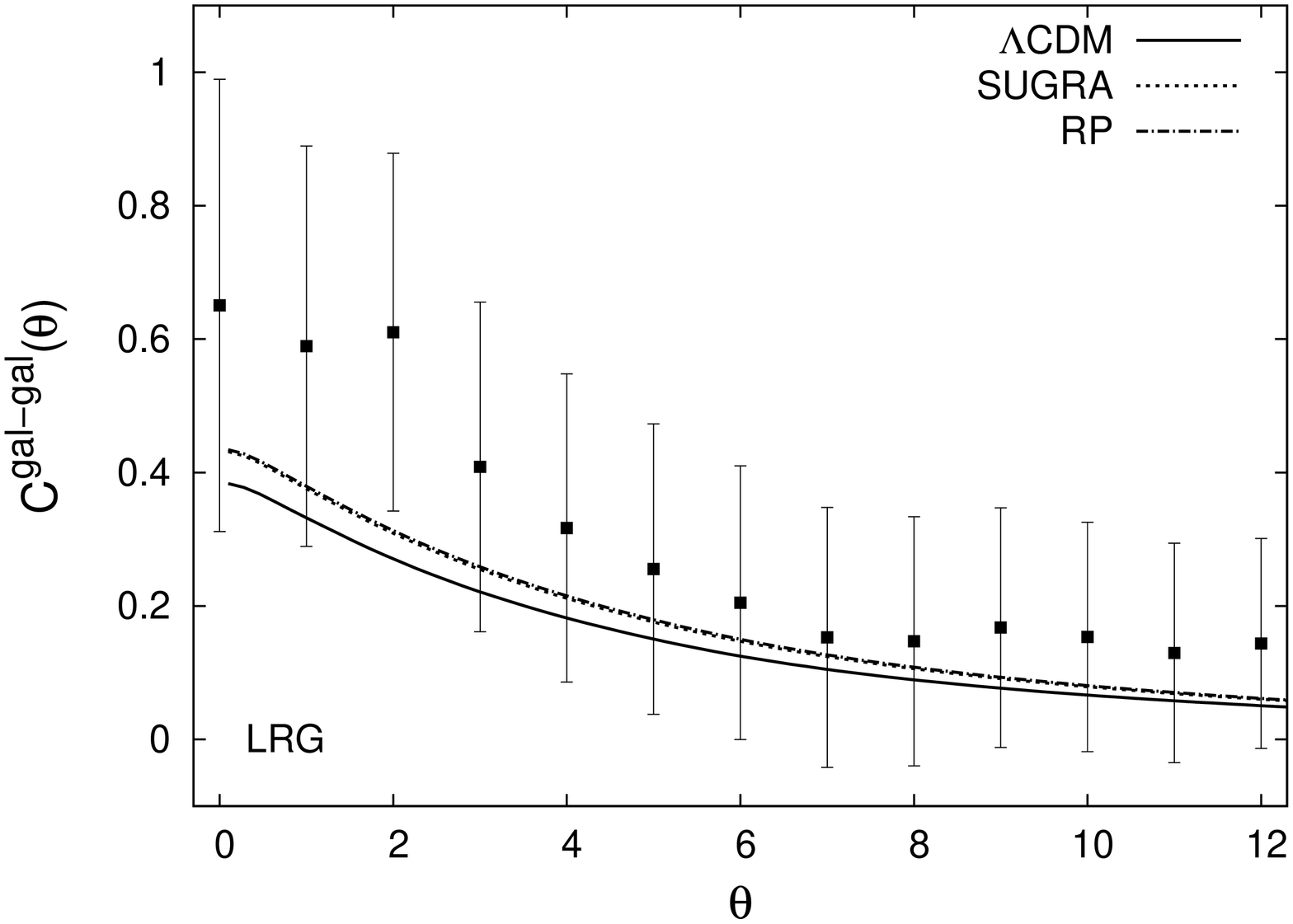}
\includegraphics[scale=0.3, angle=-90]{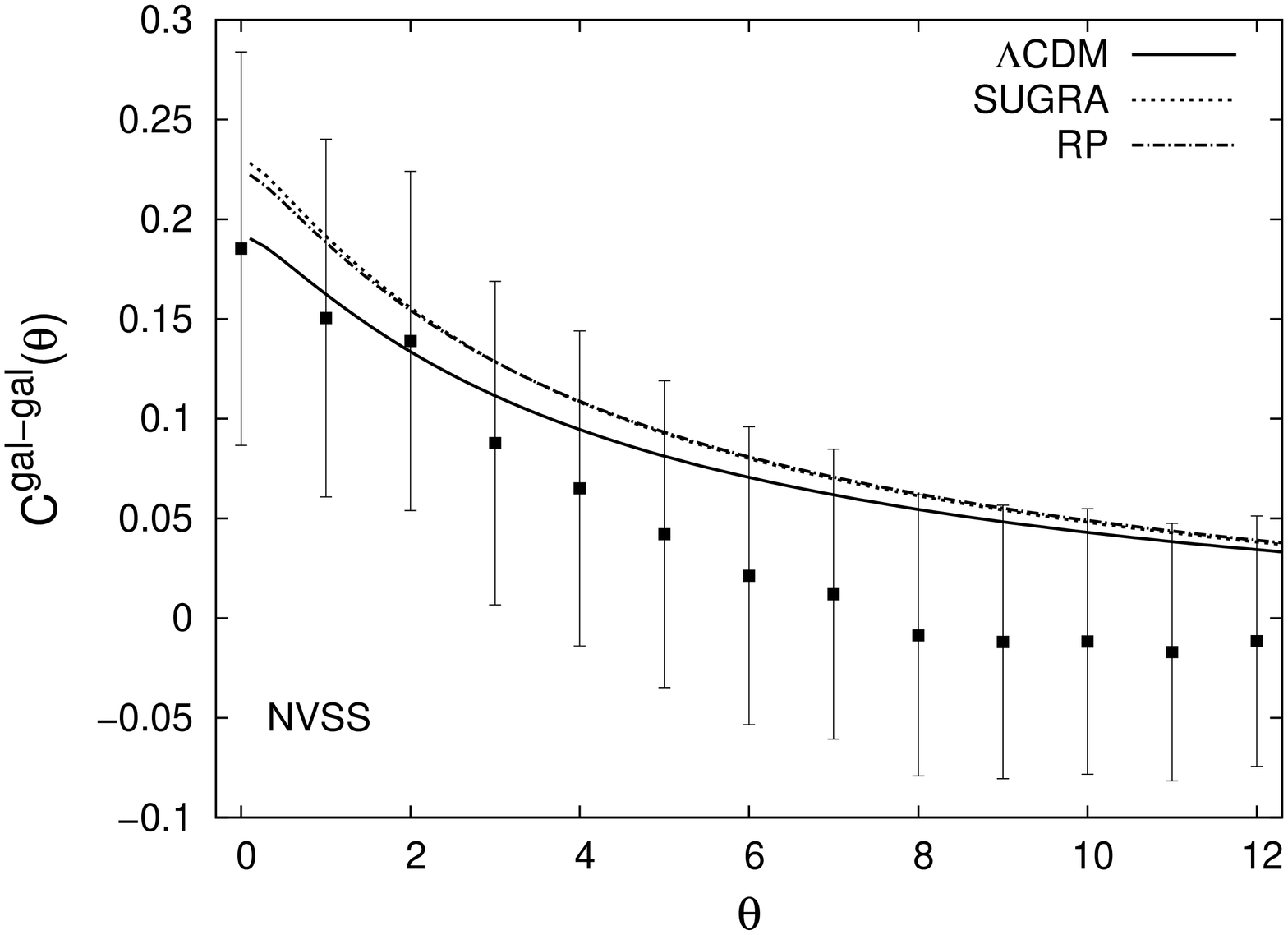}
\includegraphics[scale=0.3, angle=-90]{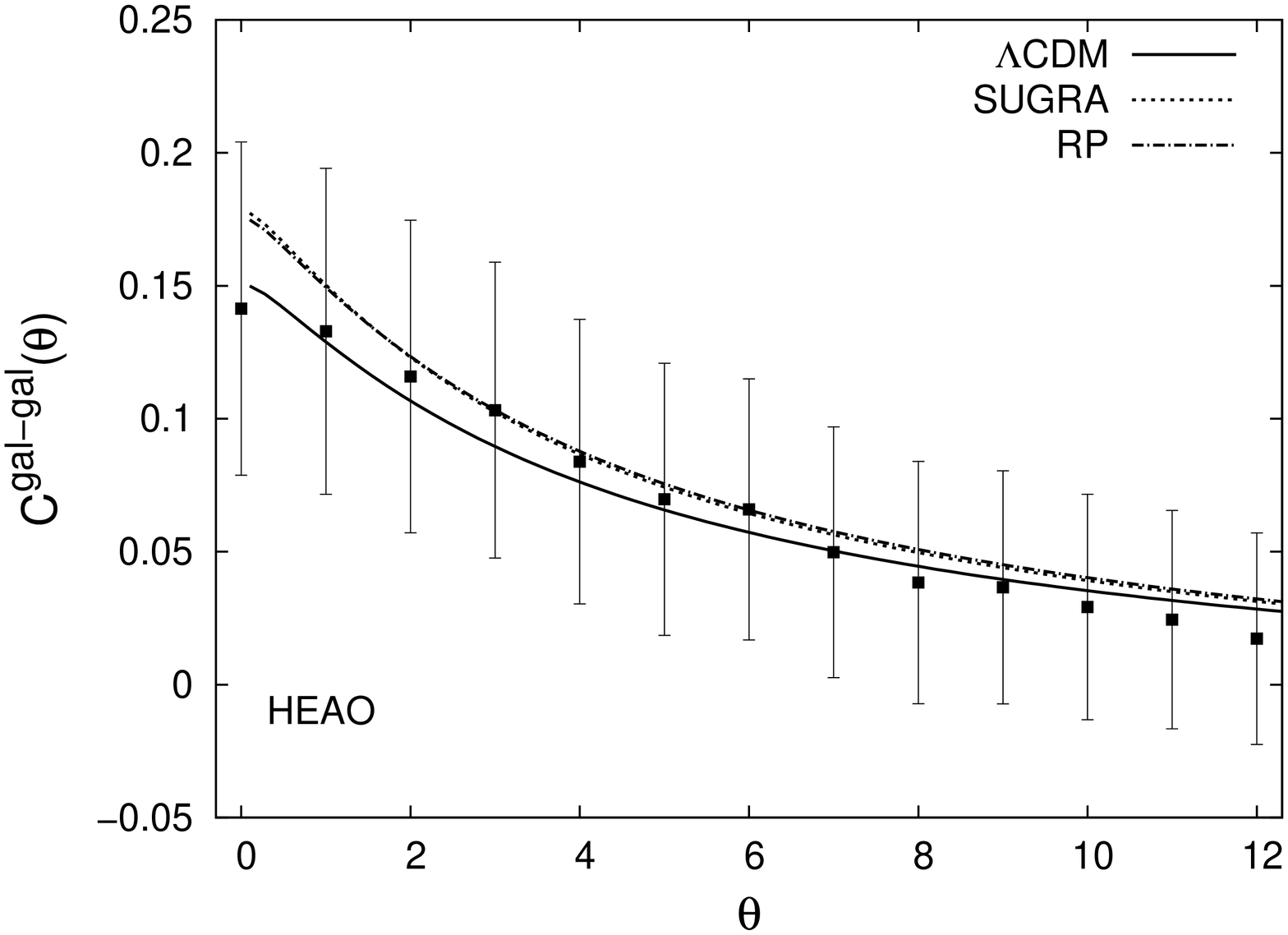}
\includegraphics[scale=0.3, angle=-90]{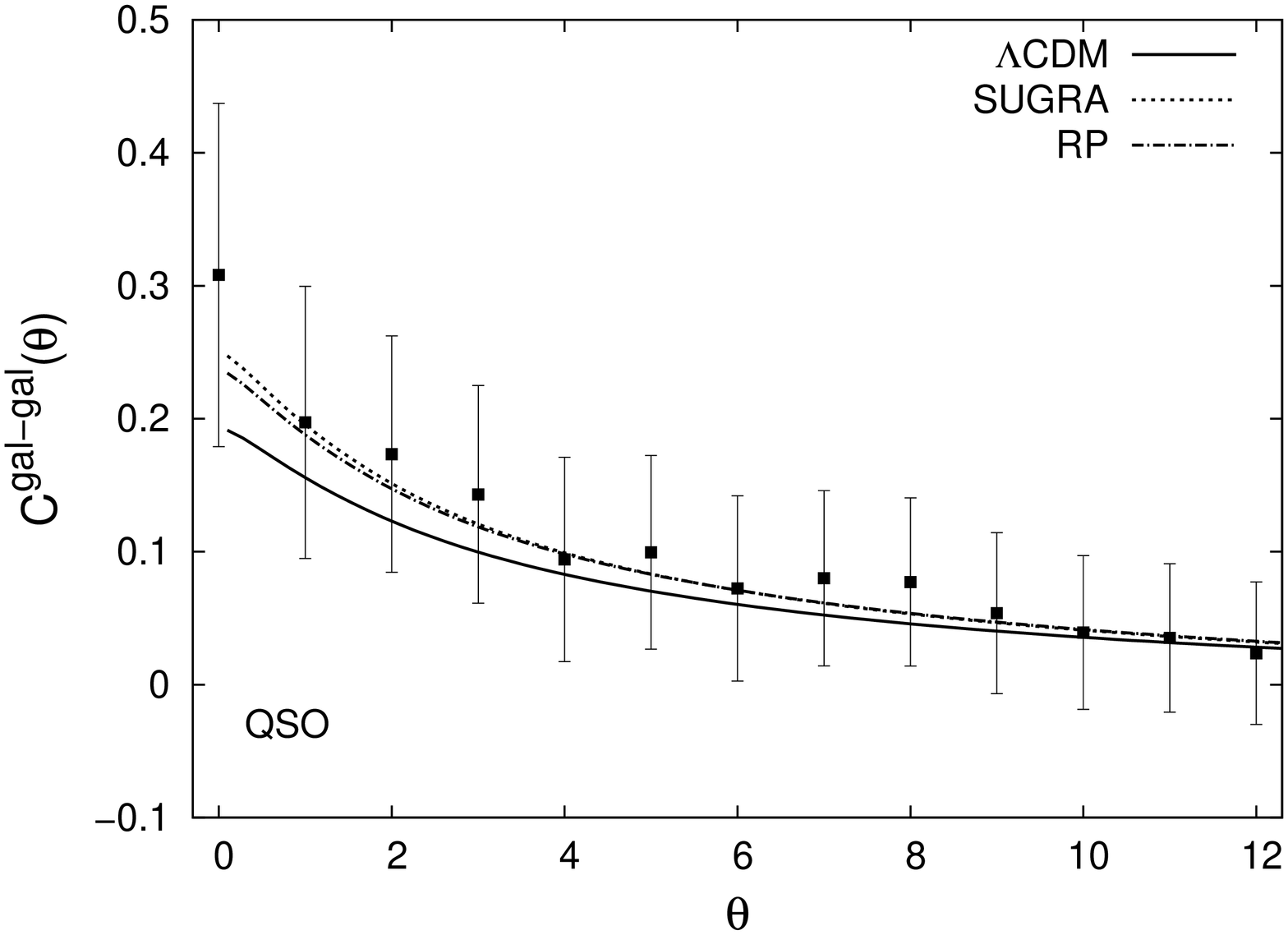}
\label{data}
\caption{The observed CCF of six different galaxy catalogues.
The curves are the theoretical predictions for the best fit 
$\Lambda$CDM, SUGRA and RP cosmologies (see text)}.
\label{f11}
\end{center}
\end{figure*}

\begin{figure*}[t]
\begin{center}
\includegraphics[scale=0.23, angle=-90]{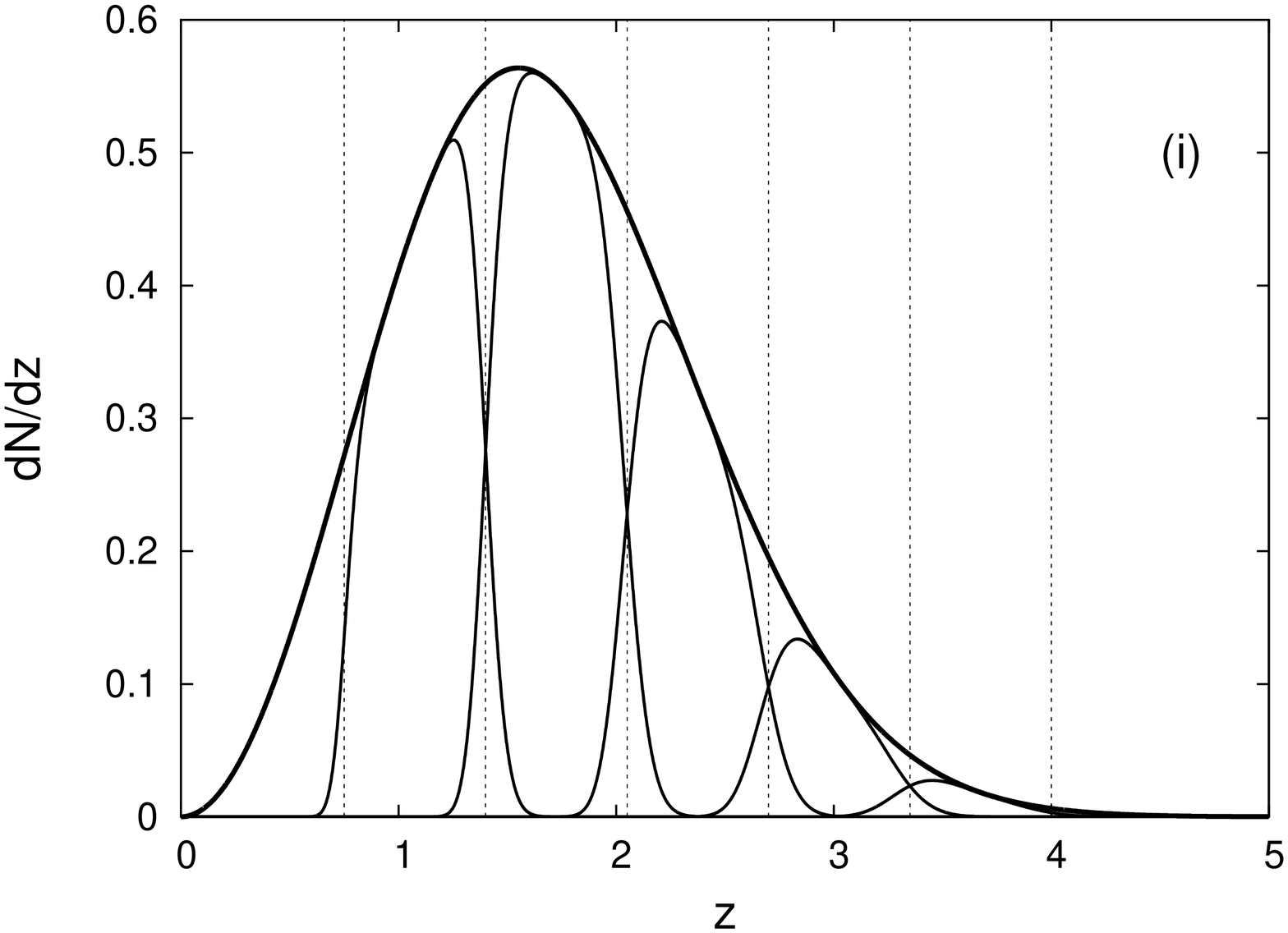}
\includegraphics[scale=0.23, angle=-90]{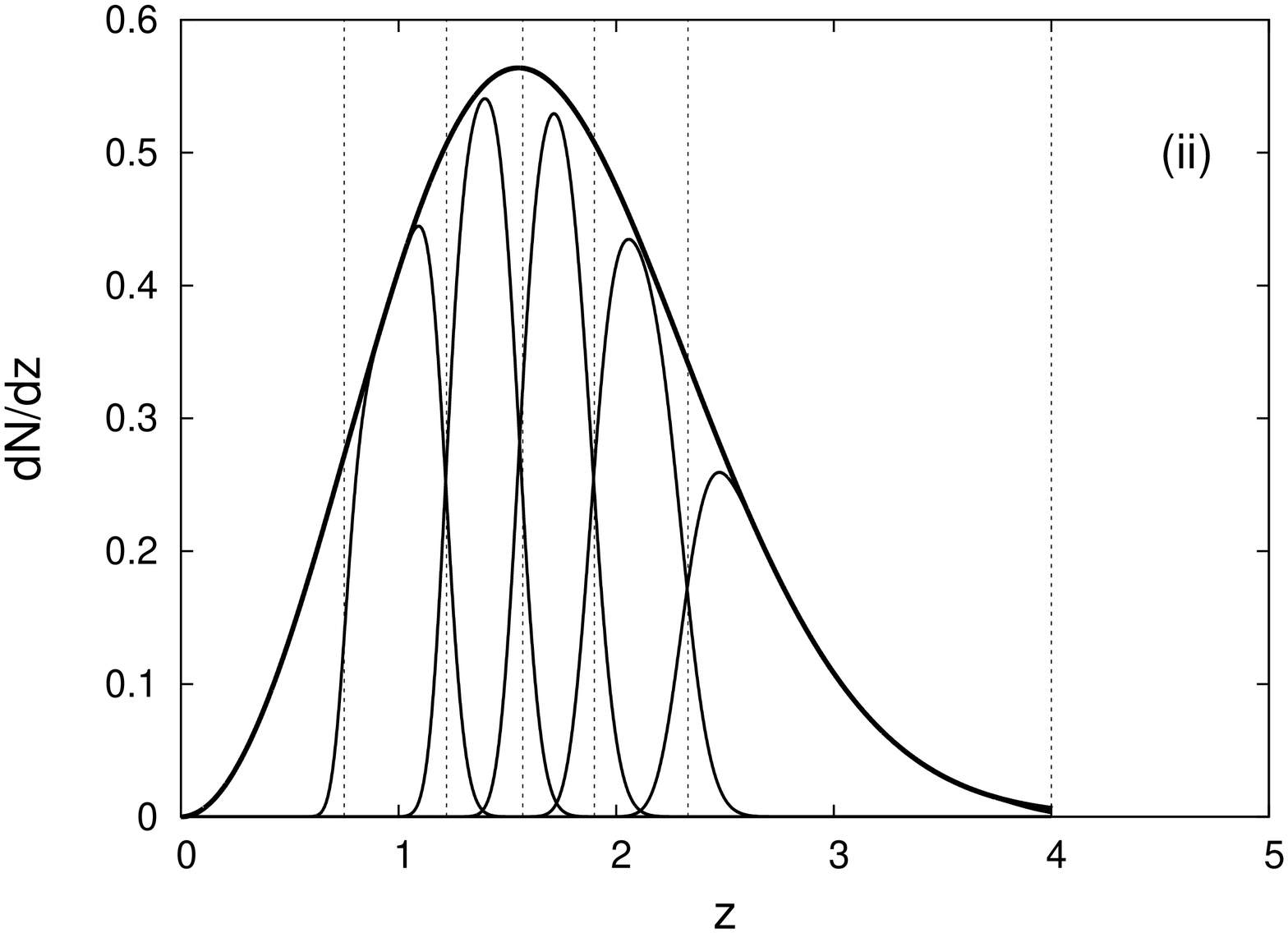}
\includegraphics[scale=0.23, angle=-90]{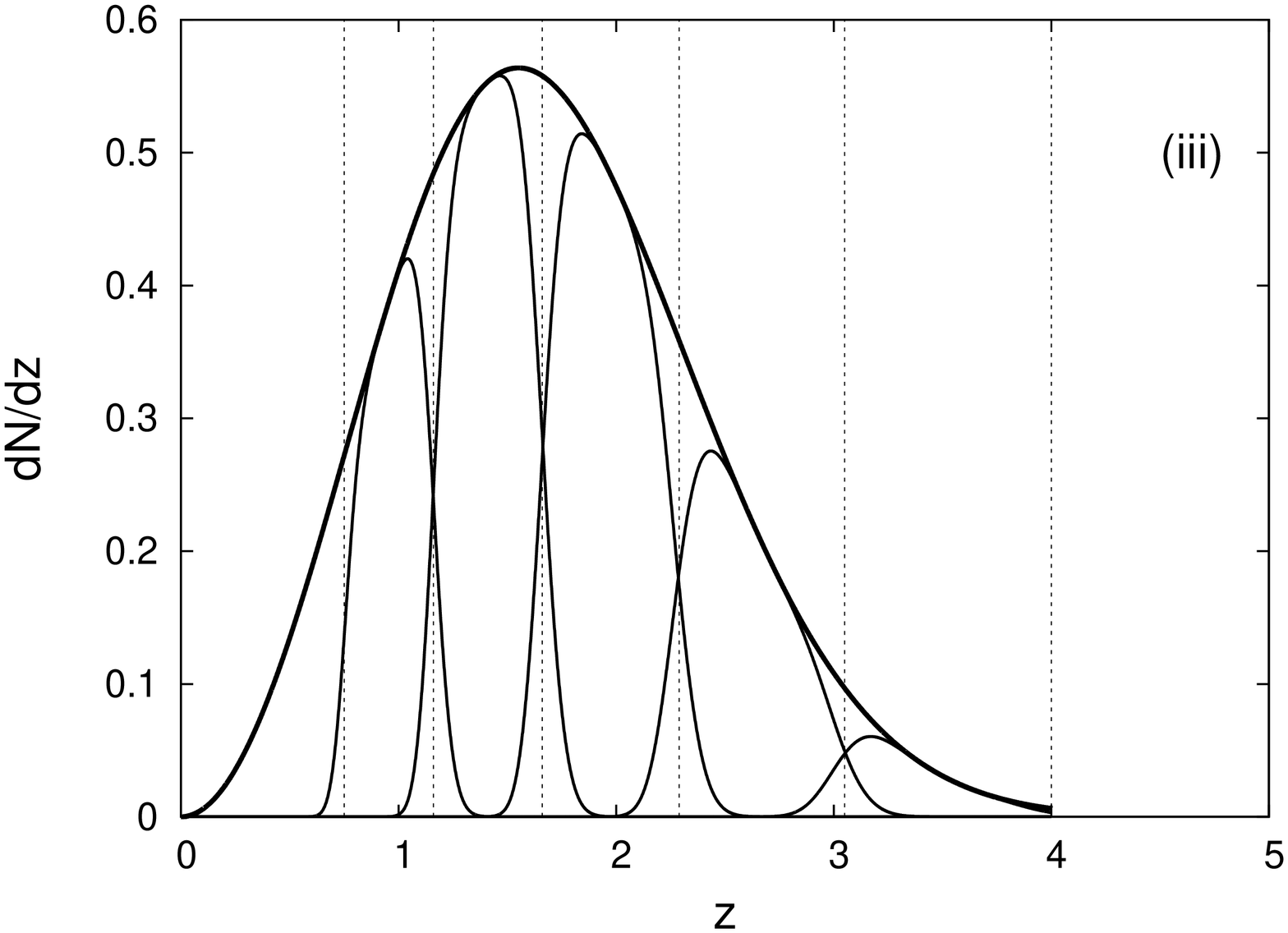}
\caption{Splitting schemes (i) (left panel), (ii) (middle panel) and 
(iii) (right panel) described in the text.}
\label{f13}
\end{center}
\end{figure*}

\section{Comparison to observational data}
\label{comparison}

Investigations of CMB--LSS correlations were made in a recent series of works which rely on 
WMAP data and a variety of LSS probes \citep{nol04,afs04,cab06,cab07,ras07,
rac08,ho08,gia08a,xia09}. 
There is an overall agreement among different groups in finding an evidence for a positive ISW effect at the $\sim 3-4 \sigma$ confidence level. It was also found a substantial agreement between the observed cross-correlations and the expected signal arising from the ISW effect in the WMAP best fit $\Lambda$CDM cosmology. Different DE mod\\ 
els were also considered in \citet{oli08,sch08,gia08b}.

In a recent work \citep{gia08a} a combined analysis of the ISW effect was performed by cross--correlating CMB map provided by the WMAP collaboration with all the relevant large scale data sets and modeling their covariance properties with different methods.

In this section, we compare our theoretical predictions 
based on the models 
described above with the
measurements made in  \citet{gia08a} by considering six different galaxy catalogues: the optical Sloan Digital Sky Survey (SDSS), the infrared 2 Micron All-Sky Survey (2MASS), the X-ray catalogue from the High Energy Astrophysical Observatory (HEAO) and the radio galaxy catalogue from the NRAO VLA Sky Survey (NVSS). In addition, given the high quality of the SDSS data, some further subsamples was extracted from it, consisting of Luminous Red Galaxies (LRG) and quasars (QSO). 

\begin{figure*}[t]
\begin{center}
\includegraphics[scale=0.23, angle=-90]{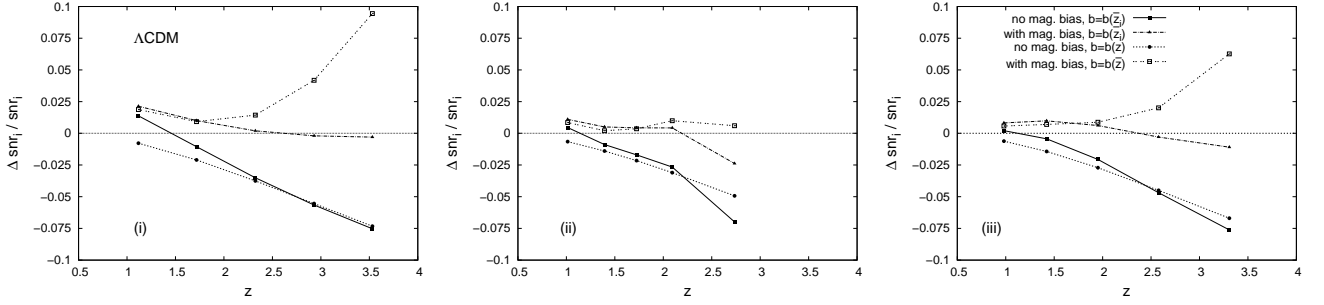}
\caption{Relative error on $snr$ in the $i$--th tomographic bin for
  the cases in Table \ref{mb} and the three different splitting
  schemes described in the text.} 
\label{snrbm}
\end{center}
\end{figure*}

\begin{figure*}[t]
\begin{center}
\includegraphics[scale=0.23, angle=-90]{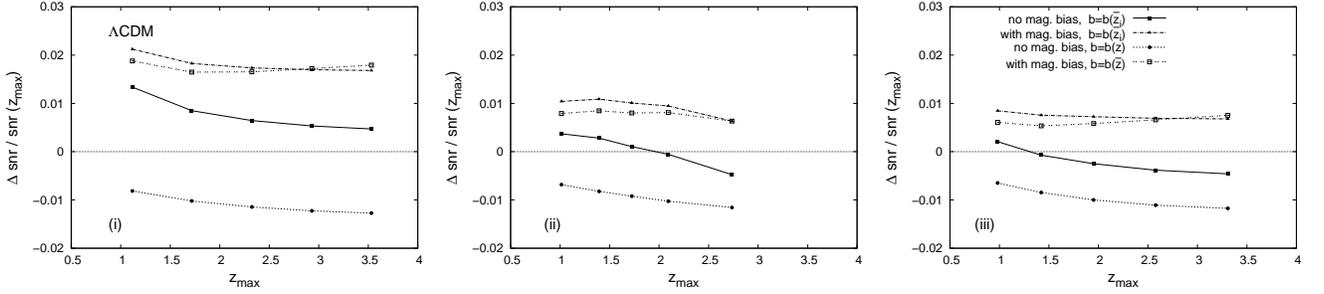}
\caption{Same as Fig. \ref{snrbm} but for the cumulative $snr$.} 
\label{snrbmc}
\end{center}
\end{figure*}

\begin{figure}[b]
\begin{center}
\includegraphics[scale=0.3, angle=-90]{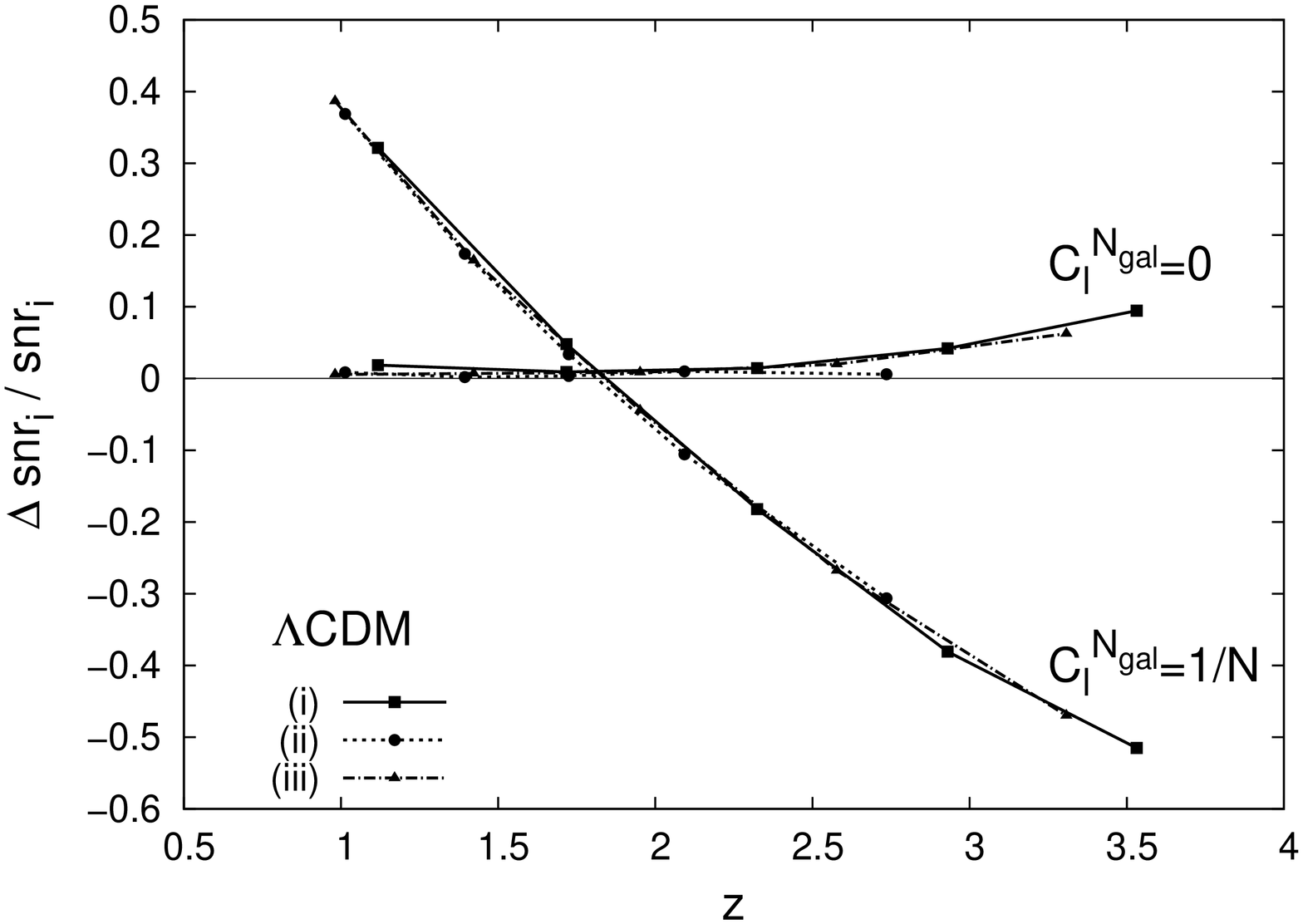}
\includegraphics[scale=0.3, angle=-90]{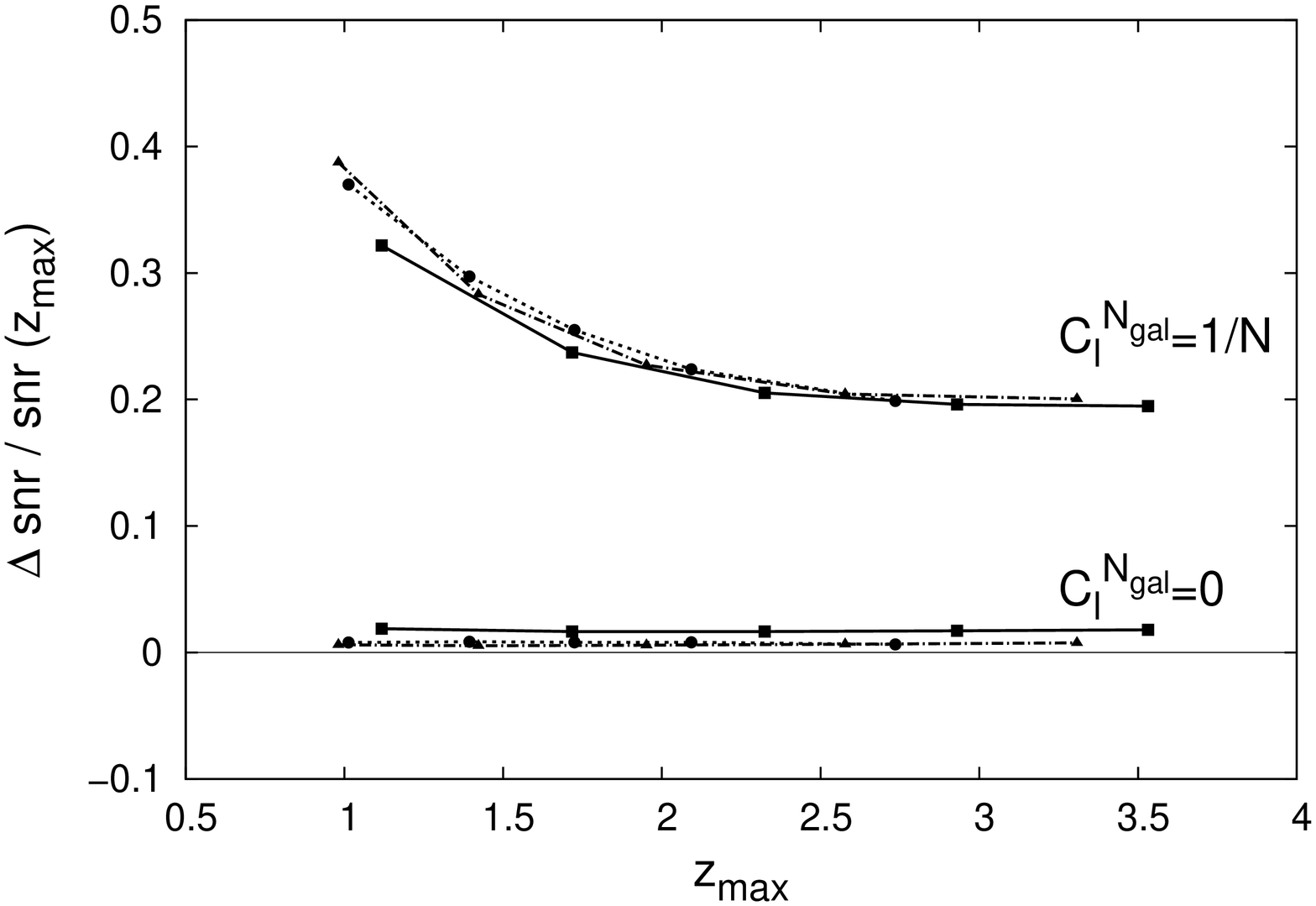}
\caption{
Relative error on $snr_i$ (top panel) and $snr(z_{max})$ (bottom
panel) for the cases in Table\ref{mb}. We assume a shot noise as for
the SDSS DR6 survey and compare with the case of negligible shot noise.}
\label{snrbmnoise}
\end{center}
\end{figure}

\begin{figure*}[t]
\begin{center}
\includegraphics[scale=0.3, angle=-90]{splitdz.ps}
\includegraphics[scale=0.3, angle=-90]{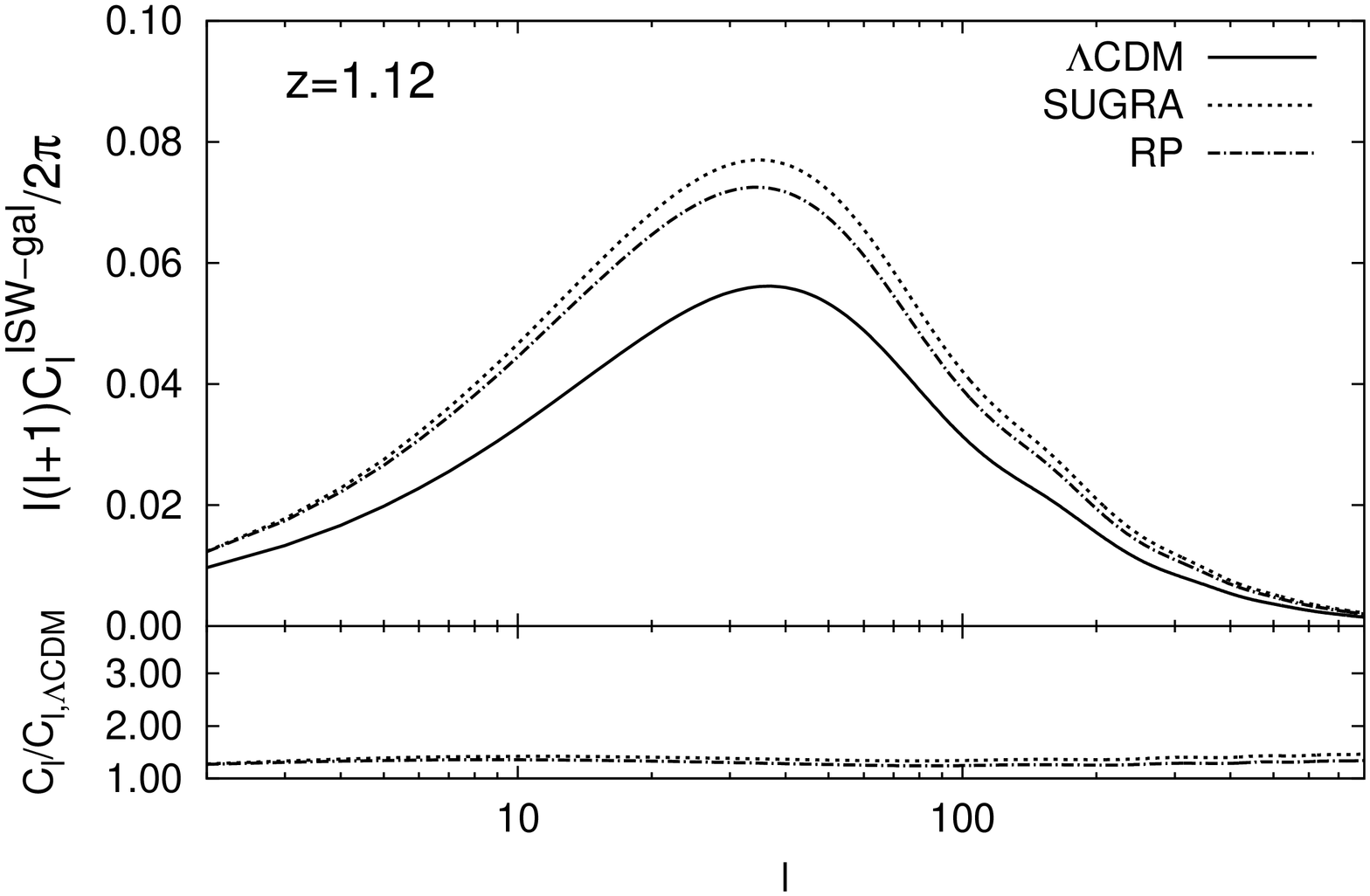}
\includegraphics[scale=0.3, angle=-90]{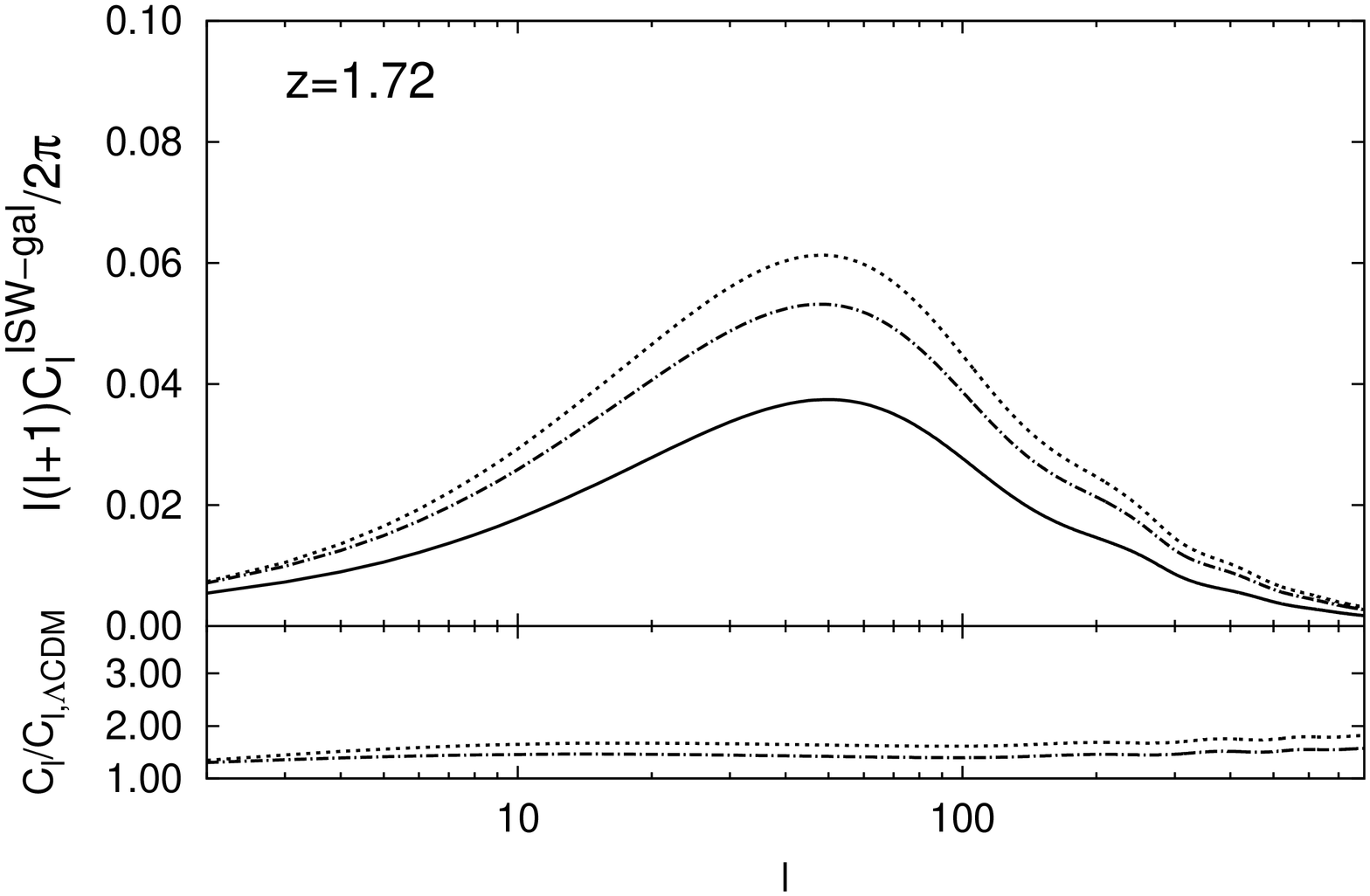}
\includegraphics[scale=0.3, angle=-90]{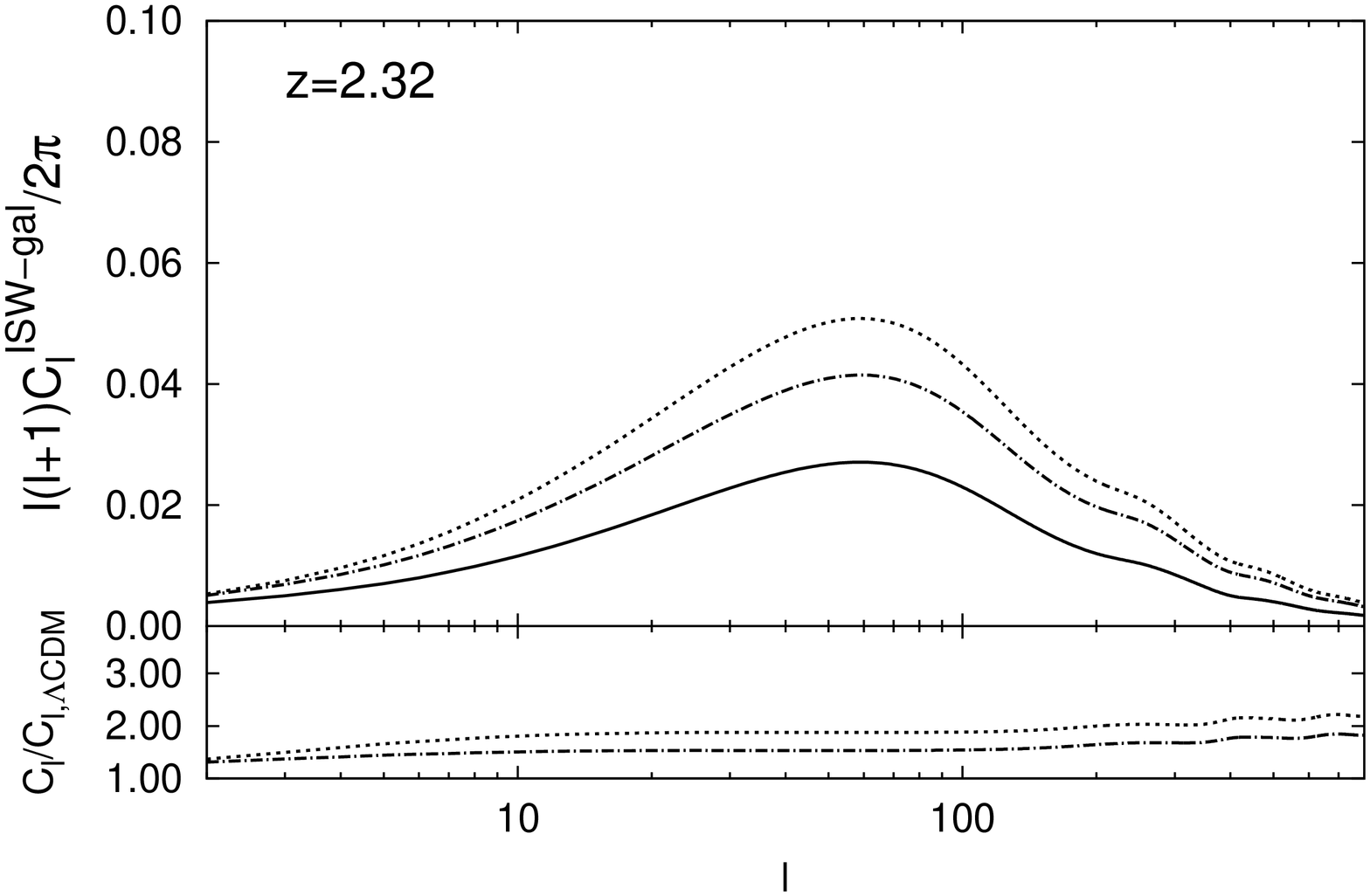}
\includegraphics[scale=0.3, angle=-90]{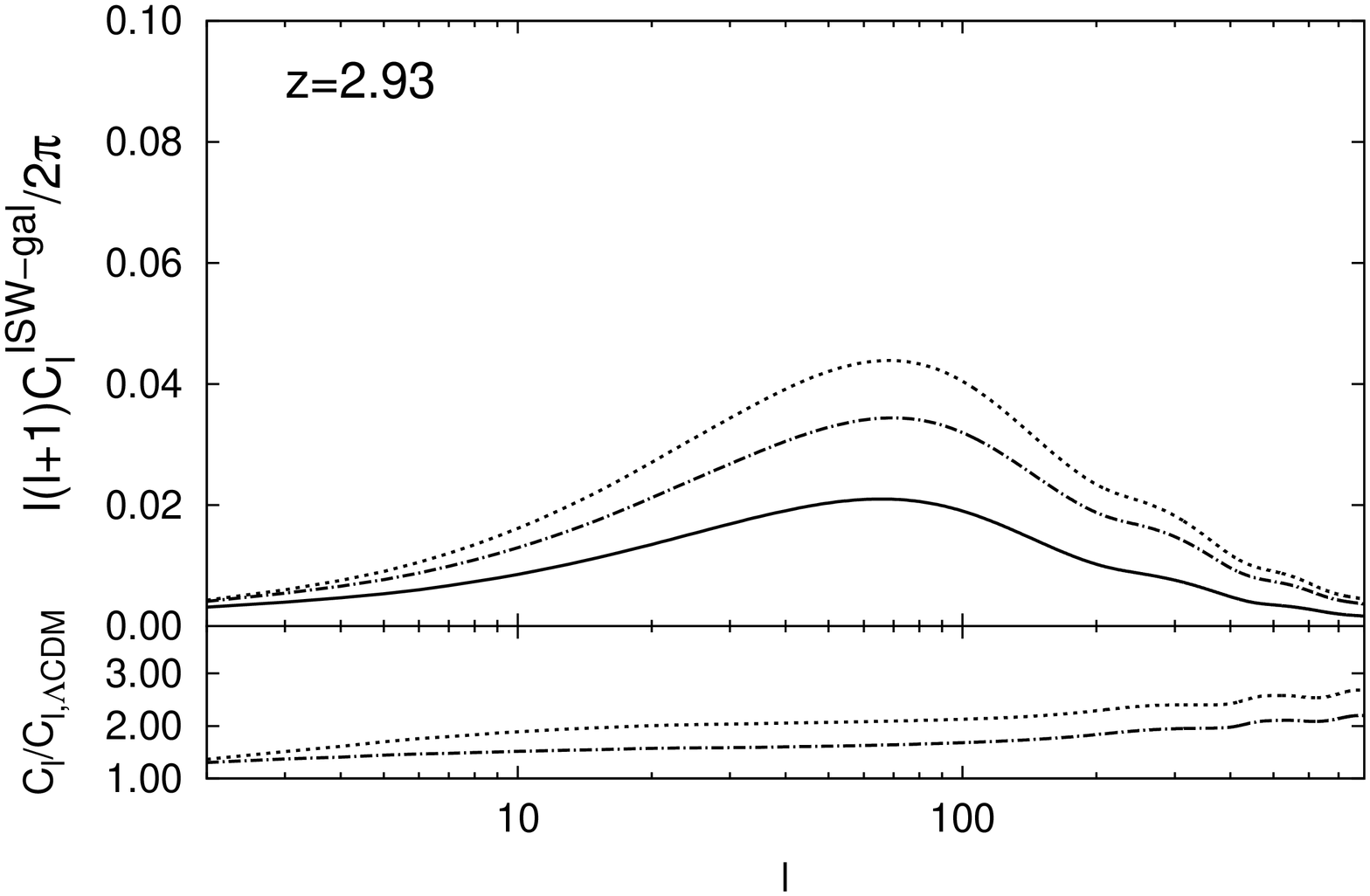}
\includegraphics[scale=0.3, angle=-90]{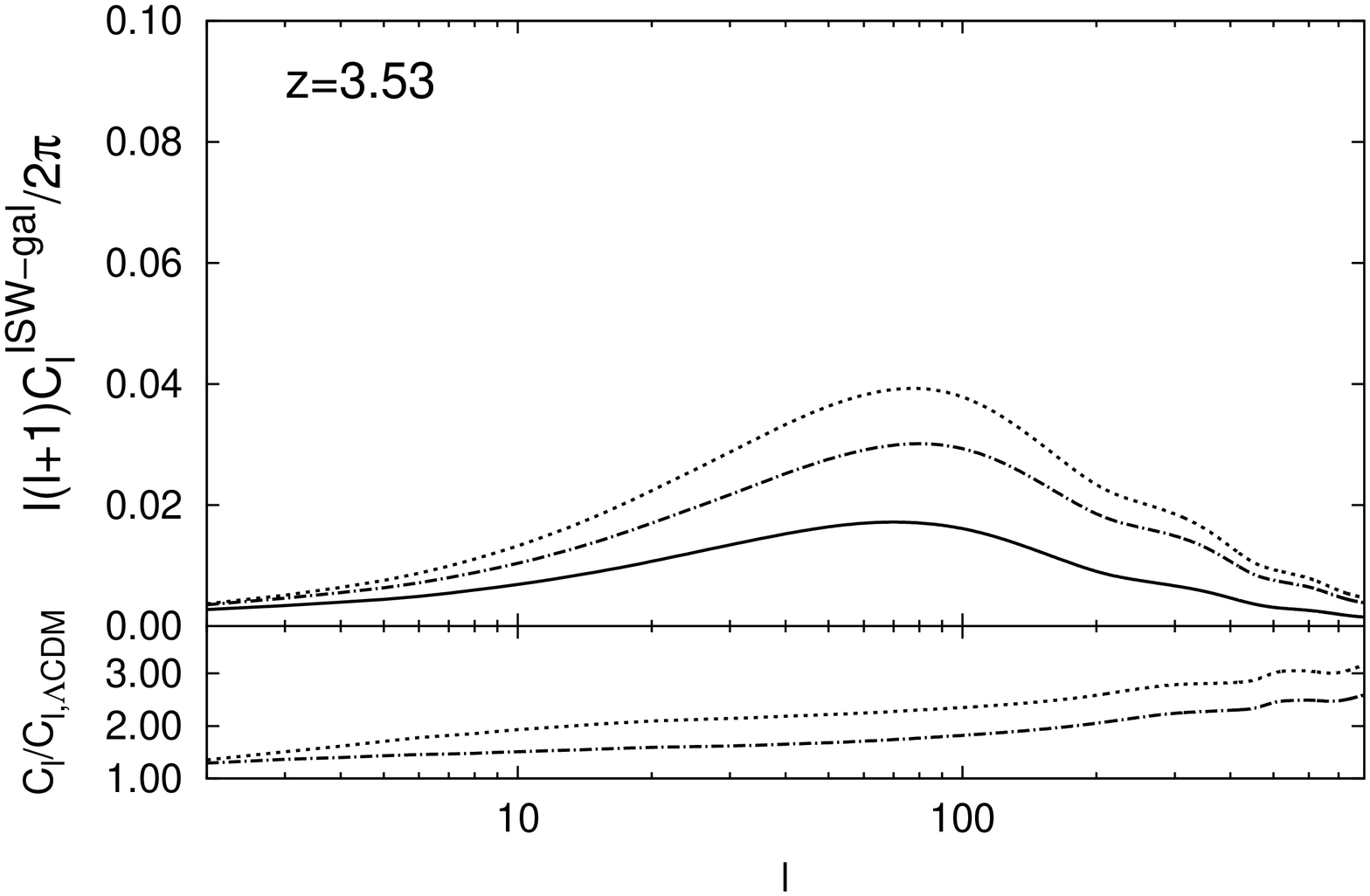}
\caption{The splitting scheme (i) (see text) is shown 
in the top panel on the left. 
Other panels show the cross--correlation signal
in the five bins considered. Lower frames of each panel display 
the ratio between the 
SUGRA and RP spectra and the $\Lambda$CDM spectrum 
}
\label{f12}
\end{center}
\end{figure*}

\begin{figure*}[htb]
\begin{center}
\includegraphics[scale=0.23, angle=-90]{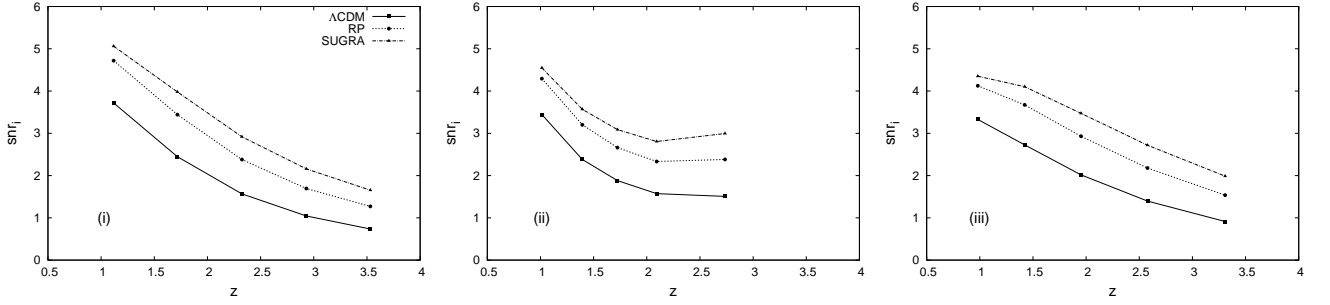}
\caption{
The expected ISW--LSS cross correlation signal to noise ratio, $snr$, for the different splitting schemes considered in the text 
as a function of the 
mean redshift of the bins for the best fit $\Lambda$CDM, RP and SUGRA
cosmologies.} 
\label{snr}
\end{center}
\end{figure*}

\begin{figure*}[htb]
\begin{center}
\includegraphics[scale=0.23, angle=-90]{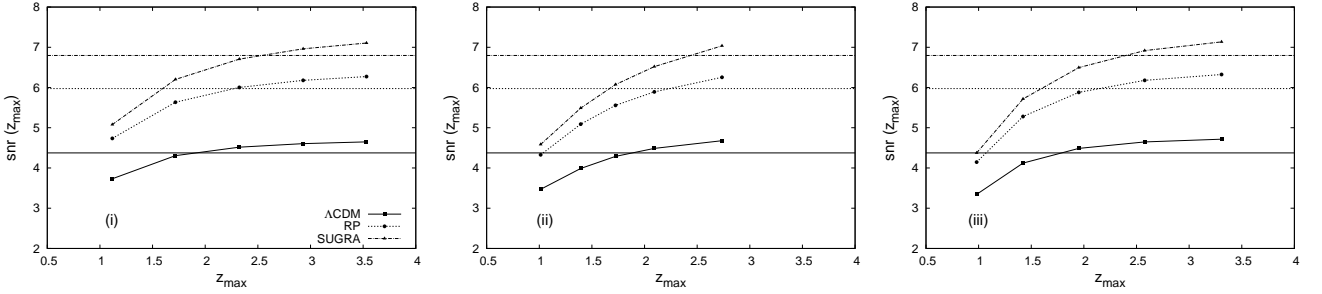}
\caption{Same as previous Figure but for the cumulative $snr$.
Horizontal lines indicate the values of $snr$ obtained by using the 
overall distribution.}
\label{snrc}
\end{center}
\end{figure*}

\begin{figure}[b]
\begin{center}
\includegraphics[scale=0.3, angle=-90]{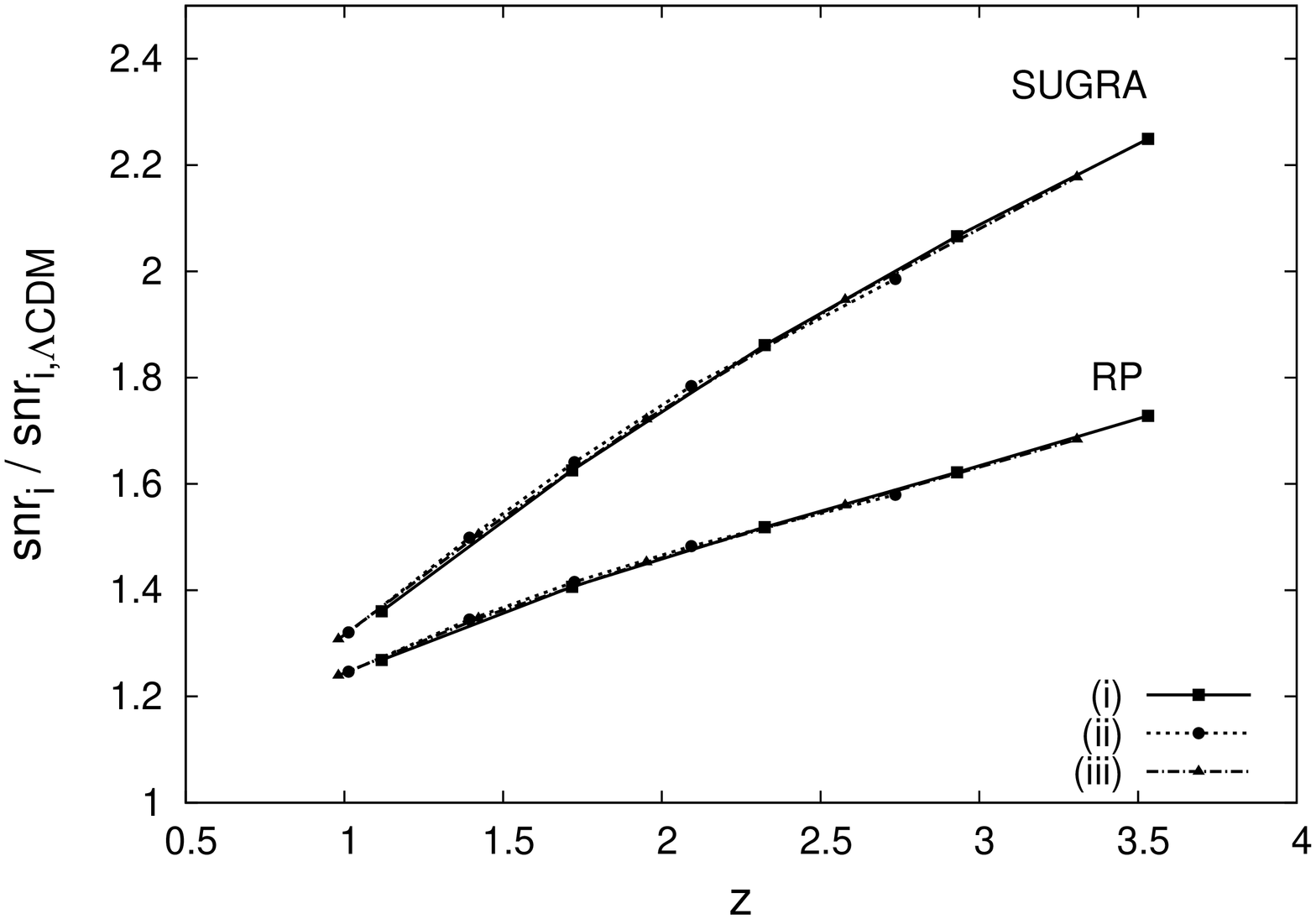}
\includegraphics[scale=0.3, angle=-90]{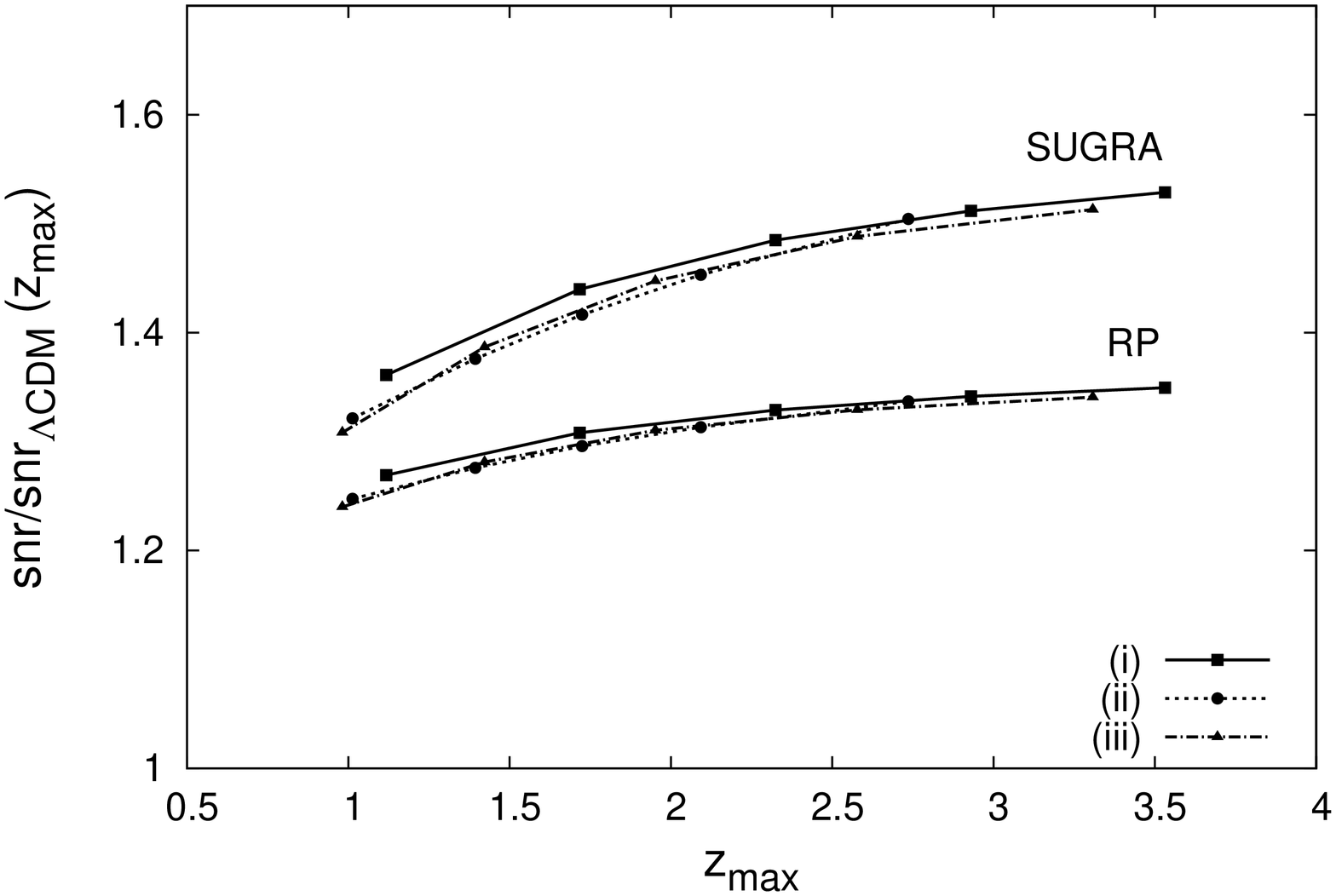}
\caption{RP and SUGRA $snr$'s of Fig. \ref{snr}  (top panel) and Fig. 
\ref{snrc} (bottom panel)
normalized by the $\Lambda$CDM ones. $snr$ obtained by using the full
distribution are also shown.\\}
\label{snr2}
\end{center}
\end{figure}

As shown in Fig. \ref{f10} their redshift distributions, $dN/dz$, span a 
redshift range $0<z<2.5$. In order of increasing mean redshift of the sample
we have:
2MASS, SDSS galaxies, LRG, NVSS, HEAO and QSO 
(for details on the catalogues see \citet{gia08a,ho08}.

Constraints from CMB, SNIa and LSS on cosmologies with coupling and
massive neutrinos have been obtained in \citet{lav09b,kri10}
by means of Monte Carlo Markov Chain technique. 
Best fit parameters from their analysis will be 
used in this and in the next section  when dealing with 
redshift tomography. 
Parameters are summarized in Table \ref{comparisontabble} where
 $\omega_{b,c}$ are the
physical baryon and cold dark matter density parameters,
$\omega_{b,c}=\Omega_{b,c} h^2$, where $h$ is the dimensionless Hubble
parameter,
 $\tau$ is the optical depth to
reionization, $n_s$ is the scalar spectral index, $A_s$ denotes the
amplitude of the scalar fluctuations at a scale of $k=0.05
\textrm{Mpc}^{-1}$, $M_\nu = \Sigma m_\nu$, assuming 3 equal neutrino masses
 $m_\nu$, $\Lambda$ denotes the energy scale in DE potentials, while
$\beta$ is the coupling parameter between DM and DE. In the following we will
use $\Lambda=10^{-6} GeV$ and $\Lambda=1 GeV$ for RP and SUGRA respectively 
which correspond to the $\sim$For the three splitting schemes described above 1--$\sigma$ limits. Best fit parameters for 
coupling and neutrino mass are approximately the same for both model, i.e.
$\beta \sim 0.1$ and $\Omega_{\nu}= M_{\nu} /h^2 93.14 eV \sim 0.01$.

For each catalogue and model, biases are shown in Table 1  
and calculated according to (\ref{rescal}) using  for $b_{\Lambda CDM}$ 
the values given in \citet{gia08a}. Given the low mean redshifts
of the catalogues we neglect the magnification bias effect which amounts to 
a few percent only in the case of quasars. 
It will be, however, considered in the next Section when dealing with redshift 
tomography and higher $z$. We will also discuss how good to approximate 
$b$ with a constat is. 

For each catalogue, we then determine the expected CCFs
for our models.

Comparison with observational data  is shown in Fig. \ref{f11} for SUGRA 
and RP models. The predictions for the $\Lambda$CDM  is also displayed.
Note that, because of know contamination from Sunyaev--Zeldovich effect in 
the 2MASS data \citep{2mass}, the four smallest angle bins should be disregarded.
While it is not possible to distinguish among the models at low redshifts,
discrepancies between coupled models and $\Lambda$CDM increase with $z$
even though RP and SUGRA models remain indistinguishable. 
In spite of this, however, current data alone seem not able to discriminate 
between coupled models and $\Lambda$CDM.

\section{Redshift tomography}

As already outlined, unlike uncoupled DE models with massless neutrinos,
both coupling and massive neutrinos causes the gravitational potentials to 
evolve even in the matter dominated epoch. Therefore, a detection of a 
non--vanishing ISW effect signal at such high redshifts would rule out 
a vast class of DE models indicating  
a possible interaction in the dark sector. 
Upcoming galaxy surveys will cover a large redshift range. One goal will be 
to use the photometric redshifts of the galaxies to 
split the survey into multiple redshift bins allowing for tomographic analysis.
 
Following the procedure of \citet{hu04}, given a galaxy distribution,
$n(z)= dN/dz$, the galaxies can be divided
into photometric bins, labelled with index $i$:
$$
n(z)=\sum_i n_i(z)
$$ 
Assuming a distribution $n(z)$ of the standard form:
\begin{equation}
n(z)=\frac{\beta}{\Gamma(\frac{m+1}{\beta})}\frac{z^m}{z^{m+1}_0}
\exp\left[-\left(\frac{z}{z_0}\right)^\beta\right]
\label{galaxies}
\end{equation}
and that the photometric redshift errors are Gaussian distributed
with an rms fluctuation $\sigma(z)$, the resulting photometric 
redshift distributions are given by:
$$
n_i(z)={1 \over 2}n(z)\left[erfc \left({z_{i-1}-z \over \sqrt{2}\sigma(z)}
\right)-erfc \left({z_{i}-z \over \sqrt{2}\sigma(z)}\right)\right]
$$

We now propose to study how a tomographic analysis is affected 
when considering different splitting schemes.
This will permit to single out the optimal splitting choices which guarantee
a signal to noise ratio, $snr$, high enough to distinguish among different 
DE models.

As we are interested in high $z$, we model the overall distribution $n(z)$ 
according to (\ref{galaxies}) 
with $m=2$, $\beta=2.2$ and $z_0=1.62$.
These values provide a good fit of the quasar distribution from SDSS DR6
considered in the previous section \citep{xia09}. We assume the shape of 
such a distribution to be approximately the same as for that expected 
from future surveys.   

%%%%%%%%%%%%%%%%%%%%%%%%%%%%%%%%%%%%%%%%%%%%%%%
\begin{table}[b]
\begin{center}
\begin{tabular}{lccc}
\\
\hline 
quasar bias  & magnification bias & {\it snr}  \\
\hline
%\\
  $b=b(\bar z)$ & no  & 4.56 \\
  $b=b(z)$  & no  & 4.31 \\
  $b=b(\bar z)$ & yes & 4.61 \\
  $b=b(z)$  & yes & 4.38 \\
\hline
\end{tabular}
\caption{Effects of ignoring galaxy bias evolution and magnification bias
on the {\it snr} for $\Lambda$CDM. 
$\bar z$ is the mean redshift of quasar distribution
used (see text).}
\label{mb}
\begin{tabular}{lccc} 
\\  
\hline 
 &quasar bias  &  magnification bias \\
\hline
%\\
 a)&$b_i=b({\bar z_i})$ & no  \\
 b)&$b_i=b(z)$  & no  \\
 c)&$b_i=b({\bar z_i})$ & yes  \\
 d)&$b_i=b({\bar z})$ & yes  \\
 e)&$b_i=b(z)$  & yes \\
\hline
\end{tabular}
\caption{Cases considered in order to discuss the effect of ignoring
galaxy bias evolution and magnification bias on $snr$ when dealing
with tomography. $\bar z_i$ and $\bar z$ are the mean redshift of the
$i$--th bin and of the overall distribution.}
\label{mb2}
\end{center}
\end{table}

\begin{figure*}[t]
\begin{center} 
\includegraphics[scale=0.3, angle=-90]{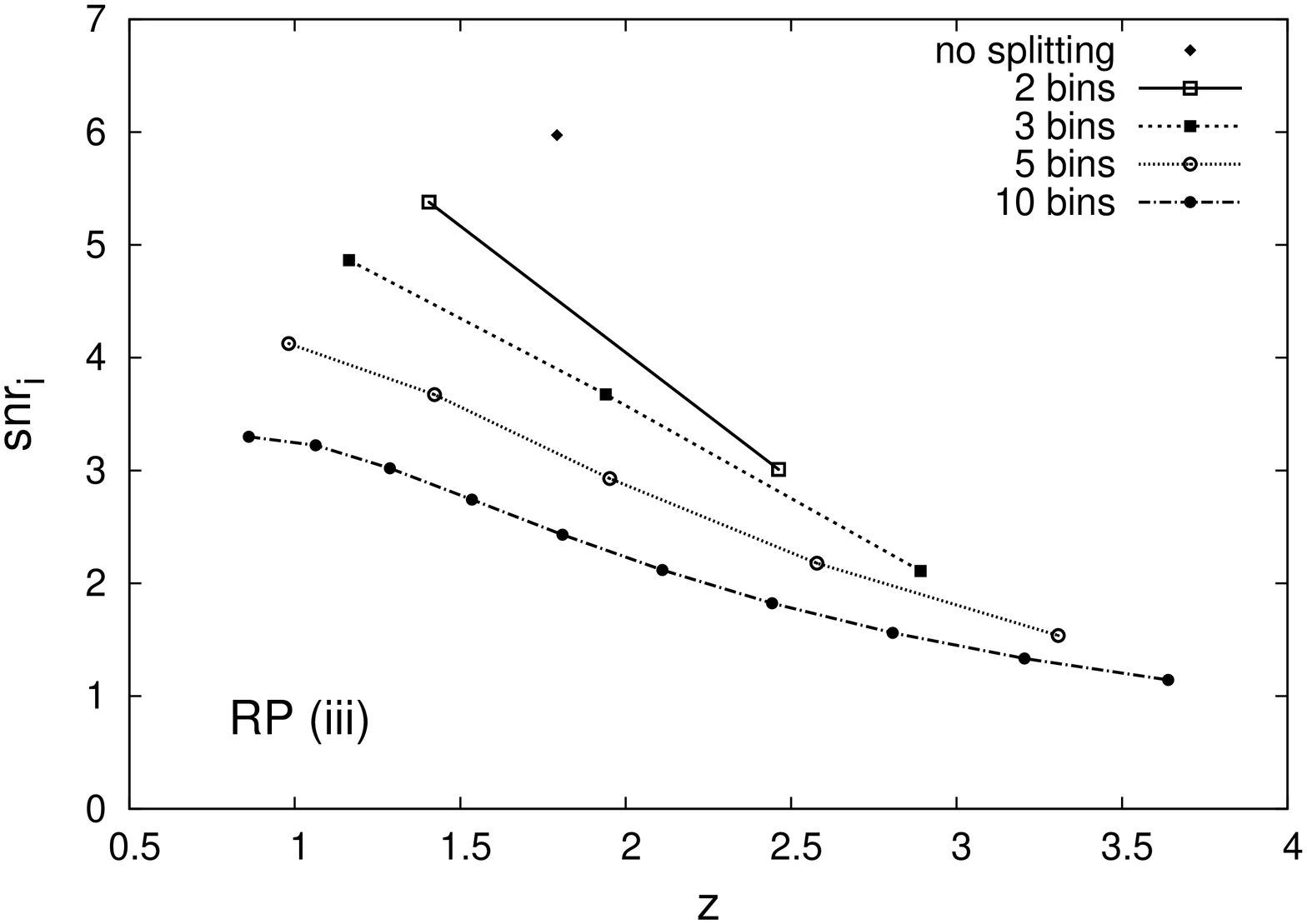}
\includegraphics[scale=0.3, angle=-90]{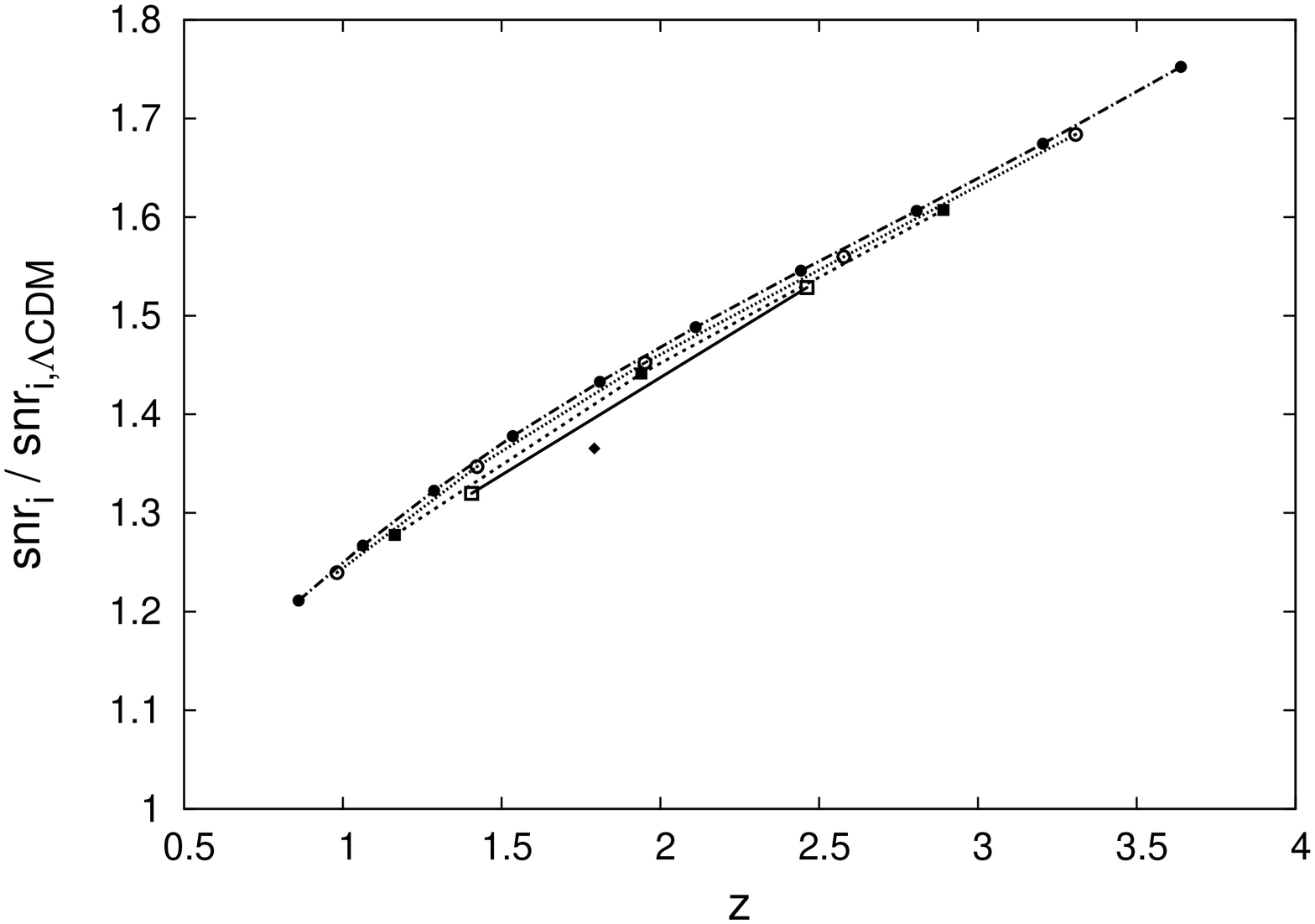}
\caption{Left panel: expected $snr$ values for the best fit RP model and splitting scheme (iii) with 2, 3,5 10 bins. $snr$ obtained by using the full
distribution is also shown. Right panel: $snr$ values of the left panel normalized by those expected in the $\Lambda$CDM case.}
\label{snr3}
\end{center}
\end{figure*}

%\begin{figure}[t]
%\begin{center}
%\includegraphics[scale=0.3, angle=-90]{snbin2.ps}
%\caption{}
%\label{snr3}
%\end{center}
%\end{figure}

Further, we assume $\sigma(z)=0.03(1+z)$ as expected from future 
experiments, and consider three different splitting schemes, 
each with $5$ bins in the redshift range from $z=0.75$ to $z=4$:

(i) bins equally spaced in $z$, with $\Delta z=z_i-z_{i-1}=0.65$;

(ii) same number of galaxies, $\Delta n$, in each bin;

(iii) bin sizes increasing proportionally to the photometric error, 
$\Delta z \propto \sigma(z)$.

The three splitting schemes are shown in  Fig. \ref{f13}. 
The thick line is the overall quasar distribution while 
the other curves are the true (spectroscopic) distributions that correspond to 
the divisions (vertical lines) in photo-z space.

We also take into account magnification bias effect which can
be important at higher $z$. In a recent analysis by \citet{ho08},
the slope of the quasar counts, $\alpha$, entering in (\ref{magb}), was found 
to be redshift dependent. They found $\alpha=0.82$ in the photometric 
redshift range $0.65<z_{photo}<1.45$ and $\alpha=0.9$ for $1.45<z_{photo}<2$.
For simplicity, we assume a constant slope $\alpha=0.9$ as in \citet{xia09}.

\subsection{Dependence on galaxy bias evolution and magnification bias}

Before comparing the different models we discuss the effects of ignoring 
magnification bias and quasar bias evolution. This is done for
the $\Lambda$CDM cosmology. Results, however, are valid for RP and SUGRA 
as well.

Notice that implications of galaxy bias evolution
on ISW measurements and parameter estimation has been considered in a pioneer work of
Schaefer et al 2009.

For the quasar bias evolution in $\Lambda$CDM, we use the empirical formula 
derived by 
\citet{cro04}: 
\begin{equation}
b(z)=0.53+0.289(1+z)^2
\label{biasfit}
\end{equation}
which provides a good fit of 
the recent observational findings by \citet{xia09}.

We first consider the overall quasar distribution and calculate the expected
signal--to--noise ratio, {\it snr}, of the cross--correlation in the following 
cases: constant quasar bias
$b=b(\bar z)$ (where $\bar z$ is the mean redshift of the survey) and  
$b=b(z)$ as given by (\ref{biasfit}). For each of them, the {\it snr} is obtain
by neglecting or considering the magnification bias correction. 
Results are summarized in table \ref{mb}. 

For Gaussian fileds, the expected $snr$ is given by (see e.g. \cite{coo02}):
\begin{eqnarray}
\nonumber
&&(snr)^2= 
f_{sky} \sum_{l_{min}}^{l_{max}} \left(2l+1\right) \times
\\
\nonumber
&&{\left(C_l^{ISW-gal}\right)^2\over
\left(C_l^{ISW-gal}\right)^2+\left(C_l^{ISW}+C_l^{N_{ISW}}\right)
\left(C_l^{gal}+C_l^{N_{gal}}\right)}\\
\label{stn}
\end{eqnarray}
where $C_l^{N_{ISW}}=C_l^T+C_l^{det}$ 
is the noise contribution to the ISW ($C_l^T$ and $ C_l^{det}$ being the total
anisotropy contribution and any detector noise contribution, 
negligible on the scales we are interested in) while
$C_l^{N_{gal}}=1/N$, is the shot noise associated with the galaxy/quasar 
catalog
($N$ is the surface density of galaxies/quasars per steradians). $f_{sky}$
is the fraction of sky common to CMB and galaxy/quasar survey maps. 

Note that (\ref{stn}) is strictly true only in the case of Gaussian fields
and full--sky coverage, $f_{sky}=1$. 
In this section, we assume that to be
the case other than a negligible $C_l^{N_{gal}}$. 
These assumptions will be relaxed in the following. Anyway, for partial 
sky coverage, one can, in first approximation, multiply for $f_{sky}$ 
the values of $snr$ here presented. 

The value of the lowest multipole, $l_{min}$,
can be approximately set to $l_{min}=\pi / 2f_{sky}$ in order to account for
the loss of low multipole modes. Anyway, although a significant 
part of the ISW signal comes from lower multipoles, we set $l_{min}=10$ 
in order 
to avoid effects of gauge correction on very large scales recently discussed by
\cite{yoo09} even if this implies a reduction of $snr$. The maximum 
multipole, $l_{max}$, is set to $l_{max}=1000$. However, we will show later
that the contribution to $snr$ from $l>400$ multipoles is 
negligible.

From Table \ref{mb}, it follows that cross--correlation measurements are 
more affected by errors when ignoring the quasar bias evolution rather 
than the magnification bias correction. In fact, in this last case,
an error of $\sim 1.6\%$ on {\it snr} is 
obtained while in this last case while
the error raises 
to $\sim 7\%$ if the quasar bias is approximated by a constat value, 
$b=b({\bar z})$ ($\bar z$ being the mean redshift of the quasar distribution)
 and it is of $\sim 4\%$ if both quasar bias evolution 
and magnification bias are neglected.

We now turn to tomography. 
According to (\ref{stn}), the signal to noise 
ratios, $snr_i$, in the {\it i}--th tomographic bin
are calculated for each of the cases listed in Table \ref{mb2}.
Fig. \ref{snrbm} shows the errors 
$\Delta snr_i$ relative to the case $e)$ for each bin of the three 
splitting schemes described above. Unlike the overall quasar distribution, 
in each bin, distributions are very narrow and
the quasar bias can be approximated by a constant, $b_i=b(\bar z_i)$ 
($\bar z_i$ being the mean redshift of the $i$--th bin),
leading to only a minor error $\lesssim 2\%$ on $snr_i$ (case $c)$).
On the other hand, ignoring magnification bias correction might be critical 
at high $z$ ($\Delta snr_i/snr_i \sim 7.5 \%$, case $b)$) and if a 
constant bias, $b=b(\bar z)$, is used for all bins, the error can 
reach $\sim 10 \%$ at higher redshifts.  

\begin{figure*}[t]
\begin{center}
\includegraphics[scale=0.3, angle=-90]{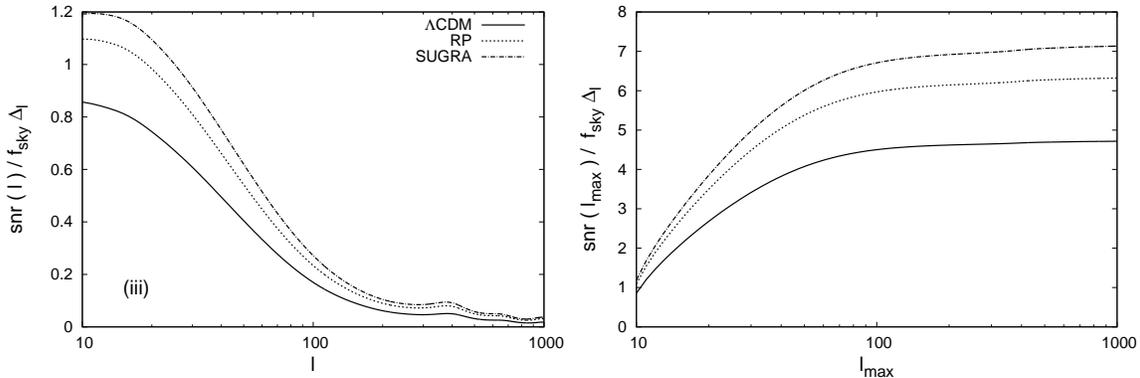}
\caption{Cumulative $snr$ from all the bins of the splitting scheme
  (iii) for $\Lambda$CDM, RP and SUGRA models. Left panel:
  contribution to $snr$ from each multipole $l$.  Right panel:
  contribution to $snr$ up to $l=l_{max}$. }
\label{snrl}
\end{center}
\end{figure*}
  
\begin{figure*}[t]
\begin{center}
\includegraphics[scale=0.24, angle=-90]{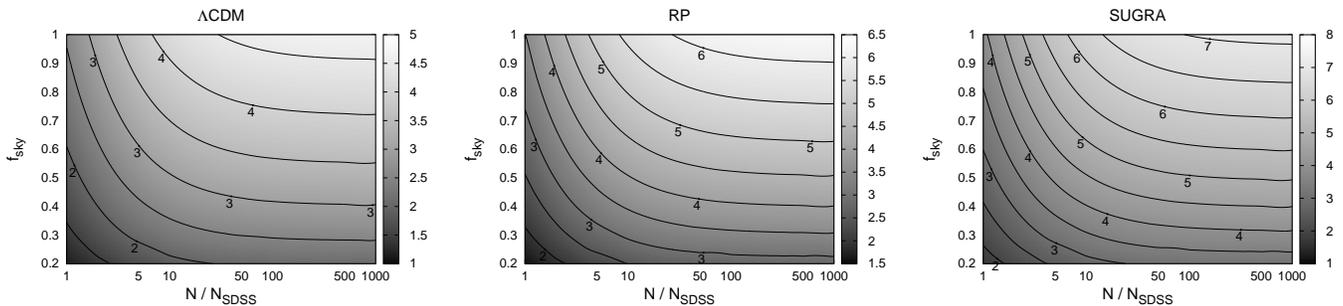}
\caption{Cumulative $snr/\Delta l$ contour levels in the plane
  $f_{sky}-N/N_{SDSS}$. Here $N_{SDSS}$ is the number density of
  quasars from the SDSS DR6 survey.}
\label{snrcontour}
\end{center}
\end{figure*}

However, since photometric redshift errors cause the bins to overlap and 
magnification of the galaxies at $z_i$ probes structures at $z<z_i$, 
cross--correlation measurements at high $z$ are quite correlated with those 
at low redshifts.
Taking into account such correlations, the net accumulated
$snr$ for measurements from all the bins up to $z_{max}$ is given by:
\begin{widetext}
\begin{equation}
\hfill
(snr(z_{max}))^2=
\sum_{z_i,z_j<z_{max}}\sum_{l_{min}}^{l_{max}}C_l^{ISW-gal_i}
\left[Cov_l^{-1}\right]_{ij}C_{l}^{ISW-gal_j}
\label{snrcov}
\end{equation}
\\
\begin{equation}
\hfill
\left[Cov_l\right]_{ij} ={ 
{C_{l}^{ISW-gal_i}C_{l}^{ISW-gal_j}+\left(C_l^{ISW}+C_l^{N_{ISW}}\right)
\left(C_l^{gal_i}C_l^{gal_j}+\delta_{ij}C_l^{N_{gal_i}}\right)
\over \left[f_{sky} \left(2l+1\right)\right]^{-1} }}
\label{cov}
\end{equation}
\\ 
\end{widetext}

The relative error on $snr(z_{max})$ is shown in Fig. \ref{snrbmc} for
the same cases as in Fig. \ref{snrbm}. In all cases, despite the
significant differences in $snr_i$ in high $z$ bins, errors are always 
$< 2\%$. This is understood because most of the cumulative {\it snr}
comes from low redshifts. 
However, in the case $e)$, errors considerably increase if the shot
noise associated to the catalog is non--negligible.   
Fig. \ref{snrbmnoise} compares the relative
errors on $snr_i$ (top panel) and $snr(z_{max})$ (bottom panel) in the
case of negligible shot noise and assuming a SDSS DR6--like survey
with a number density of quasars, $N$, of $\sim 120/deg^2$. In this last
case, the error on the cumulative $snr$ ranges from $20\%$, 
to $40\%$, depending on the number of bins considered
while the error on $snr_i$ can also reach the $50\%$ in bins at higher
redshift.

\subsection{Model comparison}

In the previous section we have shown that when dealing with narrow
tomographic bins, we can approximate the galaxy bias in each bin with  
a constant committing an error of no more than a few percent.

Then, after calculating the cross--correlation 
power spectra  for the best fit $\Lambda$CDM, SUGRA and RP models 
we apply  (\ref{rescalmag}) to each bin of the three splitting
schemes. The rescaled biases  $b(\bar z_{i})_{RP}$ and  $b(\bar z_i)_{SUGRA}$ 
are then fitted with an expression similar to (\ref{biasfit})
obtaining:
$$
b(z)=0.54+0.291(1+z)^2
$$
for RP and
$$
b(z)=0.54+0.287(1+z)^2
$$
 for SUGRA. These expressions, which will be used in the rest of the paper, 
are valid up to $z=4$ leading a galaxy
bias evolution only slightly different from that of $\Lambda$CDM.
 
In Fig. \ref{f12}, we show the cross--correlation power spectra 
for the splitting (i). 
Lower frames of each panel display the ratio between the 
SUGRA and RP spectra and the $\Lambda$CDM spectrum. 
Mean redshifts of the true bin distributions are also indicated.
A similar redshift evolution could be obtained by considering (ii) or (iii). 
As clearly visible from the figure,
a better discrimination among the models is expected at higher redshifts.

The expected $snr_i$ and $snr(z_{max})$ for the different splitting
schemes  are shown in Figs. \ref{snr} and \ref{snrc}
respectively, for the cosmologies considered. 
Despite the same qualitative behavior for all the models, 
higher $snr$ are expected in the SUGRA case; RP being in the between of 
SUGRA and $\Lambda$CDM. Horizontal lines in Fig. \ref{snrc} indicate the
$snr$ obtained by using the full distribution.

For each splitting scheme, in Figs. \ref{snr2}, we plot the ratio between 
the RP and SUGRA $snr_i$ and $snr(z_{max})$ values and those expected
in the $\Lambda$CDM case. 
A first thing to note is the overlapping, for both 
RP and SUGRA, of the three curves corresponding to the different 
splittings indicating that the three schemes considered 
perform equally well in discriminating among the models. 
A better discrimination is however achieved looking at high redshifts.

In Fig. \ref{snr3}, we investigate how $snr_i$ changes when increasing 
the number of the bins (left panel) and whether a greater number of bins 
could permit to better discriminate between models (right panel). 
Results are shown for the splitting scheme (iii) with
2,3,5,10 bins in the RP model. Similar results are obtained in the other
cases. $snr$ in the case of no splitting is also shown.
More bins, in principle, would permit to have a more detailed 
description of the redshift evolution of ISW effect. However, as clearly 
visible in the left panel, the $snr$ decreases at the increasing of the 
bin number.
In the right panel, RP model is compared to
$\Lambda$CDM. Even though, at high redshifts, tomography permits to 
distinguish among the models better than using the full distribution,
the figure shows that increasing the number of the 
bins from 2 to 10 would permit only a minor improvement in 
discriminating between the models.

\subsection{Sky coverage and shot noise}

Up to now, we have considered the ideal case of Gaussian fields,  full sky coverage and
negligible shot noise. However, after cutting out our galaxy from the analysis, future
CMB and galaxy maps are expected to cover,  at  best, a sky fraction $f_{sky}=0.7-0.8$.
In this case different multipoles are no longer independent and
(\ref{stn}), (\ref{snrcov}) and (\ref{cov}) only provide approximated
estimations and a more rigorous analysis taking into account for the
effective survey geometry is needed (\cite{cab07}, \cite{hiv02},
\cite{xia11}). 
It has been however 
shown that, under the above approximations, a better estimation 
can be obtained by binning the power spectra data in bins of
appropriate size $\Delta l$ making the bins independent. In this case, (\ref{stn}) and
(\ref{snrcov}) are increased by a multiplicative factor of $\Delta l$.
 \cite{cab07} found that $\Delta l = 20, 16, 8,
1$ works well for $f_{sky}=0.1, 0.2, 0.4, 0.8$.

In Figs. \ref{snrl} and \ref{snrcontour} we show some results  for
the cumulative $snr$  from all the bins of the splitting
(iii). Results are however the same for the other schemes. 

The left panel of Fig. \ref{snrl} shows the contribution to the 
cumulative $snr$ from each multipole while the cumulative $snr$ up to $l=l_{max}$ is
displayed in the right panel. The dependence on $f_{sky}$ and $\Delta
l$ has been removed. 
As clearly visible, most of the
cross-correlation signal comes from lower multipoles and contributions
from $l>400$ are negligible. 

\begin{figure}[t]
\begin{center}
\includegraphics[scale=0.3, angle=-90]{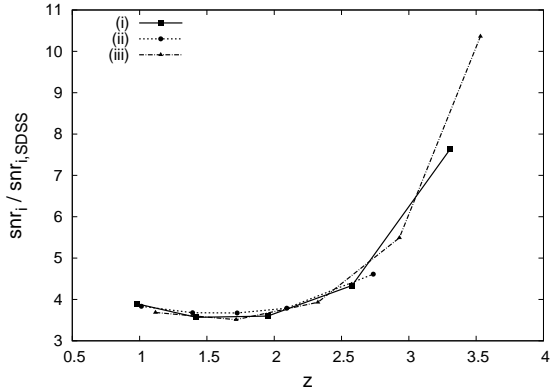}
\caption{Comparison between $snr_i$ expected from a survey with
  $f_{sky}=0.8$ and negligible shot noise and the present one. The
  three splitting schemes are considered.}
\label{snfc}
\end{center}
\end{figure}

In Fig. \ref{snrcontour} cumulative $snr$ contour levels are plotted
in the plane $f_{sky} - N/N_{SDSS}$ where $N_{SDSS}$ is the quasar
number density for a SDSS DR6-like survey. Given that such a survey
cover $\sim 20\%$ of the sky, future experiments covering a sky
fraction $f_{sky}=0.8$ will increase the cumulative $snr$ of a factor $\sim 3$
if the shot noise is reduced by $1/10$  and a factor $\sim 3.5-4$ in
the case $N=100 N_{SDSS}$.  No significant improvement is obtained
by further reducing the shot noise. 
For $f_{sky}=0.8$  and negligible shot noise, the increasing of 
$snr$ in the $i$--th tomographic bin can range from a factor of $4$ (low
$z$ bins)  up to $10$ (high $z$ bins). This is shown in Fig. \ref{snfc}.

\section{Conclusions}

In this work we have investigated ISW-LSS cross-correlation in coupled Dark Energy models with massive neutrinos.
The presence of a coupling between DM and DE as well as massive neutrinos 
change both the background and matter perturbation evolutions yielding,  
unlike the $\Lambda$CDM case, a time--variation of the gravitational potentials  
even during the matter domination.  A significant ISW signal is thus expected
also at high redshifts.

Firstly, we have investigated the dependence on the energetic scale, 
$\Lambda$, of the DE potential, the coupling strength $\beta$ and the neutrino 
mass $m_\nu$.
We considered first the uncoupled case ($\beta=0$) and massless neutrinos.
We found that, when increasing $\Lambda$, both
$C_l^{ISW-m}$ and $C^{ISW-m}(\theta)$ show an opposite behavior at low 
and high redshifts. This in fact reflects the behavior of 
$\dot\Phi + \dot\Psi$.
In the presence of coupling one can distinguish between two different behaviors
for small and large $\Lambda$'s.  In the first case, the evolution of the gravitational 
potentials and the cross--correlation signal are almost independent from 
$\Lambda$. It can be understood noticing that for small $\Lambda$, 
coupling terms in the DE field equations dominate so that  its solution 
is almost independent from $\Lambda$. 
When increasing $\Lambda$, the behavior resemble that of the uncoupled case.
Dependence on $\beta$ was also investigated and, again, the 
behavior of the cross--correlation at different redshifts reflects that of the 
ISW source. However, while coupling can affect $C_l^{ISW-m}$ 
(and $C^{ISW-m}(\theta)$) 
in an opposite fashion at high and low redshifts, massive neutrinos always 
decrease the cross--correlation signal.

Secondly, we have provided a simple expression, eq. (\ref{rescal}),
 which permits to appropriately
rescale the galaxy bias when comparing different cosmologies once the bias 
of a particular model, e.g. $\Lambda$CDM , is known and the normalization 
of the power spectrum in each model is fixed. 
We also give, a generalized version of
(\ref{rescal}) to the case when the magnification bias effect due to 
gravitational lensing is non--negligible (see eq. (\ref{rescalmag})).

Then, we compare the theoretical prediction on the cross--correlation function
for our models with the observational data obtained for six different galaxy
catalogues by \citet{gia08a}. 
We found that, while it is not possible to distinguish among the models at low redshifts,
discrepancies between coupled models and $\Lambda$CDM increase with $z$
even though RP and SUGRA models remain indistinguishable. 
In spite of this, however, current data alone seem not able to discriminate 
between coupled models and $\Lambda$CDM. 

Finally, we studied the redshift tomography.
Upcoming galaxy surveys will cover a large redshift range also providing
photometric redshifts of the galaxies with high accuracy. This will permit to 
split a survey into multiple photometric redshift bins allowing for tomographic analysis.
Here, we were  interested to study how a tomographic analysis of the ISW--LSS
cross--correlation is affected when considering different splitting  schemes and assuming photometric redshift errors as expected from future experiments. 
As we were interested in high redshifts, where our models, unlike the 
$\Lambda$CDM case, are expected to provide a significant ISW effect signal,
ISW effect was cross--correlated with quasars.
The quasar distribution was thus split in tomographic bins
according to three different schemes: (i) bins equally spaced in $z$;
(ii) same number of galaxies, in each bin;
(iii) bin sizes increasing proportionally to the photometric error. 
Cross-correlation  were then calculated in each bin.   

Our tomographic study was based on a signal--to--noise analysis.
We started our discussion investigating
the effect on $snr$ of ignoring the quasar bias evolution and
magnification bias correction for an ideal survey. 
We found that, if the overall quasar distribution is used,
 cross--correlation measurements are 
more affected by errors when ignoring the quasar bias evolution
($\Delta snr /snr \sim 7 \%$) rather 
than the magnification bias correction ($\Delta snr /snr \sim 1.6 \%$)

However, when
dealing with tomography the error on $snr_i$ ($i$ indicating the
$i$--th bin) never overcome $\sim 2.5 \%$ if the
quasar bias in each bin is approximated with an appropriated constant, 
but it can reach
$\sim 7.5 \%$, at high redshifts, when magnification bias is ignored.
Errors on the cumulative $snr$, however, always stay below the $2\%$.
On the other hand errors can increase up to $50 \%$ if the shot noise
associated to the quasar survey is set to the current values.

We then used  the tomographic analysis in order to compare different
cosmologies.
 We found that the above splitting schemes,  perform equally well in
discriminating among the models. 
A better discrimination is however achieved looking at high redshifts.

We also investigated how the expected signal to noise ratio, 
$snr$, of the cross--correlation changes when increasing 
the number of the bins and whether a greater number of bins 
could permit to better discriminate between models. Even though more bins
would allow to have more information on the redshift evolution of the ISW 
effect, the $snr$ decreases at the increasing of the 
bin number.
As a consequence, although tomography, at high redshifts, would permit
to distinguish among the models better than using the full
distribution, when comparing our models to
$\Lambda$CDM it was shown that increasing the number of the 
bins from 2 to 10 would permit only a minor improvement in the
discrimination.

Finally, we showed that future wide field surveys ($f_{sky} \sim 0.8$) can
increase the cumulative $snr$ of the cross--correlation of a factor $\sim
3$ ($3.5-4$) if the current shot noise is reduced by $1/10$ ($1/100$)
while the $snr$ of the single bins can increase up to a factor $10$ at
high redshift.

Our $snr$ analysis suggest a discrimination power of 
future ISW--LSS cross-correlation measurements able
to distinguish among different cosmologies.
However, in order to assess the discrimination, more rigorous 
analysis in terms of Fisher Matrix and Monte
Carlo Markov Chain are needed. 
They are currently 
under investigation and left for future works.

\section*{Acknowledgments} 
Tommaso Giannantonio is gratefully thanked for providing observational
data on cross--correlation and useful hints.
 DFM thanks the Research Council of Norway FRINAT grant 197251/V30 and the Abel extraordinary chair UCM-EEA-ABEL-03-2010.
DFM is also partially supported by the projects CERN/FP/109381/2009 and
PTDC/FIS/102742/2008.
\vfill
\eject

\end{document}